\documentclass[aps,prd,reprint,onecolumn,nofootinbib,superscriptaddress,amsmath,amssymb,notitlepage,longbibliography]{revtex4-1}
\usepackage{graphicx}
\usepackage{footmisc}

\newcommand{\met}{m_\eta}
\newcommand{\mpp}{m_{\pi^\pm}}
\newcommand{\mpn}{m_{\pi^0}}
\newcommand{\mpi}{m_\pi}
\newcommand{\M}{{\cal M}}

\DeclareMathOperator{\im}{Im}
\DeclareMathOperator{\re}{Re}
\DeclareMathOperator{\sign}{sign}
\DeclareMathOperator{\disc}{disc}

\newcommand{\Han}{\stackrel{\,n}{\mathcal{H}}}
\newcommand{\Ha}[1]{\stackrel{#1}{\mathcal{H}}}
\newcommand{\han}{\stackrel{\,n}{h}}

\makeatletter
\newcommand{\rim}[1]{\romannumeral #1}
\newcommand{\Rim}[1]{\expandafter\@slowromancap\romannumeral #1@}
\newcommand{\mRim}[1]{\mathrm{\Rim{#1}}}
\newcommand{\mRima}[2]{\mathrm{\Rim{#1}#2}}
\makeatother

\allowdisplaybreaks[3]

\begin{document}

\title{Analytical dispersive construction of \texorpdfstring{$\mathbf{\eta\to3\pi}$}{eta -> 3 pi} amplitude: First order in isospin breaking}

\author{Karol Kampf}
 \email{karol.kampf@thep.lu.se}
\affiliation{Department of Astronomy and Theoretical Physics, Lund University, S\"olvegatan 14A, SE 223-62 Lund, Sweden.}
\affiliation{Institute of Particle and Nuclear Physics, Faculty of Mathematics and Physics, Charles University,
V Hole\v{s}ovi\v{c}k\'ach 2, Prague, Czech Republic}

\author{Marc Knecht}
 \email{marc.knecht@cpt.univ-mrs.fr}
 \affiliation{Centre de Physique Th\'eorique, CNRS-Luminy, Case 907, F-13288 Marseille Cedex 9, France}
\thanks{Unit\'e Mixte de Recherche  (UMR 6207) du CNRS  et des Universit\'es Aix--Marseille 1 et 2 et Sud Toulon--Var, Laboratoire affili\'e \`a la FRUMAM (FR 2291)}
\author{Ji\v{r}\'{i} Novotn{\'y}}
 \email{novotny@ipnp.troja.mff.cuni.cz}
\affiliation{Institute of Particle and Nuclear Physics, Faculty of Mathematics and Physics, Charles University,
V Hole\v{s}ovi\v{c}k\'ach 2, Prague, Czech Republic}

\author{Martin Zdr{\'a}hal}
 \email{zdrahal@ipnp.troja.mff.cuni.cz}
\affiliation{Institute of Particle and Nuclear Physics, Faculty of Mathematics and Physics, Charles University,
V Hole\v{s}ovi\v{c}k\'ach 2, Prague, Czech Republic}

%\date{\today}

\begin{abstract}
Because of their small electromagnetic corrections, the isospin-breaking decays $\eta\to3\pi$ seem to be good candidates for extracting isospin-breaking parameters \mbox{$\sim(m_d-m_u)$}.
This task is unfortunately complicated by large chiral corrections and the discrepancy between the experimentally measured values of the Dalitz parameters describing the energy dependence of the amplitudes of these decays and those predicted from chiral perturbation theory. We present two methods based on an analytic dispersive representation that use the information from the NNLO chiral result and the one from the measurement of the charged $\eta\to3\pi$ decay by KLOE together in a harmonized way in order to determine the value of the quark mass ratio $R$. Our final result is $R=37.7\pm 2.2$. This value supplemented by values of $m_s/\widehat m$ or even $\hat{m}$ and $m_s$ from other methods (as sum-rules or lattice) enables us to obtain further quark mass characteristics. For instance the recent lattice value for $m_s/\widehat m\sim 27.5$ leads to $Q= 23.1\pm0.7$. We also quote the corresponding values of the current masses $m_u$ and $m_d$.
\end{abstract}

%\pacs{11.55.Fv, 12.39.Fe, 13.25.Jx, 14.65.Bt}

\maketitle

%%%%%%%%%%%%%%%%%%%%%%%%%%%%%%%%%%%%%%%%%%%%%%%%%%%%%%%%%%%%%%%%%%%%%%%%%%%%
\section{Introduction}
The masses of the light quarks are fundamental free parameters of the standard
model.
Since quarks are confined inside hadrons, there is no direct method for their
measurement.
The only method of determination is a comparison of the theoretical
prediction for some observable that depends on the quark masses with the
corresponding experimental value. For that end we need a framework, in which the
quark masses occur explicitly, and which can make predictions for such
observables with sufficient precision. Because of quark confinement and the
fact that these masses are very small in comparison to the typical hadron
scales, perturbative quantum chromodynamics (QCD) cannot play such a role
and we need to employ a non-perturbative method. Nowadays, the natural
candidates for such approaches are lattice QCD \cite{Wilson:1974sk} and chiral
perturbation theory (ChPT) \cite{Weinberg:1978kz,GL-scatlen,Gasser:1984gg}.

While the mass $m_s$ and the isospin average mass of $m_u$ and $m_d$, which is
denoted as
\begin{equation}\label{hatm}
\hat{m}=\frac{m_u+m_d}{2}\,,
\end{equation}
have become accessible through the recent lattice simulations (among others \cite{RBC-UKQCD,MILC,HPQCD,PACS-CS}) and agree well with the independent determination via QCD sum rules \cite{Maltman:2001nx,Narison:2005ny,Dominguez:2008jz}, the extraction of the individual masses $m_u$ and $m_d$ from the lattice is still polluted by various
simplifications of the electromagnetic effects that have to be made in
isospin-breaking simulations (cf.\ e.g.\ \cite{Colangelo:2010et}). Therefore, if
we want to determine the individual masses $m_u$ and $m_d$, at the moment ChPT seems to be
the more promising approach.

The most suitable processes for studies of isospin breaking within ChPT in the mesonic sector are
$\eta\to3\pi$ decays. Since these decays violate G-parity,\footnote{
Or equivalently, the decay would be forbidden as a result of isospin conservation and charge conjugation invariance (C-invariance). Indeed, the final state has to have the total isospin $I=0$ and is therefore totally antisymmetric with respect to the permutation of the three pions (the only allowed state is then $\pi^+\pi^-\pi^0$). Due to Bose symmetry, the corresponding amplitude is then totally antisymmetric under exchanges of the momenta of these three pions. On the other hand, according to C-invariance, the amplitude has to be symmetric with respect to the exchange of the momenta of the $\pi^+$ and the $\pi^-$, which implies that the amplitude is zero.} they have to proceed via isospin-breaking effects. There are two mechanisms of this breaking, either through the
electromagnetic (EM) interactions,
which are proportional to the electric charge squared,
or through the isospin-breaking mass difference between $u$ and $d$ quarks,
\begin{equation}\label{eta hamiltonian QCD}
\mathcal{H}^{IB}_{\mathrm{QCD}}(x)=\frac{m_d-m_u}{2}\left(\bar{d}(x)d(x)-\bar{u}
(x)u(x)\right).
\end{equation}

Even though the EM interactions have a sizable effect on the difference $\mpp-\mpn$
and on the pion decay constant $F_\pi$, it has turned out that their influence
on the $\eta\to3\pi$ decay amplitudes is very small
\cite{Sutherland:1966zz,Baur:1995gc,Ditsche:2008cq}. Hence,
$\mathcal{H}^{IB}_{\mathrm{QCD}}$ represents the dominant contribution and the
amplitude is proportional to $m_u-m_d$, which is usually presented in one of the
following ratios
\begin{equation}\label{RQ}
R = \frac{m_s - \hat m}{m_d-m_u}, \qquad
\frac{1}{Q^2}=\frac{m_d^2-m_u^2}{m_s^2-\hat{m}^2}
\end{equation}
that are connected via ($r=\frac{m_s}{\hat{m}}$)
\begin{equation}\label{RQ relace}
Q^2=\frac{1}{2}R(r+1).
\end{equation}

Consequently, a measurement of the decay rates of the $\eta\to 3\pi$ processes enables us a direct access to this difference (and by the use of the values $\hat{m}$ and $m_s$
from the lattice also to the individual masses of these two lightest quarks).
Of course, in order for this extraction to be possible, it is necessary to have ChPT predictions for these decay
rates with a sufficient degree of accuracy.

Achieving this is, however, a non-trivial task. The tree-level predictions,
which are equivalent to the PCAC results
(e.g.~\cite{Bell:1996mi,Cronin:1967jq,Osborn:1970nn}), would indicate a very
large difference between $m_u$ and $m_d$. Furthermore, the true energy dependence of
the amplitudes is definitely different from the trivial one that PCAC proposes. The
sizable one-loop corrections \cite{Gasser:1984pr} were still not sufficient to
correct these discrepancies. At last, the inclusion of the two-loop corrections
\cite{BG}, which are also sizable, led to the predictions for both $R$ and the Dalitz
parameters describing the energy dependence of the amplitudes
(cf.~Tables~\ref{tabulka Dalitz etac} and \ref{tabulka Dalitz eta0} below) that
were in reasonable agreement with expectations.

Nevertheless, if we study these results in greater detail, we find some hints
that the \mbox{2-loop} ChPT determination of $R$, which we are interested in, can still be
inaccurate. The feature often put forward in this respect is the discrepancy between
the experimentally measured and the predicted values of the Dalitz parameters (defined
in Sec.~\ref{Dalitz plot parametrization}), mainly for the neutral parameter $\alpha$ of (\ref{dalitz eta0}).
For a better quantification of the difference between experiment and theory, let us introduce
\begin{equation}\label{chi2}
 \chi^2 \equiv \Bigl( \frac{\text{exp} - \text{theory}}{\sigma[\text{exp}]}
\Bigr)^2\,,
\end{equation}
where the theory enters only via its central value.
Using this quantity when comparing the prediction of \cite{BG}, $\alpha = 0.013$, with
the best measurement of this observable by MAMI-C \cite{MAMI-C}, $\alpha = -0.032\pm
0.003$, we obtain indeed a huge difference of $\chi^2 = 225$.
However, there is a parameter for which this discrepancy is even more apparent, namely
$b$ from (\ref{dalitz etac}). Comparing the ChPT value $b=0.394$ with the measurement by KLOE \cite{KLOE-etac} $b=0.124\pm0.012$ produces
$\chi^2\sim 500$. This raises the question about the origin of these discrepancies, and whether
and to which extent they can also affect the determination of $R$.

As was already stressed in \cite{BG}, the explanation of this difference between theory and experiment
 can be provided by the large theoretical error bars presented there (thereby making the theoretical and the experimental values compatible).
The non-renormalizability of ChPT represents a major drawback of this theoretical framework when it becomes necessary to include higher and higher orders. Indeed, including two-loop effects to the computation means a rapid
increase\footnote{Note that $O(p^6)$ ChPT contains 102 (2+10+90) free
LECs. In order to make this theory at the given order predictive, we would thus
need to make at least 102 measurements determining these constants. Obviously,
not all of these constants appear in a given amplitude --- only a
subset of them contributes to $\eta\to3\pi$, see below (Sec.~\ref{sec:chpt}).} of the number of \emph{a priori}
unknown low energy constants (LECs) that have to be estimated before we can
get a reliable prediction. We are far from a determination of all required
LECs from experiment (or lattice), and hence for many of them we have to rely
on some estimates, predominantly of the resonance saturation type 
\cite{Ecker:1988te,Ecker:1989yg,Moussallam:1994at,Knecht:2001xc}.
This brings an unknown error into the game --- the error presented in \cite{BG} is
an estimate by the authors obtained by taking the uncertainty of the amplitudes
equal to one half of the two-loop contributions.

Both the Dalitz plot parameter discrepancy and this drawback of ChPT
affecting the predictivity of the chiral computation contributed to the development of
alternative approaches, among others the dispersive methods \cite{KWW,Anisovich:1996tx,Colangelo:2009db}
and the non-relativistic effective field theory (NREFT)
\cite{Colangelo:2006va,Gasser:2011ju,Bissegger:2007yq,Gullstrom:2008sy,Schneider:2010hs}.
In order to understand their relative advantages and disadvantages, let us recall a few basic properties they share. All these approaches are constructed as effective field theories that on the basis of some assumptions (usually represented by some expansion of the amplitude) divide the phase-space of each amplitude into the ``low-energy part'' that is included in the computation and the ``high-energy part'' that is not known or at least less known. At tree-level one simply uses
the amplitudes only in the low-energy region and is not concerned by what lies above
the cutoff. In order to work consistently one needs to introduce a mechanism
that picks up the contributions that contribute with the same importance to a
given order, usually represented by a power-counting. Then,
when computing the amplitude to the higher order, one needs to include also loop
contributions (either by means of taking into account loop Feynman diagrams, as
a unitarity contribution, or by any other method), where one has to integrate
also over the high-energy part of the intermediate amplitudes (over higher
momenta of the intermediate virtual particles). By using the ``power-counting mechanism'' or
by adding some further assumptions, part of these contributions are
considered negligible, but there always remains a part that is finite and
unknown and has to be parametrized somehow --- usually there occur new effective
parameters in the model and the old ones are renormalized or shifted. Note that in ChPT
that represents a Lagrangian effective field theory the power-counting mechanism is given by the chiral counting, which also monitors the number of LECs (effectively containing the contribution of the physics above the chiral cutoff
--- the hadronic scale) appearing at a given order.

The importance of the one-loop (and in the recent years, also of the two-loop)
$\pi\pi$ rescattering corrections has led \cite{KWW,Anisovich:1996tx,Colangelo:2009db} to abandon,
in a certain sense, strict chiral counting, instead attempting to obtain the amplitudes with two-pion rescattering
effects formally included to all orders. These approaches employ a restricted version of
unitarity (taking into account just the two-pion intermediate state), in the context of
dispersion relations, the aim being to find a numerical fixed point
solution of them. The mechanism assigning the importance to a given contribution
is therefore based on the assumption that the two-pion rescattering effects are
dominant. In the low-energy part of the amplitudes, the unitarity contribution of the physics above the threshold, where
further intermediate states contribute and where the $S$ and $P$ partial waves
of the considered amplitudes cease to be the dominant ones, are taken into account through
subtraction constants. However, in order to restrict their number to a reasonable amount, one needs to impose some
assumptions on the high-energy region (of both the physical amplitudes and of
the amplitude constructed iteratively by the numerical method). In
\cite{Anisovich:1996tx,Colangelo:2009db} these assumptions are specified by the requirement to have only
four\footnote{\label{ftnt o poctu subtrakci}Here we
classify the number of parameters appearing in the case we take masses of the
charged and the neutral pions equal, $\mpp=\mpn$. Note that from
Sec.~\ref{sec:conwithchpt} it is obvious that in two-loop ChPT results there
occur at least six independent combinations of LECs.} of them.

The methods based on the modified non-relativistic effective field theory
(NREFT)
\cite{Colangelo:2006va,Gasser:2011ju,Bissegger:2007yq,Gullstrom:2008sy,Schneider:2010hs}
implement instead of the usual chiral expansion a
combined expansion in powers of a formal non-relativistic parameter $\epsilon$ and of a
formal partial-wave $\pi\pi$ scattering-characteristics parameter $a$ (representing scattering lengths and higher threshold shape parameters). The amplitude is then computed to the two-loop order in the NREFT Lagrangian
formalism.
The power-counting scheme is therefore based on the non-relativistic expansion
together with the loop expansion (equivalent in this case to the expansion in
the pion scattering parameters). In \cite{Schneider:2010hs} the results are presented
including the orders up to $O(\epsilon^4), O(a\epsilon^5), O(a^2\epsilon^4)$, and
partially also $O(a^2\epsilon^6)$ and $O(a^2\epsilon^8)$. By assuming that the
included orders are dominant, the contribution of the intermediate states other
than the two-pion ones have to be included through four\footref{ftnt o poctu
subtrakci} parameters coming from local interaction terms.

Naturally, the reasonable question that has to be addressed in the future is
whether each set of assumptions (either of ChPT, of the dispersive approaches,
or of the modified NREFT) adequately describes the physics, and whether a possible
drawback in this respect in any of them is paid off by the other advantages it
possesses. The advantages and the disadvantages of these approaches were
nicely summarized in \cite{Kubis-talk}. We emphasize only that NREFT provides
analytic results that are easy to extent beyond the $\mpp=\mpn$ limit, while the
dispersive methods proceed numerically and their extension to full isospin
breaking was never studied. On the other hand, whereas the NREFT expansion in
$\epsilon$ is safe only inside the Dalitz region, the results of the dispersive
approaches should work also in some larger regions beyond it.
Both of them have in common that, in contrast to ChPT, they directly use the
physically measured $\pi\pi$ scattering parameters as inputs, but there remain
four\footref{ftnt o poctu subtrakci} free parameters that have to be fixed
either from matching to ChPT or to experimental data. Moreover, these decays
depend on $R$ or $Q$ merely just through the normalization, which is
factored out in both of methods. Thus, even if those
representations are fitted to experimental data, the determination of $R$ or $Q$ would
still require to match with ChPT at least at one point, thereby fixing the
normalization.

The matching is not an easy task in this context. In addition to the differences in the structures
of these results, since we are matching two different approaches with different
power-counting schemes and assumptions, we need to find the region (or as
discussed above at least one point) and the appropriate orders in both approaches in which
their results are compatible. Nevertheless, thanks to the easy
form of the one-loop ChPT amplitude, and to the fact that the physical regions of
$\eta\to3\pi$ decays are quite small, in both approaches the matching to
one-loop ChPT was obtained (cf.~\cite{Anisovich:1996tx,Colangelo:2009db} and \cite{Schneider:2010hs}).

In conclusion, in order to determine the correct value of $R$ from the $\eta\to3\pi$ decays, one cannot
avoid discussing either the accuracy of the ChPT result for the amplitude and its possible corrections (by correcting the values of the $O(p^6)$ LECs $C_i$ or by inclusion of some higher-order corrections into the ChPT calculation), or the existence of at least one point (or some region) where the current chiral result reproduces well the complete physical amplitude. For instance, the discussion of the influence of the $C_i$s on the results can be studied using directly the ChPT amplitude, but its complexity and its extreme length together with the fact that it includes
many such $C_i$s complicates such an analysis.

In \cite{kknzprep} we have worked out a method using the dispersive relations
and perturbative unitarity (i.e.~a dispersive approach) for the construction
of a representation of the $\eta\to3\pi$ decay amplitudes.
We have given up on including the $\pi\pi$ rescattering contributions to all orders, but
have instead required to obtain an analytic representation and paid care to the assumptions we
are using in the construction, thus ensuring that the ChPT result can be
obtained as a special case of our result. The method is based on very general
principles, unitarity, analyticity, crossing symmetry, and relativistic
invariance, combined with chiral counting. The fact that we require a
representation valid to two loops in the chiral counting picks up the
contributions that have to be included into the computation and tells us that at the low-energy region up to this chiral order all the other effects are taken into account effectively in terms of six subtraction
parameters\footref{ftnt o poctu subtrakci}.

The full strength (and our original motivation) of this method arises when we want to include the isospin breaking induced by the mass differences between mesons belonging to the same isomultiplet (cf.~\cite{kknzprep}). However, even in the case where we consider the leading order in the isospin breaking, for which the two-loop ChPT result is available, the representation constructed by this method can be useful. Thanks to its simple and compact analytic form, to its capabilities to include all the chiral $O(p^6)$ effects important in the kinematic decay region of $\eta\to3\pi$ into those six subtraction parameters, and to its easy correspondence to the ChPT, this representation is helpful when one is addressing the questions we have premised above, namely, whether one can obtain a reasonable agreement in the determination of the Dalitz parameters from experiment and from the NNLO ChPT amplitude with the corrected set of the $C_i$s; how such a change would influence the determination of $R$; possibly also whether there exists another simple way how to solve that disagreement.

In addition, we do not need to work only in such a close connection to the two-loop ChPT amplitude. Our representation is more general than the two-loop ChPT amplitude (simply stated in the way that the values of our parameters need not to be held at the values stemming from the ChPT), based only on the specific chiral orders of the partial waves of the amplitudes (cf.~e.g.\ \cite{nas-proceeding2}). In order to respect such chiral power-counting, we need to distinguish between various orders of our parameters. By weakening this requirement and by a simple change of their interpretation we
can perform a partial resummation that mimics a part of the previous dispersive approaches. By that we have therefore replaced the assumptions represented by the chiral counting with the assumption that the contributions we have included by this resummation are the dominant one.

In any case, such representation is suitable for fitting the experimental data. We can thus change completely the strategy and instead of trying to correct the amplitude stemming from ChPT, we use our representation as a parametrization of the data, from which we can compute the value of $R$. However, as was discussed above also in this case, we need to fix the normalization from ChPT. For that end we need to find a region where the chiral expansion of the amplitude converges fast, where the two-loop ChPT amplitude reproduces the physics well. Thanks to the form of our representation and its simple
connection to ChPT the analytic dispersive method is helpful also in this analysis, resulting with the recipe for such matching.

We want to stress that in \cite{kknzprep} the inclusion of the isospin-breaking corrections stemming from $\mpp\neq\mpn$ is presented, but as we discuss in Sec.~\ref{sec:neutral} the current experimental data do not yet allow to
perform the isospin-breaking analysis. Thanks to the planned improvement in the neutral decay measurements
(cf.~\cite{Unverzagt:2009vm}) we should add that this possibility is just behind
the
corner. In the paper we therefore work in the limit $\mpp=\mpn$ with the
exception of a few discussions of the effects appearing beyond this limit. This
discussion in full detail is however planned in our next paper \cite{kknzprep2}.

Through relation (\ref{M0Mxiso}) this limit connects the charged
$\eta\to\pi^+\pi^-\pi^0$ decay with the neutral one, $\eta\to3\pi^0$. Using our
representation on the basis of the above mentioned analyses of the charged KLOE
data \cite{KLOE-etac} (the most precise measurement of this process that exists), we can therefore determine the values of the neutral Dalitz decay parameter $\alpha$ and discuss its connection to the direct
neutral measurements (from Table~\ref{tabulka Dalitz eta0}).

The plan of our paper is as follows. After recalling some basic properties of
the amplitudes of these decays and introducing our notation in
Sec.~\ref{sec:notation}, we recall in Sec.~\ref{sec:chpt} the ChPT computation
of $\eta\to3\pi$ with the special emphasis on the contribution of $O(p^6)$ LECs
to the Dalitz plot parameters. From that analysis there follow a few combinations
of observables that should be (at least in the first approximation) safe from the
incorrect determination of these LECs. In Sec.~\ref{sec:dispersive construction}
we present the dispersive construction of our representation for the $\eta\to3\pi$ decay amplitudes. Section~\ref{sec:conwithchpt} discusses the connection between our
representation and the ChPT result. Section~\ref{sec:charged} is then devoted to
the numerical analysis of the charged decay. We start with the determination of the values of our parametrization for NNLO ChPT. Then, inspired by the result of Sec.~\ref{sec:chpt}, we study the influence of changing the $C_i$s in the NNLO ChPT amplitude in order to reproduce the charged KLOE data \cite{KLOE-etac} on the values of the physical observables we are interested in. Then in Sec.~\ref{sec:res} we perform the analysis in which the values of the dispersive parameters are set by KLOE and only the normalization is
determined from ChPT. In that section we present also the procedure of the matching that should reduce the uncertainty coming from the chiral expansion of the amplitudes. In Sec.~\ref{sec:neutral} we use these analyses also for the determination of the neutral Dalitz decay parameter $\alpha$ and discuss briefly the determination of the ratio of the neutral and the charged decay width.
Finally, our conclusions can be found in Sec.~\ref{sec:conclusions}.
We have devoted two appendices to the discussion of properties of the kinematic
functions appearing in our dispersive representation.

%%%%%%%%%%%%%%%%%%%%%%%%%%%%%%%%%%%%%%%%%%%%%%%%%%%%%%%%%%%%%%%%%%%%%%%%%%%%
\section{Basic properties}\label{sec:notation}

\subsection{Kinematics and notation}
We are interested in two decay channels of $\eta$, the charged one
$\eta\to\pi^+\pi^-\pi^0$ and the neutral one $\eta\to3\pi^0$, generically denoted as
\begin{equation}
 \eta(P) \to \pi^a (p_1) \pi^b (p_2) \pi^0 (p_3)\,.
\end{equation}
The amplitudes of these decay processes can be obtained by analytic continuation of the
amplitudes of the corresponding scattering process
\begin{equation}
 \eta(P) \pi^0(p) \to \pi^a(p_1) \pi^b(p_2)
\end{equation}
by taking $p=-p_3$.
In the following sections, we use the usual Mandelstam variables. In the
scattering region they are defined by
\begin{equation}
s=(P+p)^2,\qquad t=(P-p_1)^2,\qquad u=(P-p_2)^2 ,
\end{equation}
while in the decay region we take
\begin{equation}
s_j=(P-p_j)^2.
\end{equation}
These variables satisfy
\begin{equation}\label{soucet Mandelstamu decay}
s+t+u=3s^c=s_1+s_2+s_3,
\end{equation}
where $s=t=u=s^c$, with
\begin{equation}
3s^c=m_\eta^2 + 2 m_{1,2}^2 + m_3^2
\label{cdp}
\end{equation}
 corresponds the center of the Dalitz plot.
Here $m_3 = \mpn$, whereas $m_{1,2}=\mpn$ for $\eta\to3\pi^0$, while $m_{1,2}=\mpp$ for $\eta\to \pi^+ \pi^- \pi^0$.
Up to a convention-dependent phase factor, the crossing relation then means a substitution of the
variables $(s,t,u)$ by $(s_3,s_1,s_2)$, together with the appropriate analytic
continuation from the scattering to the decay region. Bearing this in mind, we can therefore
interchange freely between these two sets of variables.

The constraints (\ref{soucet Mandelstamu decay}) tell us that just two of the kinematic variables are independent.
We can choose them to be, for instance, $s_3=s$
and $s_1=t$. The plot of the dependence of the decay amplitudes on these variables is called Dalitz plot. The physically
allowed kinematical regions for the different crossed amplitudes are constrained by
kinematical limits arising from the condition that the energy of a real particle
has to be at least equal to its rest mass.
Therefore, for a decay process the variable $s_3$ is bounded by
\begin{equation}
(m_1+m_2)^2\le (p_1+p_2)^2=s_3=(k-p_3)^2\le (m_\eta-m_3)^2 ,
\end{equation}
whereas for a scattering in the $s$-channel
\begin{equation}
s\ge(m_\eta+m_3)^2.
\end{equation}
For a fixed value of $s_3$, we obtain bounds for the physical values of $s_1$ (and similarly for $s_2$),
$s_1^{-}(s_3)\le s_1\le s_1^{+}(s_3)$ with
\begin{equation}
s_1^\pm(s_3)=\frac{3s^c-s_3}{2}-\frac{\Delta_{\eta3}\Delta_{12}}{2s_3}\pm\frac{\lambda_{\eta3}^{1/2}(s_3)
\lambda_{12}^{1/2}(s_3)}{2s_3} \,,
\end{equation}
where ($i,j = \eta,1,2,3$)
\begin{align}
\Delta_{ij} &=m_{i}^{2}-m_{j}^{2},\\
\lambda_{ij}(s_3)&=\left(s_3-(m_{i}+m_{j})^2\right)\left(s_3-(m_{i}-m_{j})^2\right). \label{lambda}
\end{align}
Since for both cases under study $m_1=m_2$, the bounds simplify to
\begin{equation}
s_1^\pm=\frac12\left(3s^c-s_3\pm\lambda_{\eta3}^{1/2}(s_3)
\sigma(s_3)\right) ,
\end{equation}
with
\begin{equation}
\sigma(s_3)=\sqrt{1-\frac{4m_{1,2}^2}{s_3}}\,. \label{sigma}
\end{equation}

As was recalled in the Introduction, the amplitudes of the processes $\eta \to 3\pi$, $\mathcal{A}(s,t,u)$, are proportional to the difference of $m_u$
and $m_d$. We therefore pull out this factor, defining
\begin{equation}\label{ARM}
\mathcal{A}(s,t,u) = \frac{\sqrt3}{4}\frac{1}{R}\frac{1}{F_\pi^2} \mathcal{M}(s,t,u)\,,
\end{equation}
where the ratio $R$, which is defined in~(\ref{RQ}), measures the relative violations of $SU(3)$ and of $SU(2)$, and $F_\pi$ is the physical pion decay rate (in our numerical analyses of Sec.~\ref{sec:charged} in order to be in correspondence with \cite{BG}, we take\footnote{Recent analyses (e.g.\ \cite{Kampf:2009tk}) indicate a slightly smaller physical value, $F_\pi=(92.22\pm0.07)\,\mathrm{MeV}$. In order to fully include this change into our computation, redoing of the analysis \cite{BG} with new values for $F_\pi$ and for the pseudoscalar masses would be necessary. Note, however, that a mere change of this value just in this definition leads to a shift of the value of $R$ of about 0.4\%, which is negligible with respect to the other sources of errors occurring in the presented results.} $F_\pi=92.4\,\mathrm{MeV}$). 
In accordance to the notation introduced in our general
paper \cite{kknzprep}, when the distinction becomes necessary, the quantities associated to the charged ($\eta\to \pi^+\pi^-\pi^0$) or neutral decay
($\eta\to3\pi^0$) are denoted with the subscript $x$ or $0$, respectively.

In this paper we work mainly to lowest order in the isospin breaking, i.e. we consider the case where all isospin breaking is contained already
in the normalization prefactor $R$ from \eqref{ARM}, and the rest of the amplitude is computed in the isospin limit.
Then due to the isospin structure, the amplitudes $\mathcal{M}(s,t,u)$ are related by
\begin{equation}\label{M0Mxiso}
{\cal M}_0 (s,t,u) = -{\cal M}_x(s,t,u) - {\cal M}_x(t,u,s) - {\cal M}_x(u,s,t)
\end{equation}
(the minus sign is due to the Condon and Shortley phase convention) and in both $\mathcal{M}_0(s,t,u)$ and
$\mathcal{M}_x(s,t,u)$ there appears only one pion mass $\mpi$. It is why we refer to this case as the $\mpp=\mpn$ limit,
 or more loosely as the isospin limit. However, when making comparisons with the ChPT calculation of \cite{BG}, we use exactly the same values for $\eta$ and $\pi$ masses as were used there\footnote{Note that the value used $m_\eta = 547.3~\text{MeV}$ differs slightly from the current PDG value \cite{PDG}.}, and take for the isospin mass $\mpi$ in each process a different value --- in the case of the charged $\eta$ decay we take $3 \mpi^2 =2 \mpp^2 + \mpn^2$, whereas $\mpi=\mpn$ in the case of the neutral decay. So defined $m_\pi$ has the advantage that in both processes we reproduce exactly the physical location of the center of the Dalitz plot and reproduce almost exactly the physical value of the normalization $Q_\eta$ of the Dalitz variables $x$ and $y$ from (\ref{kinematic variables eta})
below. When computing the integrations over the phase space used for setting the normalization from the measured decay rate, we employ again the physical $\mpp$ and $\mpn$ masses for the determination of the phase space.

%%%%%%%%%%%%%%%%%

\subsection{Dalitz plot parametrization}\label{Dalitz plot parametrization}
The standard parametrization of a decay process $P\to3\pi$ is called a
Dalitz plot parametrization (cf.~\cite{PDG}). It is a polynomial
expansion of $|\mathcal{A}(s_1,s_2,s_3)|^2$ around the center of the Dalitz plot.
The parameters are usually normalized in order to be
dimensionless.
The variables of standard use for the charged $\eta$ decay are then
\begin{equation}\begin{split}
x=\sqrt{3}\,\frac{T_1-T_2}{Q_\eta}=\frac{\sqrt{3}\,(s_2-s_1)}{2\met
Q_\eta}\,,\qquad 
y=\frac{3T_3}{Q_\eta}-1=\frac{3}{2\met
Q_\eta}\left((\met-\mpn)^2-s_3\right)-1,
\end{split}\label{kinematic variables eta}
\end{equation}
where $T_j$ is the kinetic energy of the $j$-th pion in the $\eta$ rest-frame. For the charged decay the energy of the reaction $Q_\eta=\met-2\mpp-\mpn$ whereas for the neutral one $Q_\eta=\met-3\mpn$. In the case we use in this definition the physical values of the masses, for the charged decay the point $x=y=0$, around which we expand the amplitude, does not coincide\footnote{The $x=y=0$ point is slightly shifted in the $s_3$ direction to $s_3=s^c_x+\frac23(\mpp-\mpn)(2\met-\mpp-\mpn)$.} exactly with the center of the
Dalitz plot. However, in the isospin limit,
\begin{equation}
y=\frac{3}{2\met Q_\eta}(s^c_x-s_3)
\end{equation}
and the center of the expansion $x=y=0$ matches the center of the Dalitz plot.

The parameters relevant to the decay $\eta\to\pi^+\pi^-\pi^0$ are usually labeled according to
\begin{equation}
\label{dalitz etac}
|\M_x(s_1,s_2,s_3)|^2=|A|^2\big(1+a y+b y^2+c x+ d
x^2 + e x y+ f y^3+g x^2 y\dots\big)\,,
\end{equation}
where $A$ is the value of the amplitude $\M_x$ at the point $x=y=0$. Charge conjugation forbids the appearance of terms containing odd powers of $x$ in this expansion, and so $c=e=0$.

\begin{table*}
\begin{center}
\begin{tabular}{c@{\hspace{3.5em}}r@{$\:\pm\:$}l@{\hspace{3em}}r@{$\:\pm\:$}l@{\hspace{3em}}r@{$\:\pm\:$}l@{\hspace{3em}}r@{$\:\pm\:$}l@{\hspace{0.5em}}}
\hline\hline
&\multicolumn{2}{c@{\hspace{3.5em}}}{$a$}&\multicolumn{2}{c@{\hspace{3em}}}{$b$}&\multicolumn{2}{c@{\hspace{3em}}}{$d$}
&\multicolumn{2}{c@{\hspace{0.5em}}}{$f$}\\
\hline
Gormley et al.\cite{Gormley:1970qz} & -1.17\phantom{0} &0.02 & 0.21\phantom{0} &0.03 & 0.06\phantom{0} &0.04
&\multicolumn{2}{c@{\hspace{0.5em}}}{}\\
Layter et al.\cite{Layter:1973ti} & -1.08\phantom{0} &0.014 & 0.034&0.027 &
0.046&0.031&\multicolumn{2}{c@{\hspace{0.5em}}}{}\\
Crystal Barrel \cite{CrystalBarrel-etac} & -1.22\phantom{0} &0.07 & 0.22\phantom{0} &0.11 &
0.06\phantom{0} &0.04&\multicolumn{2}{c@{\hspace{0.5em}}}{}\\
KLOE \cite{KLOE-etac} & -1.090&0.020 & 0.124&0.012 & 0.057&0.017 & 0.14\phantom{0} &0.02\\
\hline
ChPT NNLO \cite{BG} & -1.271&0.075 & 0.394&0.102 & 0.055&0.057 &
0.025&0.160\\
\hline\hline
\end{tabular}
\caption{Values of the Dalitz plot parameters of the $\eta\to\pi^+\pi^-\pi^0$ decay
coming from various experimental and theoretical
determinations.}\label{tabulka Dalitz etac}
\end{center}
\end{table*}

The values of the parameters obtained by various experiments are listed in Table~\ref{tabulka Dalitz etac}. These values are compared with the NNLO calculation in ChPT \cite{BG}. All of
the experiments find the values of $c$ and $e$ compatible with zero.
From the table it is obvious that the precision of the determination from KLOE
\cite{KLOE-etac} exceeds significantly the precision of all the other experiments, which
are more than ten years older. It is also up to now the only experiment that has
determined the parameter $f$ with a reasonable precision.

At this point let us also mention the \emph{linear\/} Dalitz parametrization for the amplitude itself (cf. Appendix A of \cite{BG}):
\begin{equation}\label{lindalnab}
 \M_x(s,t,u) = A ( 1 + \bar a y + \bar b y^2 + \bar d x^2  + \bar f y^3 + \bar g
x^2 y + \ldots)\,,
\end{equation}
where the parameters $\bar{a}, \bar{b}, \dots$ can now be complex in general. (We have already omitted the terms violating the charge conjugation symmetry of the amplitude.) The parameters of (\ref{dalitz etac}) can be expressed in terms of these \emph{linear\/} Dalitz parameters --- the relations are simple to obtain by squaring (\ref{lindalnab}) and by comparing the terms with the same powers of $x$ and $y$.

At leading order, the parametrization of the $\eta\to3\pi^0$ differential decay rate depends only on the
kinematical variable
\begin{equation}
z=\frac{3}{2\met^2(\met-3m_\pi)^2}\sum_j(s_j-s^c_0)^2
=\frac{3}{2\met^2(\met-3m_\pi)^2}
\left(s_1^2+s_2^2+s_3^2-3(s^c_0)^2\right),
\end{equation}
which denotes the distance from the center of the Dalitz plot, normalized to one at
the edge of the decay region. However,
higher orders corrections do not preserve this accidental rotational symmetry, and we
need again $x$ and/or $y$ from relations (\ref{kinematic variables eta}). Note that the relation
\begin{equation}
z=x^2+y^2
\end{equation}
holds. The Dalitz plot parametrization for this process reads
\begin{equation}\label{dalitz eta0}
\frac{|\M_0(s_1,s_2,s_3)|^2}{|\M_0(s^c_0)|^2}=1+2\alpha z+2\beta
y(3z-4y^2)+\gamma z^2+\dots\,.
\end{equation}
The factor of $2$ in front of $\alpha$ and $\beta$ is a mere convention to stress the
connection with the direct \emph{linear\/} Dalitz parametrization of the amplitude
itself (see (\ref{lindalnab}) above). For a better visualization of the violation of the rotation symmetry in $(x,y)$- plane at higher orders, it is convenient to introduce the polar coordinates (cf. also \cite{Schneider:2010hs}), $x=\rho
\cos\phi$, $y=\rho\sin\phi$ with distance $\rho^2=z$, for which we have
$y(3z-4y^2) = \rho^3 \sin(3\phi)$.

\begin{table}
\begin{center}
\begin{tabular}{c@{\hspace{3em}}r@{$\:\pm\:$}l@{\hspace{1em}}}
\hline\hline
&\multicolumn{2}{c@{\hspace{1em}}}{$\alpha$}\\
\hline
Crystal Barrel \cite{CrystalBarrel-eta0}&-0.052\phantom{0}&0.020\\
Crystal Ball \cite{CrystalBall}&-0.031\phantom{0}&0.004\\
WASA/CELSIUS \cite{Celsius}&-0.026\phantom{0}&0.014\\
WASA/COSY \cite{WASAatCOSY}&-0.027\phantom{0}&0.009\\
Crystal Ball {@} MAMI-B \cite{MAMI-B} &-0.032\phantom{0}&0.003\\
Crystal Ball {@} MAMI-C \cite{MAMI-C} &-0.0322&0.0025\\
KLOE \cite{KLOE-eta0} &-0.0301&0.0050\\
\hline
ChPT NNLO \cite{BG} &0.013\phantom{0}&0.032 \\
\hline\hline
\end{tabular}
\caption{Experimental and theoretical values of the slope parameter $\alpha$ of the
$\eta\to3\pi^0$ decay.}\label{tabulka Dalitz eta0}
\end{center}
\end{table}

Various experimental and theoretical determinations of the parameter
$\alpha$ are given in Table~\ref{tabulka Dalitz eta0}. Note the sign discrepancy
between the ChPT determination (with however large error bars) and the experimental measurements, which we
will briefly address in Sec.~\ref{sec:neutral}. Up
to now, no experiment has so far published any constraint on the other parameters, such as
$\beta$.

In the case we work to first order in isospin breaking, the isospin relation (\ref{M0Mxiso}) translates into the following
relations between the neutral Dalitz parameters and the parameters of the \emph{linear\/} parametrization (\ref{lindalnab}) (cf.\ again Appendix~A of \cite{BG})
\begin{align}
\alpha&= \frac12 (\re \bar b + \re \bar d),\label{iso-alpha-bar}\\
\beta&= \frac14 (\re \bar g - \re \bar f).\label{iso-beta-bar}
\end{align}
They can be rewritten in terms of Dalitz parameters of the charged decay. However, there still remains a dependence on the imaginary parts of the \emph{linear\/} parameters,
\begin{align}
\alpha&=\frac14\left(b+d-\frac{a^2}{4}\right)-\frac{\left(\im \bar{a}\right)^2}{4},\label{iso-alpha}\\
\beta&=\frac18\left(g-f+\frac{a}{2}(b-d)-\frac{a^3}{8}\right)+\frac{\im \bar{a}}{4}
\left(\im \bar{b}-\im \bar{d}-a \frac{\im\bar a}{4}\right).
\label{iso-beta}
\end{align}

\subsection{Adler zero}\label{sec:adler}
The isospin-breaking part of the QCD Hamiltonian density (\ref{eta hamiltonian QCD}) can be written as (in this subsection $\lambda^a$ are Gell-Mann SU(3) matrices)
\begin{equation}
\mathcal{H}^{IB}_{\mathrm{QCD}}(x)=(m_d-m_u)S_3(x),
\end{equation}
where
\begin{equation}
S_3(x)=\frac 12\overline{q}(x)\lambda^3 q(x).
\end{equation}
Therefore, to first order in $m_d-m_u$, the amplitudes of the isospin-breaking processes that are described by this Hamiltonian
are connected to Green functions with one insertion of zero momentum scalar density $S_3$ (calculated in the limit $m_u=m_d$). In
the $SU(2)\times SU(2)$ chiral limit $m_u=m_d=0$, pions are genuine Goldstone bosons. For the corresponding amplitudes $\langle f+\pi
^a(p)|S_3(0)|i\rangle $ with a pion in the final state, we can thus derive the $SU(2)\times SU(2)$ soft-pion theorem in the general form
\begin{equation}
\langle f+\pi ^a(p)|S_3(0)|i\rangle \rightarrow 0\quad\mathrm{for}\quad
p\rightarrow 0.
\end{equation}
The derivation of the theorem proceeds in the usual way, except that now, because of the insertion
of $S_3(0)$ transforming under the axial $SU(2)$ rotation nontrivially as
\begin{equation}
\delta_5^aS^3(0)=\frac{\mathrm{i}}2\overline{q}(x)\{\lambda^3,\lambda
^a\}\gamma_5 q(x),
\end{equation}
it only  holds provided $a=1,2$. (For $a=3$ there occurs an additional contribution from $\delta_5^3 S^3(0)$, which does not vanish.)

For the decay $\eta \to \pi^{+}(p_{+})\pi^{-}(p_{-})\pi^0(p_0)$, this means that the amplitude $\mathcal{M}_x(s,t,u)$ defined in (\ref{ARM}) vanishes in the $SU(2)\times SU(2)$ chiral limit for either $p_{+}=0$ or $p_{-}=0$, i.e.\ it develops two Adler zeroes \cite{Adler:1965ab,Adler:1965ga} $s=u=0$, $t=m_\eta^2$ and $s=t=0$, $u=m_\eta^2$.
As a consequence, expanding the amplitude $\mathcal{M}_x(s,t,u)$ beyond the $SU(2)\times SU(2)$ chiral limit in the independent variables $s$ and $\left(t-u\right) $ around the points where $s=0$, $(t-u)^2=m_\eta^4$, or more specifically around the points
\begin{equation}\begin{split}
\overline{s} &=0, \\
\overline{t} &=\frac 12(m_\eta ^2+3m_\pi ^2\pm m_\eta ^2), \\
\overline{u} &=\frac 12(m_\eta ^2+3m_\pi ^2\mp m_\eta ^2),
\end{split}\end{equation}
according to (here we use the $t\leftrightarrow u$ symmetry of the amplitude)
\begin{equation}
\mathcal{M}_x(s,t,u)=\sum_{i,j\geq 0}c_{ij}s^i((t-u)^2-m_\eta ^4)^j,
\label{A_zero_expansion}
\end{equation}
we can restate the above $SU(2)\times SU(2)$ theorem in the form
\begin{equation}
\lim_{m_u,m_d\rightarrow 0}c_{00}=0.
\end{equation}
Since the position of the Adler zero is determined up to $O(m_\pi^2)$ corrections, an analogous statement remains true also for similar expansion coefficient $c_{00}^{(\xi,\zeta)}$ corresponding to an expansion around the
points with $s=\xi m_\pi^2$, $(t-u)^2=\left( m_\eta ^2+\zeta m_\pi ^2\right)^2$, namely,
around the points
\begin{equation}\label{adlerxi}\begin{split}
s_{(\xi,\zeta)} &=\overline{s}+\xi m_\pi ^2, \\
t_{(\xi,\zeta)} &=\overline{t}\pm \frac 12(\zeta\mp \xi)m_\pi ^2, \\
u_{(\xi,\zeta)} &=\overline{u}\mp \frac 12(\zeta\pm \xi)m_\pi ^2,
\end{split}\end{equation}
where $\xi,\zeta$ are reasonably small and behave as $O(1)$ for $m_u,m_d\rightarrow 0$.
For the value of the amplitude at these points we therefore obtain
\begin{equation}
\mathcal{M}_x(s_{(\xi,\zeta)},t_{(\xi,\zeta)},u_{(\xi,\zeta)})=c_{00}^{(\xi,\zeta)}=O(m_\pi^2)
\end{equation}
and its absolute value is expected to be numerically small.

Note, however, that the remaining coefficients $c_{ij}$ are not protected by such a factor $m_\pi ^2$, and the same is true
also for the value of the amplitude at points far from $(\overline{s},\overline{t},\overline{u})$, where
$\mathcal{M}_x(s,t,u)$ can be enhanced by a factor $m_\eta^2/m_\pi^2$ with respect to
$\mathcal{M}_x(\overline{s},\overline{t},\overline{u})$.
Note also that a small numerical value of $c_{00}$ (or $c_{00}^{(\xi,\zeta)}$ in general) does not
necessarily imply that its chiral expansion shows better convergence than
the one of any other $c_{ij}$, in the sense that for the ratio of two subsequent
corrections the relation
\begin{equation}
\frac{c_{00}^{(n+1)}}{c_{00}^{(n)}}\ll \frac{c_{ij}^{(n+1)}}{c_{ij}^{(n)}}
\end{equation}
does not necessarily hold.

\subsection{Isospin violation and cusp}\label{sec:cusp}
In the case we go beyond the first order in the isospin breaking, in addition to the complications that the two $\eta\to3\pi$ decay amplitudes are no longer connected by (\ref{M0Mxiso}), and that the expressions are more complicated due to the fact that there appear two different masses of pions, in the processes with two neutral pions in the final state there occurs an interesting phenomenon called cusp. This effect is caused by different charged and neutral pion masses and is connected with the contributions of $\pi^+\pi^-$ intermediate states rescattering back to $\pi^0\pi^0$. Such  a state generates a square root singularity, which resides at $4\mpp^2$, lying above the physical threshold, $4\mpn^2$, and the unitarity cusp is a result of the interference between the part of the amplitude containing this singularity and the
rest without it.

It is obvious that the cusp emerges only in the case when isospin breaking is included also in $\M(s,t,u)$ and that its
strength is sensitive to $\pi^+\pi^-\to\pi^0\pi^0$ scattering at threshold (mainly to the scattering length of this process). This property can be used for a determination of the scattering lengths from the measurement of the cusp \cite{Cabibbo:2004gq,Batley:2005ax,Abouzaid:2008js}.

Let us try to estimate the relative sizes of the cusps in various processes where a pseudoscalar, namely $K^+$, $K_L$ or $\eta$, decays into three pions. (This discussion is inspired by \cite{CI} and \cite{DiLella-talk}.) Because the pion rescattering part will be approximately the same for all the processes, we may consider the notion of ``visibility'' of the cusp in these processes by comparing the relative ratios between the cusps and the regular parts of the amplitudes,
\begin{equation}
\gamma(P^c)=\left.\frac{\kappa_c |\mathcal{A}_{P^c\to\pi^+\pi^-\pi^c}|\
|\mathcal{A}_{P^c\to\pi^0\pi^0\pi^c}|}{|\mathcal{A}_{P^c\to\pi^0\pi^0\pi^c}|^2}
\right\rvert_{\text{cusp}}
=\left.\kappa_c\frac{|\mathcal{A}_{P^c\to\pi^+\pi^-\pi^c}|}{|\mathcal{A}_{P^c\to\pi^0\pi^0\pi^c}|}
\right\rvert_{\text{cusp}},
\end{equation}
where $|\mathcal{A}|$ is the absolute value of the matrix element of the indicated
process and $\kappa_c$ is a multiplicity factor corresponding to that process,
equal to $2$ in the case the decaying particle is charged (two possible
$\pi^+\pi^-\to\pi^0\pi^0$ scatterings are then possible), and to $1$ in the other
cases. These ratios have to be evaluated at the cusp point $s=4\mpp^2$.

Using the measured relative decay rates and the values of Dalitz parameters from
\cite{PDG}, we obtain for these processes,
\begin{equation}
\gamma(K^\pm)\sim7.3,\qquad \gamma(K_L)\sim0.45,\qquad \gamma(\eta)\sim0.93.
\end{equation}
From that we can estimate that the effect is approximately 16 (8) times more pronounced in the $K^\pm$
decay with respect to $K_L$ $(\eta)$ decay.

First indications of the cusp effect were already observed also in the $\eta\to3\pi^0$ decay (cf.\ e.g.\ \cite{MAMI-C}). This effect however appears at the edge of the decay region\footnote{In the $(x,y)$ plane, the cusp is located on the segment
$y=y(4\mpp^2)\sim0.773$ and on two other segments obtained by $s\leftrightarrow t$ and $s\leftrightarrow u$ (i.e.\ obtained by rotation of the original one by $\pm120^\circ$ around the center of the Dalitz plot). Its position thus does not respect the accidental rotation symmetry, and depending on its direction in the $(x,y)$ plane, the corresponding value of $z$ changes from $0.597$ to $0.883$ as $\frac{0.597}{\cos(\phi-\phi_0)}$, with $\phi_0=0^\circ,\pm120^\circ$.} and is therefore not simple to measure.

For the time being, because of this lack of data, we shall not  pursue the discussion about the cusp here (even though our representation describing also this effect is prepared \cite{kknzprep, nas-proceeding1}). Instead we will work in the strict isospin limit beyond the trivial order at which $\eta\to3\pi$ decay is forbidden, i.e. $\M_x$ or $\M_0$ is taken in the isospin symmetry limit.

%%%%%%%%%%%%%%%%%%%%%%%%%%%%%%%%%%%%%%%%%%%%%%%%%%%%%%%%%%%%%%%%%%%%%%%%%%%%
\section{Chiral perturbation theory}\label{sec:chpt}
Let us briefly recapitulate the ChPT calculation of $\eta\to3\pi$ decays.
As was discussed in the Introduction, direct electromagnetic corrections to these processes are very small, and thus they
proceed mainly through the part~(\ref{eta hamiltonian QCD}) of the QCD Lagrangian. The leading order (LO) calculation was performed in \cite{Bell:1996mi,Cronin:1967jq,Osborn:1970nn}, which in our notation\footnote{In this work for the various chiral orders we follow the convention of \cite{BG}, where the amplitudes are at a given order simplified using Gell-Mann-Okubo relations, physical decay constants and physical pseudoscalar masses and the so induced differences are included into higher orders.} takes the very simple form
\begin{equation}
\M^{(2)}(s,t,u) = \frac{4}{3} \mpi^2 - s.
\end{equation}
The next-to-leading order (NLO) was provided in~\cite{Gasser:1984pr}. Its form is discussed in Sec.~\ref{sec:etaoneloop} below. The $O(p^6)$ corrections were studied quite recently in~\cite{BG}.
From these three successive orders one can see that $\eta\to3\pi$ thus represents a case where the chiral corrections are large~\cite{Bijnens:2006zp}. Moreover, it seems that also the two-loop ChPT result supplemented with the existing LECs determination of \cite{BG} is not working very well as we have demonstrated on the example of Dalitz parameters in the Introduction.

\subsection{Contribution of the constants \texorpdfstring{$C_i$}{Ci} to Dalitz parameters}
\label{sec:Ci}
In the NNLO result there occurs a great deal of $O(p^6)$ low-energy constants $C_i$ which are only estimated from resonance saturation. Hence, the first question one has to ask is whether the discrepancy with experiment cannot be accounted for by the unsatisfactory knowledge of the $O(p^6)$ low-energy constants.

Let us thus study the contribution of $O(p^6)$ LECs ($C^r_i$) to the Dalitz parameters of the individual decay modes.
There are several possibilities how to determine these parameters from the computed amplitude $\M(s,t,u)$. For instance, we can expand $|\M(s,t,u)|^2$ to the order $O(p^6)$, and then make the Taylor expansion at the center of the Dalitz plot. This would result in the linear dependence of the Dalitz parameters on the $C_i$s. Provided we did not chiral expand first and instead made a fit of the modulus squared of the complete $O(p^6)$ amplitude to the Dalitz parametrization (as it was done in \cite{BG}), we would get a more complicated dependence on the $C_i$s including also quadratic and mixed terms.
Such contributions should be, however, suppressed by the chiral counting. Nevertheless, they can bring sizable changes in the final numerical predictions. In order to obtain the linear contribution only, we follow the first procedure.

We start with the neutral decay mode. The explicit dependence of $\alpha$ on the $C_i$s was already given in~\cite{Schneider:2010hs},
\begin{equation}
 \alpha_C = \frac{16 m_\eta^2 (m_\eta-3 m_\pi)^2}{3F_\pi^4}\, {\cal C}_\alpha
\end{equation}
with
\begin{equation}
{\cal C}_\alpha = C^r_5 + C^r_8 + 3 C^r_9 + C^r_{10} - 2 C^r_{12} + 2 C^r_{22}
+ 3 C^r_{24} + C^r_{25}.
\end{equation}
Further, by a careful investigation of the $O(p^6)$ polynomial of the amplitude calculated in \cite{BG},
we realize that there is no contribution of the $C_i$s to the second neutral Dalitz parameter (it is connected with relation (\ref{iso-beta-bar}) and the fact that in the charged decay $\bar f_C = \bar g_C$ as stated below),
\begin{equation}
\beta_C =0.
\end{equation}

In the case of the charged decay we summarize first the contributions of the $C_i$s to the \emph{linear} coefficients $\bar a$, $\bar b$, $\bar d$, $\bar f$, $\bar g$ defined in~(\ref{lindalnab}) that are directly connected with the amplitude. These parameters can be in general complex but since we deal only with the linear contribution of the $C_i$s, they contribute only to their real values.

By a simple algebra one obtains
\begin{equation}
\bar a_C = - \frac{8 m_\eta (m_\eta-3 m_\pi)}{3 F_\pi^4 (m_\eta^2-m_\pi^2)}\,
\left(m_\eta^4 {\cal C}_{a}^\eta -6 m_\eta^2 m_\pi^2 {\cal C}_{a}^{\eta\pi}
-3 m_\pi^4 {\cal C}_{a}^\pi\right),
\end{equation}
where we have slightly more complex structure
\begin{align}
\begin{split}
{\cal C}_{a}^\eta &= C^r_1-2C^r_3-6C^r_4-2C^r_5-3C^r_6+C^r_8+C^r_{10}
+6C^r_{11}-20C^r_{12}-18C^r_{13}+9C^r_{14}-36C^r_{16}+9C^r_{17}
+45C^r_{18}\\
&-81C^r_{19}-54C^r_{20}+8C^r_{22}+9C^r_{24}+C^r_{25}
-18C^r_{27}-36C^r_{28}-54C^r_{31}-54C^r_{32}-108C^r_{33},
\end{split}\\
\begin{split}
{\cal C}_{a}^{\eta\pi}&= C^r_1-2C^r_3-2C^r_4+C^r_5+2C^r_6-C^r_{10}-4C^r_{11}
-4C^r_{12}+12C^r_{13}-4C^r_{14}-3C^r_{15}-12C^r_{16}+2C^r_{17}
+6C^r_{18}\\
&+4C^r_{22}+3C^r_{24}-C^r_{25}-2C^r_{26}-6C^r_{27}
-12C^r_{28}+4C^r_{29},
\end{split}\\
\begin{split}
{\cal C}_{a}^\pi &= C^r_1- 2C^r_3+ 2C^r_4 + 3C^r_6 + 3C^r_8 - 3C^r_{10} - 6C^r_{11}
+ 12C^r_{12} + 18C^r_{13}+ 3C^r_{14} - 6C^r_{15} + 12C^r_{16} - 9C^r_{17}
+ 3C^r_{18}\\
&- 27C^r_{19} - 18C^r_{20} - 3C^r_{24}- 3C^r_{25} - 4C^r_{26}
+ 6C^r_{27} + 12C^r_{28} + 8C^r_{29} - 18C^r_{31} - 18C^r_{32} - 36C^r_{33}.
\end{split}
\end{align}
Similarly, we have
\begin{equation}
\bar b_C = \frac{8 m_\eta^2 (m_\eta-3 m_\pi)^2}{3 F_\pi^4 (m_\eta^2-m_\pi^2)}\,
\left(m_\eta^2 {\cal C}_{b}^\eta + m_\pi^2 {\cal C}_{b}^\pi\right),
\end{equation}
where
\begin{align}
{\cal C}_{b}^\eta &= C^r_1 -2 C^r_3+ 2 C^r_4+4C^r_5+3 C^r_6+C^r_8+6 C^r_9+ C^r_{10} 
-6 C^r_{11} -2 C^r_{12}+18 C^r_{13}+2 C^r_{22}+6C^r_{24}+4 C^r_{25},
\\
{\cal C}_{b}^\pi &= 3 C^r_1 -6 C^r_3+6C^r_4-2 C^r_5+3 C^r_6+C^r_8-6 C^r_9-5 C^r_{10}
-6 C^r_{11}+10 C^r_{12}+18 C^r_{13}-10
C^r_{22}-6 C^r_{24}+4 C^r_{25},
\end{align}
\begin{equation}
\bar d_C = -\frac{8 m_\eta^2 (m_\eta-3 m_\pi)^2}{3 F_\pi^4
(m_\eta^2-m_\pi^2)}\,\left(m_\eta^2 {\cal C}_{d}^\eta + m_\pi^2 {\cal C}_{d}^\pi\right),
\end{equation}
with
\begin{align}
{\cal C}_{d}^\eta  &= C^r_1 -2 C^r_3+2 C^r_4+3 C^r_6-3C^r_8-6 C^r_9-3 C^r_{10}
-6 C^r_{11}+6 C^r_{12}+18 C^r_{13}-6 C^r_{22}-6 C^r_{24},
\\
{\cal C}_{d}^\pi  &= 3 C^r_1-6 C^r_3+6 C^r_4+2C^r_5+3 C^r_6+5 C^r_8+6 C^r_9
-C^r_{10}-6 C^r_{11}+2 C^r_{12}+18 C^r_{13}-2 C^r_{22}+6C^r_{24}+8 C^r_{25},
\end{align}
and
\begin{equation}
\bar f_C =\frac{16 m_\eta^3 (m_\eta-3 m_\pi)^3}{3 F_\pi^4 (m_\eta^2-m_\pi^2)}\, {\cal C}_f
\end{equation}
with
\begin{equation}
{\cal C}_f = C^r_1-2 C^r_3 - 2 C^r_4.
\end{equation}

Finally, the contribution of $C_i$s to the parameter $\bar{g}$ is the same as in the case of the parameter $\bar{f}$,
\begin{equation}\label{f rovna se g}
\bar g_C = \bar f_C.
\end{equation}

\newpage

Using these relations in the same spirit as in \cite{Bijnens:2009zd}, we can construct the
combinations of physical (or quasi-physical) quantities which do not depend on any $C_i$:
\begin{enumerate}
 \item $\left(\bar b + \bar d - 2 \alpha \right)\bigr|_C=0$   $\Rightarrow$
    \begin{equation}\label{eqrel1}
      \text{rel}_1 \equiv \left(4 (b+d)-a^2 - 16 \alpha\right)\bigr|_C=0
    \end{equation}
 \item $\left(\bar f - \bar g\right)\bigr|_C= 0$ $\Rightarrow$
    \begin{equation}\label{eqrel2}
     \text{rel}_2 \equiv \left(a^3-4 a b+4 a d+8 f-8 g\right)\bigr|_C =0
    \end{equation}
 \item $\beta\bigr|_C = 0$
 \item ${\cal C}_b^\eta + {\cal C}_b^\pi -{\cal C}_d^\eta - {\cal C}_d^\pi =0$
\end{enumerate}
Let us discuss them in more detail (in a reverse order).
The last expression, of course, does not represent any combination of physical quantities, and so it is not possible to use it directly in connection with any observable. It could be, however, useful for lattice simulations, where one can vary the meson
masses. On the contrary the third relation, stating that the second neutral Dalitz parameter $\beta$ does not depend on any $C_i$, represents a simple possibility, open to future experiments, how to check the ChPT results unaffected by the error stemming from the estimates of $C_i$. Now let us turn our attention to the
relations $\text{rel}_1$ and $\text{rel}_2$. The latter was mentioned in \cite{Bijnens:2009zd}, while the first one was implicitly stated in \cite{BG}.
In fact, $\text{rel}_1$ is a simple consequence of the isospin relation (\ref{iso-alpha}) stating that the $C_i$s do not contribute to $\im\bar{a}$
and should thus be valid for any real contributions to the Dalitz parameters appearing there (not only for the contributions of the $C_i$s).

\begin{table*}
\centering
\begin{tabular}{@{\hspace{1em}}c@{\hspace{2em}}r@{$\:\pm\:$}l@{\hspace{2em}}r@{$\:\pm\:$}l@{\hspace{2em}}r@{$\:\pm\:$}l@{\hspace{2em}}r@{$\:\pm\:$}l@{\hspace{2em}}r@{$\:\pm\:$}l@{\hspace{0.5em}}}
\hline\hline
&\multicolumn{2}{c@{\hspace{2em}}}{KLOE} & \multicolumn{2}{c@{\hspace{2em}}}{ChPT}& \multicolumn{2}{c@{\hspace{2em}}}{ChPT$_g$}& \multicolumn{2}{c@{\hspace{2em}}}{NREFT$_i$}& \multicolumn{2}{c@{\hspace{0.5em}}}{NREFT}\\\hline
$a$
& $-1.09\phantom{0}$  & $0.02$
& $-1.271$  &  $0.075$
& $-1.262$  &  $0.079$
& $-1.279$  &  $0.012$
& $-1.213$  &  $0.014$
\\
$b$
& $0.124$ & $0.012$
& $0.394$  &  $0.102$
& $0.407$  &  $0.108$
& $0.361$  &  $0.021$
& $0.308$  &  $0.023$
\\
$d$
& $0.057$ & $0.017$
& $0.055$  &  $0.057$
& $0.081$  &  $0.089$
& $0.053$  &  $0.003$
& $0.050$  &  $0.003$
\\
$f$
& $0.14\phantom{0}$  & $0.02$
& $0.025$  &  $0.160$
& $0.009$  &  $0.170$
& $0.089$  &  $0.018$
& $0.083$  &  $0.019$
\\
$g$
& \multicolumn{2}{c@{\hspace{1em}}}{$\sim0$}
& \multicolumn{2}{c@{\hspace{1em}}}{$0$}
& $-0.07\phantom{0}$  &  $0.19$
& $-0.043$  &  $0.002$
& $-0.039$  &  $0.002$
\\\hline
$\alpha$
& $-0.030$ & $0.005$
& \multicolumn{4}{c}{$0.013\pm0.016$}
& $-0.024$ &  $0.005$
& $-0.025$ &  $0.005$
\\
$\beta$ &
\multicolumn{2}{c}{}
& \multicolumn{4}{c}{$-0.002\pm 0.025\phantom{\pm}$}
& $-0.004$ & $0.001$
& $-0.004$ & $0.001$
\\\hline\hline
\end{tabular}
\caption{Dalitz parameters for charged and neutral decay mode of $\eta\to 3\pi$
calculated in ChPT at order $O(p^6)$ \cite{BG}, NREFT \cite{Schneider:2010hs} (in addition to their final value given in the ultimate column,
in the penultimate column captioned with NREFT$_i$ we list also the values without isospin-breaking effects included) and as
measured by KLOE \cite{KLOE-etac,KLOE-eta0}. ChPT$_g$ represents a fit redone with $g$
included (but without any higher parameters), cf.\ fit NNLOq in \cite{BG}.}
\label{tab:inputdalitz}
\end{table*}

For the comparison of ChPT results \cite{BG} with the values measured by KLOE \cite{KLOE-etac}
we can use not only the values of the Dalitz parameters summarized in Table~\ref{tab:inputdalitz} but also the combinations of these parameters (\ref{eqrel1}) and (\ref{eqrel2}) that are (at least in a first approximation) $C_i$-independent.
It means that the influence of all physics beyond the pseudoscalar domain (mainly scalar and vector resonances) on these combinations is hidden in $O(p^4)$ LECs $L_i$, which are phenomenologically much better under control than the $C_i$s, thereby providing a clearer theoretical output. We should remark, however, that the independence of all these relations on the $C_i$s occurs only in the case we take $\mpp=\mpn$. Away from this limit the situation can be different and these combinations can still have non-negligible dependence on the $O(p^6)$ LECs.

\begin{table*}
\centering
\begin{tabular}{@{\hspace{1em}}c@{\hspace{2em}}r@{$\:\pm\:$}l@{\hspace{2em}}r@{$\:\pm\:$}l@{\hspace{2em}}r@{$\:\pm\:$}l@{\hspace{2em}}r@{$\:\pm\:$}l@{\hspace{2em}}r@{$\:\pm\:$}l@{\hspace{1em}}}
\hline\hline
&\multicolumn{2}{c@{\hspace{2em}}}{KLOE} & \multicolumn{2}{c@{\hspace{2em}}}{ChPT}& \multicolumn{2}{c@{\hspace{2em}}}{ChPT$_g$}& \multicolumn{2}{c@{\hspace{2em}}}{NREFT$_i$} & \multicolumn{2}{c@{\hspace{1em}}}{NREFT}\\\hline
rel$_1$
& $0.02$  &  $0.12$
& $-0.03$  &  $0.72$
& $ 0.15$  &  $0.79$
& $0.41$  &  $0.12$
& $0.35$  &  $0.13$
\\
$\chi^2$
&\multicolumn{2}{c@{\hspace{2em}}}{} & \multicolumn{2}{c@{\hspace{2em}}}{$0.1$} & \multicolumn{2}{c@{\hspace{2em}}}{$1.2$} & \multicolumn{2}{c@{\hspace{2em}}}{$10$} & \multicolumn{2}{c@{\hspace{1em}}}{$7.6$}
\\\hline
rel$_2$
& $0.12$  &  $0.21$
& $-0.13$  &  $1.4$
& $0.28$  &  $2.1$
& $0.54$  &  $0.19$
& $0.44$  &  $0.20$
\\
$\chi^2$
&\multicolumn{2}{c@{\hspace{2em}}}{} & \multicolumn{2}{c@{\hspace{2em}}}{$1.4$} & \multicolumn{2}{c@{\hspace{2em}}}{$0.6$} & \multicolumn{2}{c@{\hspace{2em}}}{$4.0$} & \multicolumn{2}{c@{\hspace{1em}}}{$2.4$}
\\\hline\hline
\end{tabular}
\caption{Two relations defined in~(\ref{eqrel1}) and (\ref{eqrel2}) compared
using different models. The quantity $\chi^2$ defined in~(\ref{chi2}) was computed for the theoretical frameworks by taking for the experiment the values from $\text{exp}\equiv\text{KLOE}$ as given in the first column.}
\label{tab:howwell}
\end{table*}

The values of these combinations that use the data from Table~\ref{tab:inputdalitz} are presented in Table~\ref{tab:howwell}.
This table indicates that even though the central values of the individual Dalitz parameters determined by ChPT and KLOE
differ, the central values of these two combinations are in a good agreement, which indicates that ChPT is not working at all that badly. Unfortunately the large errors quoted there somehow put down the importance of any conclusions. However, one should bear in mind that the values quoted in Table~\ref{tab:howwell} were computed just using the values and the error bars presented in Table~\ref{tab:inputdalitz} that were attributed mainly from the fitting procedures and are thus strongly correlated. This can affect the positions of the central values by small changes, but primarily the error bars of these combinations are then overestimated. Note that the errors of the Dalitz parameters from ChPT are enhanced also by large systematic uncertainties of the amplitudes entering these fitting procedures. Such uncertainties were caused mainly by uncertainties of the $C_i$s, which should be substantially eliminated in these combinations. We also observe another artifact of the fitting procedure when comparing the values denoted by ChPT and ChPT$_g$ that differ just by the truncation of the Dalitz parametrization at $f$ and $g$, respectively. The combination $\text{rel}_1$ is according to relation (\ref{iso-alpha}) equal to $4(\im\bar{a})^2$, which should be therefore positive. The value denoted by ChPT does not possess this property even though that the value of $4(\im\bar{a})^2$ obtained by a direct fit of the original amplitude in \cite{BG} reproduces well the value given in the column ChPT$_g$.

A similar effect can occur also for the KLOE values since in \cite{KLOE-etac} the value\footnote{Note the different notation of this Dalitz parameter in \cite{KLOE-etac} --- for the parameter denoted in this text by $g$, KLOE uses symbol $h$.} of $g$ was not presented (only its compatibility with zero). As an illustration, we remind the reader that if we added to the values of $a,b,d,f$ measured by KLOE the value of $g=-0.02$ ($g=-0.04$), we would obtain an exact match of the so defined experimental\footnote{Naturally, repeating the KLOE fit with $g$ included would also change the values of the further parameters (cf.\ again the difference between the values from ChPT and ChPT$_g$).} value of $\text{rel}_2$ with the value from ChPT$_g$ (NREFT).

In these two tables we have also studied predictions of NREFT \cite{Schneider:2010hs}. Since that method is built in a different way than ChPT, the combinations of the observables appearing in $\mathrm{rel}_1$ and $\mathrm{rel}_2$ have no special significance there. However, they are still valid combinations of observables and so nothing prevents us from using them for comparison of the predictions from any theory with the experiment. The lesser agreement of NREFT and KLOE in $\mathrm{rel}_1$ was already pointed out in \cite{Schneider:2010hs} in terms of different values of $\im \bar{a}$ stemming from the representation of \cite{Schneider:2010hs} and the one coming from the KLOE measurement and the relation (\ref{iso-alpha}).
Together with the slight inconsistency also in $\text{rel}_2$ depending only on the parameters of the charged decay,
this indicates that there is a problem either on the side of the current determination from the KLOE group or on the side of the NREFT representation.

We conclude this discussion with the statement that a new measurement of the charged Dalitz parameters (possibly taking into account these two relations) would therefore be highly desirable. Before that, we are not able to answer the question whether it is possible to reproduce the physical Dalitz plot distribution with a better determination of the LECs $C_i$ or whether the discrepancy between the ChPT-computed and the experimentally measured distributions has some other origin (slow convergence of the chiral counting, \dots). In addition, should the experimental value confirm the values inconsistent with the predictions of \cite{Schneider:2010hs}, even if one accepts the explanation for the discrepancy of the neutral parameter $\alpha$ proposed in \cite{Schneider:2010hs}, the issue of the discrepancy for the charged parameter $b$ would remain open.

But for now, inspired just by the quite good consonance of the current KLOE and the ChPT values of the $C_i$-independent relations, we would expect that by finding the right values of the $C_i$s we would reproduce (at least partially) better the physical values of the Dalitz parameters. The natural question can arise now whether it would be possible to find
an elaborate determination of such $C_i$s going beyond the crude estimate of the simple resonance saturation model used
in~\cite{BG}.

Let us start with $\alpha$. Its resonance saturation is simpler as there are no vector resonance
contributions. For the simple scalar resonance model used in~\cite{BG} we obtain (cf.~\cite{Schneider:2010hs})
\begin{equation}
 \alpha|_C = \frac{16 m_\eta^2 (m_\eta-3 m_\pi)^2 c_d c_m}{3 F_\pi^2 M_S^4}
\approx 0.005
\end{equation}
that is positive. However, the minimal chiral symmetry breaking introduced in \cite{Cirigliano:2003yq} changes $c_m$ into $c_m + c_d e_m$, and especially for standard hierarchy ($e_m<0$) one can thus produce a negative contribution to $\alpha$.
Using the same numbers as obtained from the phenomenological study in \cite{Cirigliano:2003yq}, where they distinguish two models, one representing the inverted hierarchy (the model called A) and one representing the standard hierarchy (called B), we obtain
\begin{equation}
\alpha|_{C\,[A]} = 0.002,\; \text{or}\quad
\alpha|_{C\,[B]} = -0.0005\,,
\end{equation}
which lead to the final values $\alpha=0.01$ and $\alpha=0.007$, respectively.

The situation for the charged decay mode looks more complicated. The transition from the amplitude to the Dalitz parametrization leads to many mixing terms and the dependence on the $C_i$s is not linear. Even though, as already mentioned, such higher terms are theoretically suppressed by chiral counting,
in practice they can turn out to be more substantial than anticipated (it is true especially for model $B$). In order
to get more reliable results we perform a full fit to the Dalitz distribution in exact correspondence with \cite{BG}, with the exception that we fit a polynomial of the third order (i.e.\ including $g$), which corresponds to ChPT$_g$ in Table~\ref{tab:inputdalitz}. The vector resonance saturation employed here is based on the model and the phenomenology constraints from~\cite{Kampf:2006bn}. The resulting fits to the Dalitz parametrization are summarized in Table~\ref{tab:dalitz}.
\begin{table}
\centering
\begin{ruledtabular}
\begin{tabular}{crccccc}
model & $|A|^2$ & $a$ & $b$ & $d$ & $f$ & $g$ \\\hline
ChPT$_g$
&534
&-1.26
&0.41
&0.081
&0.009
&-0.072
\\
simple
& 516
& -1.39
& 0.47
& 0.10\phantom{0}
& 0.025
&-0.088
\\
model $A$
&  723
& -1.31
& 0.41
& 0.081
& 0.024
&-0.069
\\
model $B$
&  1835
& -1.19
& 0.33
& 0.052
& 0.020
&-0.040
\end{tabular}
\end{ruledtabular}
\caption{The comparison of the impact of the different models for resonance saturation on
the Dalitz parameters. For vector resonances we have taken the model from~\cite{Kampf:2006bn},
different cases for scalar resonances are: the simple one from \cite{BG}, and the models $A$ and $B$ from \cite{Cirigliano:2003yq}.}
\label{tab:dalitz}
\end{table}
It is clear from this table that, as we have anticipated, the $C_i$s have a bigger effect
than expected from mere chiral counting. They also have an impact on the normalization $|A|^2$, which
in the case of model $B$ is far from being negligible. Let us note at this
point a few things concerning the resonance saturation. It is obvious from
Table~\ref{tab:dalitz} that model $B$ would produce an unrealistic increase
of the amplitude (thereby also of $R$ or $Q$). It does not, however, mean that
this model for scalar resonances is disqualified. Higher resonances, representative of the physics beyond the pseudo-Goldstone bosons, contribute to both $L_i$s and $C_i$s (when talking about NNLO). One cannot just
keep their influence on $C_i$s ignoring their presence in $L_i$s and thus merge
inconsistently two models, i.e.\ in our case the model used in ``fit 10'' of
\cite{Amoros:2001cp} and the models $A$ or $B$. One can always try to be as ``harmless'' as
possible with any extension of the simple resonance saturation and try to
preserve the original values of $L_i$s (as was to some extent possible for the chiral
symmetry breaking construction done in \cite{Kampf:2006bn}), hoping that the new effects
induced by the new resonance terms will not change considerably the original and phenomenologically successful ``fit 10''.
But generally this is not guaranteed.

The detailed analysis based on the current experimental data which would take into account simultaneously and consistently
various resonance estimates for both $O(p^4)$ and $O(p^6)$ LECs is beyond the scope of this paper (however such a project is under investigation \cite{Bijnens:2009hy,Bijnens:2011tb}).
Instead we present another representation that can be used for analyzing the data without addressing the values of the individual $C_i$s.

%%%%%%%%%%%%%%%%%%%%%%%%%%%%%%%%%%%%%%%%%%%%%%%%%%%%%%%%%%%%%%%%%%%%%%%%%%%%
\section{Dispersive construction}\label{sec:dispersive construction}
The dispersive construction to be presented below is based on the reconstruction theorem
\cite{Stern:1993rg,Knecht:1995tr,Zdrahal:2008bd}, which takes into account only the most general properties of the
amplitude, namely, relativistic invariance, unitarity, analyticity and crossing, supplied with chiral counting
(e.g.\ expansion in powers of momenta and of masses of the pseudoscalars). This framework provides the most general form of the amplitude under consideration in the low-energy domain, up to a remainder of the chiral order $O(p^{8})$. Such a construction requires at the same time the scattering amplitudes related to the original one by two-particle unitarity and by crossing. (Contributions to the unitarity condition arising from intermediate states with more than two pseudo-Goldstone particles only start at $O(p^8)$ --- cf.\ \cite{Stern:1993rg,Knecht:1995tr,Zdrahal:2008bd}.) These amplitudes are constructed along the same lines.
The details of the construction, including a full isospin breaking arising from $\mpn\neq\mpp$, will be published elsewhere \cite{kknzprep} (cf.~also \cite{nas-proceeding2}). In this work we
concentrate on the qualitative description of the result.

The dispersively constructed scattering amplitudes of the pseudo-Goldstone bosons (pGB) take the following general form
\begin{equation}
\mathcal{A}(s,t,u)=\mathcal{N}\left( \mathcal{P}(s,t,u)+\mathcal{U}(s,t,u)\right) +O(p^{8}).  \label{tvar amplitudy}
\end{equation}
Here $\mathcal{N}$ is an overall normalization and $\mathcal{P}(s,t,u)$ is a
third order polynomial with the same symmetry properties with respect to $s$, $t$ and $u$ as the complete amplitude $\mathcal{A}(s,t,u)$. The coefficients of these polynomials for the independent amplitudes related by
two-particle unitarity in all the crossed channels are identical for all the amplitudes and are the only free
parameters entering the game. The non-analytic unitarity part $\mathcal{U}(s,t,u)$, which takes into account the contribution of the two-particle intermediate states in all the crossed channels, is then a known function of these parameters. In the low-energy region, intermediate states containing more than two pGB states contribute only to the $O(p^8)$ remainder, while intermediate states involving other hadronic states contribute to the coefficients of the subtraction polynomial.

In the case of the amplitudes concerning one $\eta$ and three pion states, there are several two-pGB intermediate states to consider: $\pi\pi$, $KK$, $\pi\eta$. Since we shall only be concerned by the decay region, only the nearest singularity, coming from the cut produced by the $\pi\pi$ intermediate state, will be close enough to affect sizably the amplitude. The contribution from the remaining states ($KK$, $\pi\eta$) can be expanded in a polynomial, which is included in ${\cal P}(s,t,u)$ (see also the discussion at the beginning of the next section). Of course, such an approximation would not be appropriate\footnote{However, the presented construction can be extended also to include the unitarity cuts from the other two-pGB intermediate states which are relevant in the scattering region \cite{Zdrahal:2008bd}. This then, however, brings into a game more free parameters (describing such intermediate processes).} to describe the $\eta\pi\to\pi\pi$ amplitude in the scattering region. In conclusion, for our purposes the only relevant related amplitude is therefore the $\pi \pi $ scattering one.

For the charged $\eta \rightarrow 3\pi $ decay channel the polynomial $\mathcal{P}_x(s,t,u)$ can be expressed in terms of six free parameters corresponding to the $t-u$ symmetric expansion at the center of the Dalitz plot
\begin{equation}
\mathcal{P}_x(s,t,u)=A_x \met^{2}+B_x(s-s^{c})+C_x(s-s^{c})^{2}
+D_x \left[(t-s^{c})^{2}+(u-s^{c})^{2}\right]+E_x (s-s^{c})^{3}
+F_x\left[(t-s^{c})^{3}+(u-s^{c})^{3}\right],
\label{polstu}
\end{equation}
which is closely related to the traditional PDG parametrization of the Dalitz plot distribution. We take the overall normalization as
\begin{equation}
{\cal N}_\eta = \frac{\sqrt3}{4}\frac{1}{R} \frac{1}{F_\pi^2}\,,
\end{equation}
so we have simply (cf. (\ref{ARM}))
\begin{equation}\label{vyjadreni M}
\M_x(s,t,u) = {\cal P}_x(s,t,u) + {\cal U}_x(s,t,u).
\end{equation}
Let us make one remark concerning the Dalitz plot parametrization. Between the polynomial~(\ref{polstu}) and the \emph{linear} parametrization~(\ref{lindalnab}) there is a simple connection. However, the dependence of $\bar a$, $\bar b$, $\bar d$, $\bar f$, $\bar g$ on parameters $A_x$, $B_x$, $C_x$ and $D_x$ is complicated by the presence of these four parameters also in the unitarity part ${\cal U}_x(s,t,u)$ (see below). The direct correspondence can be, however, established for the dependence of $\bar f$, $\bar g$ on $E_x$ and $F_x$ with very simple form
\begin{equation}
\bar f- \bar g \sim E_x + 2 F_x,\label{Eplus2F}
\end{equation}
which we will need in Sec.~\ref{sec:ObOC} (the exact connection will not be needed).

For the related
$\pi ^{+}\pi ^{-}$ $\rightarrow \pi ^{0}\pi ^{0}$ scattering amplitude (which is the only independent one in
the isospin conservation case) we choose the following parametrization of
the polynomial part in terms of the subthreshold parameters \cite{Stern:1993rg,Knecht:1995tr}
\begin{multline}
\mathcal{P}_{\pi \pi }(s,t,u) =\frac{1}{3}\alpha_\pi M_{\pi }^{2}+\beta_\pi \left(s-\frac{4}{3}M_{\pi }^{2}\right) +\frac{\lambda_{1}}{F_{\pi }^{2}}\left(s-2M_{\pi }^{2}\right)^{2}
+\frac{\lambda _{2}}{F_{\pi }^{2}}\left[ \left(t-2M_{\pi
}^{2}\right)^{2}+\left(u-2M_{\pi }^{2}\right)^{2}\right]\\
+\frac{\lambda _{3}}{F_{\pi }^{4}}\left(s-2M_{\pi }^{2}\right)^{3}
+\frac{\lambda _{4}}{F_{\pi }^{4}}\left[\left(t-2M_{\pi }^{2}\right)^{3}+\left(u-2M_{\pi }^{2}\right)^{3}\right]
\end{multline}
and the overall normalization $\mathcal{N}_{\pi \pi }=F_{\pi}^{-2}$.
The unitarity part of the $\eta \rightarrow 3\pi $ decay amplitude $\mathcal{U}_x$ is then a function of a subset of the above polynomial parameters, namely
\begin{equation}
\mathcal{U}_x=\mathcal{U}_x\left(A_x,B_x,C_x,D_x;\alpha_\pi ,\beta_\pi ,\lambda _{1},\lambda _{2}\right).
\end{equation}
The general form of $\mathcal{U}(s,t,u)$ for the process $\text{AB}\rightarrow \text{CD}$ reads
\begin{equation}
\mathcal{U}_{\text{AB}\rightarrow \text{CD}}(s,t,u)
=W_{S}^{0}(s)+W_{T}^{0}(t)+W_{U}^{0}(u) 
+(t-u)W_{S}^{1}(s)+(s-u)W_{T}^{1}(t)+(t-s)W_{U}^{1}(u),
\end{equation}
where the discontinuities of the functions $W_{S,T,U}^{0,1}(s)$ are given in terms of the right-hand cut discontinuities of the $S$ and the $P$ partial waves
$S_{\ell}$, $T_{\ell}$ and $U_{\ell}$ $(\ell=0,1)$ of the processes in the $s$-, the $t$- and the $u$-channels, respectively, as\footnote{Note that in our case we need to continue these discontinuities analytically and they become complex (cf.\ \cite{kknzprep,BronzanKacser,Kacser,KWW}).}
\begin{align}
\disc W_{S}^{0}(s) &= 16\pi\, \Bigg(\disc S_{0}(s)
+\Delta_{\text{AB}}\Delta _{\text{CD}} \disc \frac{S_{1}(s)}{\lambda _{\text{AB}}^{1/2}(s)\lambda
_{\text{CD}}^{1/2}(s)}\Bigg),
\\
\disc W_{S}^{1}(s)&=48\pi s \disc \frac{S_{1}(x)}{\lambda
_{\text{AB}}^{1/2}(x)\lambda _{\text{CD}}^{1/2}(x)}\,,
\end{align}
where $\Delta_{ij}$ and $\lambda_{ij}(s)$ were defined in (\ref{lambda}). Similar relations for $W_{T,U}^{0,1}(s)$ can be obtained by an appropriate permutations of $\text{A},\ldots ,\text{D}$. The right hand cut discontinuities  are iteratively constructed from the generalized two-particle partial-wave
unitarity relations  as described in \cite{Stern:1993rg,Knecht:1995tr,Zdrahal:2008bd}.  The functions
$W_{S,T,U}^{0,1}(s)$ are then reconstructed by means of
appropriately subtracted dispersion relation. Note that such a subtraction prescription is an indivisible part of the definition
of the polynomial part of the amplitude. The first iteration
reconstruct the  amplitude at $O(p^4)$ while the second one yields
the $O(p^6)$ results.

For the $\eta \rightarrow \pi ^{+}\pi ^{-}\pi ^{0}$ decay the above general
form simplifies since there are only two independent masses in the problem
and the amplitude is $t-u$ symmetric. We get
\begin{equation}
\mathcal{U}_x(s,t,u) =W_{S}^{0}(s)+W_{T}^{0}(t)+W_{T}^{0}(u)
 +(s-u)W_{T}^{1}(t)+(s-t)W_{T}^{1}(u),
\label{UetaW}
\end{equation}
where the subscripts $S$, $T$ refer to the $\eta \pi ^{0}\rightarrow
\pi ^{+}\pi ^{-}$ and the $\eta \pi ^{+}$ $\rightarrow \pi ^{+}\pi
^{0}$ channels, respectively.
The relevant discontinuities can be rewritten schematically as
\begin{equation}
\disc W_S^0(s) =\sum_{i=1}^{5}\mathcal{F}_{i}(s)\sum_{k=-1}^{4}a_{k}^{(i)}s^{k}
\end{equation}
and similarly for $W_T^0$ (with coefficients $\bar a_k^{(i)}$), while
\begin{equation}
\disc W_T^1(s)
=\sum_{i=1}^{5}\mathcal{F}_{i}(s)\Bigg(
\sum_{k=-1}^{2}\bar b_{k}^{(i)}s^{k}
+\frac{1}{\lambda (s)}\sum_{k=-1}^{5}\bar c_{k}^{(i)}s^{k}+\frac{1}{x\sigma (s)^{2}}
\sum_{k=0}^{3}\bar d_{k}^{(i)}s^{k}\Bigg).
\end{equation}
Here $a_{k}^{(i)}, \bar a_{k}^{(i)}, \ldots , \bar d_{k}^{(i)}$ are known polynomials of the parameters $\{A_x,B_x,C_x,D_x;\alpha_\pi ,\beta_\pi ,\lambda_{1},\lambda _{2}\}$ and the masses $\mpi$, $\met$;
$\lambda (s)=\lambda_{\pi \eta }(s)$, $\sigma (s)=\lambda _{\pi \pi }(s)/s$ and $\mathcal{F}_{i}(s)$ represents a set of elementary functions listed in Appendix A.

The corresponding functions $W_{S,T,U}^{0,1}(s)$ are now expressed
in terms of the dispersion integrals (the Hilbert transforms)
$\mathcal{G}_{i}(s)$ of these functions, i.e.
\begin{equation}
\mathcal{G}_{i}(s)=\frac{s^{k_{i}}}{\pi }\int_{4m_{\pi }^{2}}^{\infty }\frac{dx}{x^{k_{i}}}\frac{\mathcal{F}_{i}(x)}{x-s}
\end{equation}
with a suitable number $k_{i}$ of subtractions, $0\leq k_{i}\leq 1$.
The $S$-wave contributions in the $s$- and the $t$-channels are given by
\begin{align}
W_{S}^{0}(s)&=\sum_{i=1}^{5}\mathcal{G}_{i}(s) \sum_{k=-1}^{4}a_{k}^{(i)}s^{k},\\
\qquad W_{T}^{0}(s)&=\sum_{i=1}^{5}\mathcal{G}_{i}(s) \sum_{k=-1}^{4}\bar{a}_{k}^{(i)}s^{k}.
\label{WS0}
\end{align}
The $P$-wave contribution in the $t$-channel is more complicated,
\begin{multline}
W_{T}^{1}(s) =\sum_{i=1}^{5}\biggl\{ \mathcal{G}_{i}(s)
\sum_{k=-1}^{2}b_{k}^{(i)}s^{k}+\mathcal{G}_{i}^{(\lambda
)}(s)\sum_{k=-1}^{3}c_{k}^{(i)}s^{k}
+\mathcal{G}_{i}^{(\sigma
)}(s)\sum_{k=0}^{3}d_{k}^{(i)}s^{k}\biggr\}\\
+\sum_{i=1}^{5}\overline{\mathcal{G}}_{i}^{(\lambda
)}(s)~c_{-1}^{(i)}s^{-1}
+\sum_{i=1}^{5}\widetilde{\mathcal{G}}_{i}^{(\lambda
)}(s)~c_{4}^{(i)}s^{4}+\sum_{i=1}^{5}\widehat{\mathcal{G}}_{i}^{(\lambda
)}(s)~c_{5}^{(i)}s^{5},\label{WT1}
\end{multline}
where
(in the following formulae $m_{\pm }=\met\pm \mpi$)
\begin{align}\label{derGi-first}
\mathcal{G}_{i}^{(\lambda)}(s) &= \frac{1}{m_{+}^{2}-m_{-}^{2}}
\bigg( \frac{\mathcal{G}_{i}(s)-\mathcal{G}_{i}(m_{+}^{2})}{s-m_{+}^{2}}
-\frac{\mathcal{G}_{i}(s)-\mathcal{G}_{i}(m_{-}^{2})}{s-m_{-}^{2}}\bigg),
\\
\mathcal{G}_{i}^{(\sigma )}(s) &= \frac{\mathcal{G}_{i}(s)-\mathcal{G}_{i}(4M_{\pi }^{2})}{s-4M_{\pi }^{2}},\\
\overline{\mathcal{G}}_{i}^{(\lambda )}(s) &=\mathcal{G}_{i}^{(\lambda)}(s)-\mathcal{G}_{i}^{(\lambda )}(0), \\
\widetilde{\mathcal{G}}_{i}^{(\lambda )}(s) &=\frac{1}{s}\frac{1}{m_{+}^{2}-m_{-}^{2}}
\bigg(m_{+}^{2}\frac{\mathcal{G}_{i}(s)-\mathcal{G}_{i}(m_{+}^{2})}{s-m_{+}^{2}}
-m_{-}^{2}
\frac{\mathcal{G}_{i}(s)-\mathcal{G}_{i}(m_{-}^{2})}{s-m_{-}^{2}}\bigg),
\\
\widehat{\mathcal{G}}_{i}^{(\lambda )}(s) &=\frac{1}{s^{2}}\Biggl( \mathcal{G}_{i}(s)+\frac{1}{m_{+}^{2}-m_{-}^{2}}
\biggl( m_{+}^{4}\frac{\mathcal{G}_{i}(s)-\mathcal{G}_{i}(m_{+}^{2})}{s-m_{+}^{2}}-m_{-}^{4}
\frac{\mathcal{G}_{i}(s)-\mathcal{G}_{i}(m_{-}^{2})}{s-m_{-}^{2}}\biggr) \Biggr).
\label{derGi-last}
\end{align}
The dependence of these functions on $\mathcal{G}_i(s)$ ensures the correct discontinuity of the function $W_T^1(s)$ and in addition is dictated by the requirement that the appropriate behavior in the chiral limit \cite{Knecht:1995tr,kknzprep} is reproduced.

The explicit form of the functions $\mathcal{G}_{i}(s)$ as well as the properties of the Hilbert transform are discussed in Appendices A and B. Here we only illustrate the above general procedure by means of the explicit result of the first iteration corresponding to the $O(p^4)$ part of the amplitude, and briefly discuss the $O(p^6)$ result.

\subsection{\texorpdfstring{$\eta\to3\pi$}{eta -> 3 pi} at one-loop order}
\label{sec:etaoneloop}
At the one-loop order our dispersive representation (\ref{vyjadreni M}) of the amplitude $\M_x(s,t,u)$ simplifies substantially. The polynomial $\mathcal{P}_x(s,t,u)$ is only of the second order,
\begin{equation}
\mathcal{P}_x(s,t,u)=A_x \met^{2}+B_x(s-s^{c})+C_x(s-s^{c})^{2}
+D_x \left[(t-s^{c})^{2}+(u-s^{c})^{2}\right],
\end{equation}
and of all the functions ${\cal F}_i(s)$ and their Hilbert transforms ${\cal G}_i(s)$ that were introduced in the previous section only the case $i=1$ occurs in the unitarity part (\ref{UetaW}). Besides we only need the first term from (\ref{WT1}).

The single function appearing at $O(p^4)$ is thus
\begin{equation}
\mathcal{G}_1(s)=\frac{s}{\pi}\int_{4M_{\pi }^{2}}^{\infty }\frac{dx}{x}\frac{\sigma(x)}{x-s}=\frac{1}{\pi}\left(2+\sigma(s)\log\frac{\sigma(s)-1}{\sigma(s)+1}\right).
\end{equation}
The form of ${\cal F}_1$ was chosen in order to ensure the relation ${\cal G}_1 = (16 \pi)\bar J(s)$ \cite{Knecht:1995tr} (known also as Chew-Mandelstam function \cite{Chew:1960iv}).

The form of the unitarity part (\ref{UetaW}) at the $O(p^4)$ order is extremely simple in this formalism. For the polynomials introduced in~(\ref{WS0}) and~(\ref{WT1}) in the case of the charged decay $\eta\to \pi^+\pi^-\pi^0$ we find
\begin{align}
16\pi F_\pi^2 a^{(1)}_0  &= \frac16 A_x (7 \alpha_\pi - 16 \beta_\pi) \mpi^2 \met^2
- \frac2 9B_x (\alpha_\pi-\beta_\pi) \mpi^2 ( 3\mpi^2 + \met^2),
\label{a01op4}\\
16\pi F_\pi^2 a^{(1)}_1  &= 2 A_x \beta_\pi \met^2 + \frac16 B_x \bigl(4 \alpha_\pi \mpi^2
- \beta_\pi (7 \mpi^2 +\met^2)\bigr),
\\
16\pi F_\pi^2 a^{(1)}_2 &= \frac12 B_x \beta_\pi
\end{align}
for the polynomials of S-wave in $s$ channel (the polynomials that are not displayed here are identically zero). And then similarly for S-wave in $t$-channel
\begin{align}
16\pi F_\pi^2\bar a^{(1)}_0  &= \frac13 A_x (\alpha_\pi + 2 \beta_\pi) \mpi^2 \met^2
 + \frac{1}{18}B_x (\alpha_\pi+2\beta_\pi) \mpi^2 ( 3\mpi^2 + \met^2),
\\
16\pi F_\pi^2\bar a^{(1)}_1  &= -\frac12 A_x \beta_\pi \met^2 
- \frac{1}{12} B_x \bigl(2 \alpha_\pi \mpi^2 + \beta_\pi (7 \mpi^2 +\met^2)\bigr),
\\
16\pi F_\pi^2\bar a^{(1)}_2 &= \frac14 B_x \beta_\pi
\end{align}
and finally the polynomials of the P-wave contributions that are not zero are given by
\begin{align}
16\pi F_\pi^2 \bar{b}^{(1)}_0  &= - \frac13 B_x \beta_\pi \mpi^2,\\
16\pi F_\pi^2 \bar{b}^{(1)}_1  &= - \frac{1}{12} B_x \beta_\pi.
\end{align}

\subsection{\texorpdfstring{$\eta\to3\pi$}{eta -> 3 pi} at two-loop order}
The whole amplitude at two loops, or equivalently at $O(p^6)$ order, is of course more complicated. We employ here the full form of the polynomial (\ref{polstu}). The non-trivial part follows from the same general form (\ref{UetaW}) with the functions $W_{S,T,U}^{0,1}(s)$ from (\ref{WS0})--(\ref{WT1}), but contrary to the one-loop situation, where we have only one function ${\cal G}_1(s)$, we have to deal with five basic functions ${\cal G}_i$, together with five derived types~(\ref{derGi-first})--(\ref{derGi-last}).
Let us explicitly write down the first coefficient (which stands in front of ${\cal G}_1$ in the $s$-channel of S partial wave and thus together with~(\ref{a01op4}) represents the full $a_0^{(1)}$ at $O(p^6)$):
\begin{widetext}\begin{equation}
\begin{split}
(16\pi)^3&\Delta a_0^{(1)} = \frac{4\met^2}{27F_\pi^4}A_x\mpi^2
\bigl[
 \left(1152 \alpha_\pi ^2-2619 \alpha_\pi  \beta_\pi +3130 \beta_\pi ^2\right)\mpi^2-45 \beta_\pi  (\alpha_\pi
   +2 \beta_\pi )\met^2 
   +1152 \pi ^2
   \left(17 \lambda _1+18 \lambda _2\right) \mpi^2\bigr]\\
&+\frac{4}{81F_\pi^4} B_x \mpi^2 \bigl[
\beta_\pi
    (61 \alpha_\pi -514 \beta_\pi )\met^4+ \left(-459 \alpha_\pi
   ^2+1170 \alpha_\pi  \beta_\pi -1996 \beta_\pi ^2\right)\mpi^2 \met^2\\
&\qquad
-3 \left(447 \alpha_\pi ^2-987
   \alpha_\pi  \beta_\pi +2150 \beta_\pi ^2\right) \mpi^4
   -4608 \pi ^2 \left(2 \lambda _1+3
   \lambda _2\right) \mpi^2 \left(3 \mpi^2+\met^2\right)\bigr]\\
&+\frac{128}{27F_\pi^2} \pi ^2 C_x
   \mpi^2 \bigl[7 \alpha_\pi  \left(3 \mpi^2+\met^2\right)^2-2 \beta_\pi
   \left(81 \mpi^4+30 \mpi^2 \met^2+17
   \met^4\right)\bigr]\\
&+\frac{256}{27F_\pi^2} \pi ^2 D_x\mpi^2 \bigl[7
   \alpha_\pi  \left(3 \mpi^2+\met^2\right)^2-\beta_\pi  \left(171 \mpi^4+42
   \mpi^2 \met^2+43 \met^4\right)\bigr].
\end{split}
\end{equation}\end{widetext}
From this example one can infer the general structure of all other parameters $a_{k}^{(i)}, \ldots , \bar d_{k}^{(i)}$. The full form can be obtained from the authors upon request.
%%%%%%%%%%%%%%%%%%%%%%%%%%%%%%%%%%%%%%%%%%%%%%%%%%%%%%%%%%%%%%%%%%%%%%%%%%%%

\section{Connection with ChPT: order-by-order correspondence}
\label{sec:conwithchpt}
Let us briefly comment on the connection of the dispersive construction with
the standard ChPT expansion. In analogy to the dispersive one, the $O(p^{6})$ ChPT
amplitude can also be split into a polynomial part and a non-analytic
unitarity part. The former corresponds to the tree-level counterterm
contributions as well as to the chiral logs and sunset graphs, while the
latter takes explicitly into account the nontrivial contributions of the
loops. Though this splitting is not unambiguous and depends on the particular
definition of the nontrivial part of the loop graphs, the unitarity part
has to reproduce the correct discontinuities of the amplitude as required by
(generalized) unitarity and corresponding to the two-particle intermediate
states. Along with the pure pion loop contributions also the higher
intermediate states are taken into account, namely, the graphs with kaons and
$\eta $ inside the loops. However, below the $\pi \eta $ threshold the
contributions of discontinuities corresponding to the $\pi \eta $, $KK$ and $\eta \eta $ intermediate states are analytic and can therefore be expanded in powers of $s$, $t$, $u$. Sufficiently far below these thresholds one can show that their effects can be approximated by means of only the terms up to the third order (cf.\ \cite{nas-proceeding3} and the numerical estimate of such error made in Sec.~\ref{sec:ObOC}). As a result we should obtain in this region an
approximate ChPT amplitude with the same structure as our dispersively
constructed amplitude (recall that both of them include the higher non-Goldstone
intermediate states contributions only effectively through the low-energy and the subtraction constants, respectively). The only difference is that the polynomial part of the $O(p^{6})$ ChPT amplitude is generally
complex due to the contribution of the sunset diagram\footnote{Note that this diagram ($\eta\to3\pi\to3\pi$) does not contribute to the unitarity cut of the $\eta\pi\to\pi\pi$ amplitude but instead its contribution in the decay region is analytic and can be expanded into polynomial. This polynomial can be complex since $m_\eta$ is unstable ($m_\eta>3m_\pi$).}  with three
intermediate pions which develop nonzero imaginary part. However, it has been
found to be tiny in \cite{BG,nas-proceeding3} and therefore can be neglected. We reverify this observation in Sec.~\ref{sec:ObOC}.

These common features of both amplitudes suggest that the $O(p^{6})$ ChPT
amplitude ${\mathcal{A}}_{\mathrm{ChPT}}(s,t,u)$, which we write in the form
\begin{equation}
{{\mathcal{A}}}_{\mathrm{ChPT}}(s,t,u)={\cal N}_\eta\bigg( \mathcal{M}_{\mathrm{ChPT}}^{(2)}(s,t,u)
+\mathcal{M}_{\mathrm{ChPT}}^{(4)}(s,t,u)+\mathcal{M}_{\mathrm{ChPT}}^{(6)}(s,t,u)\bigg) ,
\end{equation}
can be reproduced as a special case of the dispersively constructed one. This can be quantified as follows in terms of what we call \emph{order-by-order fit}. The ChPT amplitude in our dispersive parametrization is then represented by expressing particular chiral orders of our subtraction constants $A_x, \dots, F_x$ and $\alpha_\pi, \dots, \lambda_{2}$ in terms of the LECs of ChPT, quark masses and chiral logarithms. Such expressions are then useful when one wants to organize the chiral result and to identify the renormalization-scale invariant combinations of LECs on which the amplitude depends. For the aims of the current work, it is however sufficient to perform this matching numerically and obtain the numerical values of our subtraction constants using the procedure described in the following lines (note that the same procedure would remain valid also if we wanted to obtain the analytic expressions, but instead of fitting the numerical results we would just compare expressions coming from ChPT with the ones of the analytic dispersive construction).

Let us formally split the parameters $A_x$, \ldots ,$F_x$
of our amplitude into their $O(p^{2})$, $O(p^{4})$ and $O(p^{6})$ parts, i.e.
\begin{align}
A_x &=A_x^{(2)}+\Delta A_x^{(4)}+\Delta A_x^{(6)},\\
B_x &=B_x^{(2)}+\Delta B_x^{(4)}+\Delta B_x^{(6)},\\
C_x &=C_x^{(4)}+\Delta C_x^{(6)},\\
D_x &=D_x^{(4)}+\Delta D_x^{(6)},\\
E_x &\equiv E_x^{(6)},\\
F_x &\equiv F_x^{(6)}.
\end{align}
This induces a following splitting of the polynomial part of the
amplitude
\begin{equation}
{\mathcal{P}}_x(s,t,u)={\mathcal{P}}_x^{(2)}(s,t,u)+{\mathcal{P}}_x^{(4)}(s,t,u)+{\mathcal{P}}_x^{(6)}(s,t,u),
\end{equation}
where
\begin{align}
{\mathcal{P}}_x^{(2)}(s,t,u) &=A_x^{(2)}\met^{2}+B_x^{(2)}(s-s^{c}), \\
{\mathcal{P}}_x^{(4)}(s,t,u) &=\Delta A_x^{(4)}\met^{2}+\Delta
B_x^{(4)}(s-s^{c})+C_x^{(4)}(s-s^{c})^{2}+D_x^{(4)}\left[(t-s^{c})^{2}+(u-s^{c})^{2}\right],  \\
\begin{split}
{\mathcal{P}}_x^{(6)}(s,t,u) &=\Delta A_x^{(6)}\met^{2}+\Delta
B_x^{(6)}(s-s^{c})
+\Delta C_x^{(6)}(s-s^{c})^{2}
+\Delta D_x^{(6)}\left[ (t-s^{c})^{2}+(u-s^{c})^{2}\right]\\
&+E_x^{(6)}(s-s^{c})^{3}+F_x^{(6)}\left[(t-s^{c})^{3}+(u-s^{c})^{3}\right].
\end{split}\label{P6}
\end{align}
Note that the unitarity part ${\mathcal{U}}_x(s,t,u)$ splits by construction naturally into the genuine one-loop $O(p^{4})$ and the remaining $O(p^{6})$ parts that correspond to the first and the second iteration of the generalized
unitarity relations, respectively, (see \cite{kknzprep} for more details),
\begin{equation}
{\mathcal{U}}_x={\mathcal{U}}_x^{(4)}\left(A_x,B_x;\alpha_\pi ,\beta_\pi \right)
+\mathcal{U}_x^{(6)}\left(A_x,B_x,C_x,D_x;\alpha_\pi ,\beta_\pi
,\lambda _{1},\lambda _{2}\right).
\end{equation}
The unitarity part ${\mathcal{U}}_x^{(4)}$ has been given in Sec.~\ref{sec:etaoneloop}, where we have written out the explicit dependence on the polynomial parameters of the $\eta \rightarrow 3\pi $ and $\pi \pi
\rightarrow \pi \pi $ amplitudes. The $O(p^{6})$ part consists further of the
genuine two-loop part and the one-loop part
\begin{equation}
{\mathcal{U}}_x^{(6)}(A_x,B_x,C_x,D_x;\alpha_\pi ,\beta_\pi
,\lambda _{1},\lambda _{2})
={\mathcal{U}}_{\mathrm{2-loop}}^{(6)}(A_x,B_x;\alpha_\pi ,\beta_\pi )
+{\mathcal{U}}_{\mathrm{1-loop}}^{(6)}(A_x,B_x,C_x,D_x;\alpha_\pi ,\beta_\pi ,\lambda _{1},\lambda _{2}).
\end{equation}

The $O(p^{2})$ ChPT amplitude ${\mathcal{M}}_{\mathrm{ChPT}}^{(2)}(s,t,u)$
is now exactly reproduced by ${\mathcal{P}}_x^{(2)}(s,t,u)$ with
\begin{equation}
A_x^{(2)} =\frac{\met^{2}-m_{\pi }^{2}}{3\met^{2}},\qquad B_x^{(2)} =1.\label{ABLO}
\end{equation}
The imaginary part of the $O(p^{4})$ ChPT amplitude below the $\pi \eta $
threshold is fixed by unitarity and therefore there holds exactly
\begin{equation}
\im {\mathcal{M}}_{\mathrm{ChPT}}^{(4)}(s,t,u)=\im \mathcal{U
}_x^{(4)}(A_x^{(2)},B_x^{(2)};\alpha_\pi ^{(2)},\beta_\pi ^{(2)}),
\end{equation}
where
\begin{equation}
\alpha_\pi ^{(2)}=\beta_\pi ^{(2)}=1 \label{abLO}
\end{equation}
are the leading order ChPT values of the $\pi \pi \rightarrow \pi \pi $
subthreshold parameters. Hence, up to a polynomial of the second order
in $s$, $t$ and $u$, the amplitudes ${\mathcal{M}}_{\mathrm{ChPT}}^{(4)}(s,t,u)$ and ${\mathcal{U}}_x^{(4)}(A_x^{(2)},B_x^{(2)};\alpha_\pi ^{(2)},\beta_\pi ^{(2)})$ coincide (here we have tacitly assumed
that the higher two-particle intermediate states contributions to $\mathcal{M}_{\mathrm{ChPT}}^{(4)}(s,t,u)$ has been expanded to the second order in $s$, $t$ $\ $and $u$ as described above) and we can therefore write
\begin{equation}
{\mathcal{M}}_{\mathrm{ChPT}}^{(4)}(s,t,u)={\mathcal{P}}_x^{(4)}(s,t,u)+{\mathcal{U}}_x^{(4)}(A_x^{(2)},B_x^{(2)};\alpha_\pi ^{(2)},\beta_\pi
^{(2)})
\end{equation}
for appropriate $\Delta A_x^{(4)}$, $\Delta B_x^{(4)}$, $C_x^{(4)}$ and $D_x^{(4)}$. These parameters are found numerically by fitting the difference
\begin{equation}
\Delta ^{(4)}(s,t,u)={\mathcal{M}}_{\mathrm{ChPT}}^{(4)}(s,t,u)-{\mathcal{U}}_x^{(4)}(A_x^{(2)},B_x^{(2)};\alpha_\pi ^{(2)},\beta_\pi ^{(2)})
\end{equation}
to the second order polynomial ${\mathcal{P}}_x^{(4)}(s,t,u)$. When these $O(p^{4})$ parameters are fixed, we proceed similarly to the $O(p^{6})$
order. We compute the $O(p^6)$ corrections to the unitarity part,
\begin{equation}\begin{split}
{\mathcal{V}}_x^{(6)}(s,t,u) &={\mathcal{U}}_x^{(4)}(\Delta A_x^{(4)},\Delta B_x^{(4)};\alpha_\pi ^{(2)},\beta_\pi ^{(2)})
+{\mathcal{U}}_x^{(4)}(A_x^{(2)},B_x^{(2)};\Delta \alpha_\pi ^{(4)},\Delta \beta_\pi
^{(4)})\\
&+{\mathcal{U}}_{\mathrm{1-loop}}^{(6)}(A_x^{(2)},B_x^{(2)},C_x^{(4)},D_x^{(4)};\alpha_\pi ^{(2)},\beta_\pi ^{(2)},\lambda
_{1}^{(4)},\lambda _{2}^{(4)})
+{\mathcal{U}}_{\mathrm{2-loop}}^{(6)}(A_x^{(2)},B_x^{(2)};\alpha_\pi ^{(2)},\beta_\pi ^{(2)}),
\end{split}\end{equation}
where in addition to the parameters known from the previous steps there appear the NLO corrections of the subthreshold parameters of $\pi\pi\to\pi\pi$ scattering that are needed as inputs to this procedure. The discontinuities originating from the $\pi \pi $ intermediate states in \mbox{$s$-}, $t$- and $u$-channels of ${\mathcal{M}}_{\mathrm{ChPT}}^{(6)}(s,t,u)$ and of this ${\mathcal{V}}^{(6)}_x(s,t,u)$ coincide (modulo a power expansion of the higher-intermediate-state
contributions to the third power as discussed above). Finally, we fit the difference
\begin{equation}
\Delta ^{(6)}(s,t,u)={\mathcal{M}}_{\mathrm{ChPT}}^{(6)}(s,t,u)-{\mathcal{V}}_x^{(6)}(s,t,u)
\end{equation}
to the third order polynomial ${\mathcal{P}}_x^{(6)}(s,t,u)$ and set the remaining $O(p^6)$ parameters. In this way, all
the parameters of the polynomial part of the amplitude are numerically
determined and the $O(p^{6})$ ChPT amplitude ${\mathcal{A}}_{\mathrm{ChPT}}(s,t,u)$ is represented now as ${\mathcal{A}}_{\mathrm{ChPT}}(s,t,u) \rightarrow {\mathcal{A}}_{\mathrm{ord}}^{\mathrm{disp}}(s,t,u)$, where
\begin{equation}
{\mathcal{A}}_{\mathrm{ord}}^{\mathrm{disp}}(s,t,u)={\mathcal{P}}_x^{(2)}(s,t,u)+{\mathcal{P}}_x^{(4)}(s,t,u)+{\mathcal{P}}_x^{(6)}(s,t,u)
+{\mathcal{U}}_x^{(4)}(A_x^{(2)},B_x^{(2)};\alpha_\pi ^{(2)},\beta_\pi
^{(2)})+{\mathcal{V}}^{(6)}(s,t,u).
\end{equation}

By construction, the chiral orders of the various contributions to $\mathcal{A}_{\mathrm{ChPT}}^{\mathrm{disp}}(s,t,u)$ were strictly respected --- for instance the genuine two-loop unitarity corrections depend only on
the leading order parameters $A_x^{(2)},B_x^{(2)}$ and
$\alpha_\pi ^{(2)},\beta_\pi ^{(2)}$. However, the known general form of
the dispersively constructed amplitude ${\mathcal{A}}(s,t,u)$ can
be further used in order to go beyond the strict chiral expansion
and partially resum also the higher chiral-order contributions.
This representation that we call \emph{resummed fit} can be achieved by means of inserting the full parameters
$A_x$, \ldots ,$F_x$ obtained by the above
\emph{order-by-order fit} and the full $O(p^{4})$ $\pi \pi$ subthreshold
parameters (or even the experimental values of the $\pi\pi$ subthreshold parameters from \cite{DescotesGenon:2001tn})
into the unitarity part of the amplitude, i.e.\ to define
\begin{equation}
{\mathcal{A}}_{\mathrm{res}}^{\mathrm{disp}}(s,t,u)={\mathcal{P}}_x^{(2)}(s,t,u)+{\mathcal{P}}_x^{(4)}(s,t,u)+{\mathcal{P}}_x^{(6)}(s,t,u)
+{\mathcal{U}}_x^{(4)}(A_x,B_x;\alpha_\pi ,\beta_\pi )
+{\mathcal{U}}_x^{(6)}(A_x,B_x,C_x,D_x;\alpha_\pi ,\beta_\pi ,\lambda
_{1},\lambda _{2}).
\end{equation}
The difference ${\mathcal{A}}_{\mathrm{res}}^{\mathrm{disp}}-{\mathcal{A}}_{\mathrm{ord}}^{\mathrm{disp}}$ is of order $O(p^{8})$ and contains effectively
contributions of the one and the two-loop graphs with higher-order counterterms.
It might be therefore treated as a rough estimate of the convergence of the
chiral expansion.

Let us note that we could also use another parametrization of the relevant $\pi \pi $ scattering amplitude based on the scattering lengths and effective
ranges instead of the subthreshold parameters (see \cite{kknzprep,nas-proceeding2} for details) and
repeat the above construction along the same lines. In such a case the
amplitude ${\mathcal{A}}_{\mathrm{ord}}^{\mathrm{disp}}(s,t,u)$
has to be numerically the same as before, namely, the parameters $A_x$, \ldots ,$F_x$ should be the same. However, the amplitude will now
depend on the scattering lengths and the effective ranges of the $\pi \pi $
scattering taken up to the order $O(p^{4})$. Provided we then use the
experimental values of these parameters in the \emph{resummed} amplitude $\mathcal{A}_{res}^{\mathrm{disp}}(s,t,u)$, we can interpret the result as a partial
resummation of the two-particle rescattering in the final state. The
numerical effect of such a resummation might be even larger than within the
previous parametrization, because the scattering lengths are known to have
much worse convergent chiral expansion than the subthreshold parameters.

\section{Analysis of the charged decay: \texorpdfstring{$\eta\to\pi^+\pi^-\pi^0$}{eta -> pi(+) pi(-) pi(0)}}\label{sec:charged}
We have prepared everything to employ the dispersive representation for our analysis of the process $\eta\to3\pi$.
It proceeds as follows. We start with the NNLO result of ChPT \cite{BG}. We determine the values of our parameters that reproduce the ChPT result, thereby checking also that the correspondence between these two frameworks holds using the \emph{order-by-order fit} as outlined in the previous section. Our further analysis is motivated by the conclusion of Sec.~\ref{sec:Ci} that the observed mismatch between the ChPT $O(p^6)$ predictions of the Dalitz parameters and their experimental determination by KLOE might be caused by the incorrect determination of the $O(p^6)$ LECs $C_i$ of ChPT. We therefore study the dispersive representation of ChPT with the values of the $C_i$s undetermined and try to find the values of their combinations that reproduces the experimental data. Finally, after that we change completely the strategy and fit directly our dispersive representation to the experimental data. Such a fit gives us the $\eta\to3\pi$ amplitude up to the normalization that is determined from the matching with ChPT in the region where we can believe the ChPT result. In all the cases we are interested in the distribution we obtain and then by comparing the decay widths computed from these distributions (by integration of the square of the amplitude over the physical phase space) with the experimentally measured one, we obtain the value of $R$.

In principle, this could be done for both the charged and the neutral $\eta\to3\pi$ decay. However, as was discussed in Sec.~\ref{Dalitz plot parametrization}, no current experiment determined more than just one Dalitz parameter describing the neutral decay, thus we concentrate mainly on the charged one. Even in the charged sector the experimental situation is poor --- only KLOE \cite{KLOE-etac} published just 4+1 Dalitz parameters (the last one claimed to be compatible with zero) describing the amplitude. From these values of the Dalitz parameters we have constructed a distribution in the physical region (in the similar way as done in \cite{Kupsc:2009zza}) and all our experimental fits are fits to such KLOE-like distributions, in our analysis we therefore depend fully on these KLOE measurements.

\subsection{Order-by-order correspondence: obtaining numerical ChPT distribution}
\label{sec:ObOC}
\begin{figure}
\begin{center}
\includegraphics{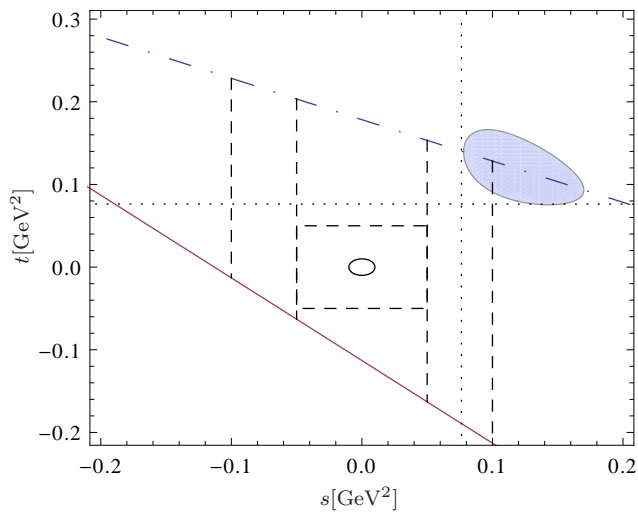}
\end{center}
\caption{Domain of applicability of the dispersive method together with the
physical region (ellipse). Full magenta line
represents the $\pi\eta$ threshold in the $u$-channel (the only $\pi\eta$ threshold visible for this range of $s$ and $t$). The blue dot-dashed line represents the axis of the $t-u$ symmetry of the amplitude: e.g.\ the Adler zero in $s,t=0$ (represented by a circle) has its counterpart for $s=0$ and $t=m_\eta^2 + 3 m_\pi^2$.
Dotted lines denote the $4 m_\pi^2$ thresholds in the $s$- and the $t$-channels.
Dashed lines define different regions in the matching procedure (see main
text).}
\label{fig:domains}
\end{figure}

As was discussed in the previous section we can obtain the approximate chiral $O(p^6)$ amplitude as a special case of our dispersive parametrization with some particular values of our parameters. The correspondence between such amplitude and the result of ChPT has to be almost identical neglecting only small effects descending from expansion of the two-kaon and the $\pi\eta$ contributions and a tiny imaginary part produced by sunset-like diagrams. In principle, working in the $(s,t)$ plane they should agree in the region%
\footnote{Note that even though we are talking about the expansion for small Mandelstam variables (e.g.\ $s$ and $t$), it does not simply mean that the smaller these variables are, the better agreement between these theoretical frameworks we obtain. The amplitude depends on three kinematic variables $s,t,u$ which are connected by relation~(\ref{soucet Mandelstamu decay}). So keeping two of them small, the third one is shifted up by the $\met$ mass.}
for small $s,t$ under the $\pi\eta$ thresholds in all the crossed channels. Although much bigger deviation should be visible only after $KK$ threshold (the contribution of $\pi\eta$ is very small) we stick on this as a strict limit of our method.
Influence of systematic uncertainties is studied using different regions in our matching procedure (see below).
The physical and the matching regions together with the $\pi\eta$ threshold are depicted in Fig.~\ref{fig:domains}.

We match the amplitude along the lines of the previous section really order by order.
The correspondence for LO and the imaginary part of NLO can be verified analytically, having $A_x^{(2)}$ and $B_x^{(2)}$
from (\ref{ABLO}) and $\alpha_\pi^{(2)}=\beta_\pi^{(2)}=1$. After that we have proceeded with the matching numerically.
From the NLO real part of the amplitudes we have fitted the parameters $A_x^{(4)}$, $B_x^{(4)}$, $C_x^{(4)}$, $D_x^{(4)}$
(in the notation of Sec.~\ref{sec:conwithchpt}, $A_x^{(4)} = A_x^{(2)}+\Delta A_x^{(4)}$, etc.). After that we have verified the matching of the imaginary NNLO amplitudes and finally from the real part of the NNLO amplitudes fitted the parameters
$A_x^{(6)}, \dots, F_x^{(6)}$ (again $A_x^{(6)} = A_x^{(4)}+\Delta A_x^{(6)}$; this superscript is used for the NNLO values of $A_x, \dots$ just to distinguish these values from the ones of the \emph{overall fit} from Sec.~\ref{sec:res}).

Concerning the $\pi\pi$ part we follow closely the determination of its subthreshold parameters as established in~\cite{Knecht:1995tr}. For the particular values we have used~(\ref{abLO}) for the leading order and set the NLO values to be
\begin{equation}\begin{split}
      \alpha_\pi^{(4)} = 1.044\,,&\qquad
      \beta_\pi^{(4)}  = 1.083\,,\\
      \lambda_1^{(4)}  = -1.43\times 10^{-3}\,,&\qquad
      \lambda_2^{(4)}  = 8.5\times 10^{-3}.
\end{split}\end{equation}
The fits were performed in the following regions\footnote{Since the amplitude is $t\leftrightarrow u$ symmetric, one can fit it only in the region below the $t-u$ line.} (all numbers in GeV$^2$; cf.~Figure~\ref{fig:domains}):
\begin{itemize}
 \item set 0: the physical region;
 \item set 1: the square region around the Adler zero $(s,t)\in (-0.05,0.05)$;
 \item set 2: the triangle region between the lines $s=4 m_\pi^2$, $t=4 m_\pi^2$ and the $\pi\eta$ threshold;
 \item set 3: $s\in (-0.05,0.05)$, $t$ between the $\pi\eta$ threshold and the $t\leftrightarrow u$ axis;
 \item set 4: $s\in (-0.1,0.1)$, $t$ between the $\pi\eta$ threshold and the $t\leftrightarrow u$ axis.
\end{itemize}
Distances between the points in grids are constant in both the $s$ and the $t$ directions,
and the approximate total number of them is the following: set 1: 300, set 2: 900, set 3:
1600 and set 4: 4400. Further, for the physical region (set 0) we have chosen the same points that were used
in~\cite{BG}, i.e.\ 174 points, which is a very similar number to the KLOE's
number of bins (154, cf.\ also discussion in \cite{Kupsc:2009zza}).
The different regions with the different numbers of points were set in order to have systematic and statical errors under control.

For the fits we have used \textsf{MINUIT} package with the weights of the individual points set to $\Delta O(p^6)/2$.
Results for the NLO parameters are summarized in Table~\ref{tab:sub4} and the ones for the NNLO parameters then in
Table~\ref{tab:sub6}. The error bars quoted for the individual parameters are results of \textsf{MINUIT}.

\begin{table*}
\centering
\begin{tabular}{@{\hspace{1em}}c@{\hspace{2em}}r@{$\:\pm\:$}l@{\hspace{2em}}r@{$\:\pm\:$}l@{\hspace{2em}}r@{$\:\pm\:$}l@{\hspace{2em}}r@{$\:\pm\:$}
l@{\hspace{2em}}r@{$\:\pm\:$}l@{\hspace{0.5em}}}
\hline\hline
& \multicolumn{2}{c@{\hspace{2em}}}{set 0} & \multicolumn{2}{c@{\hspace{2em}}}{set 1} & \multicolumn{2}{c@{\hspace{2em}}}{set 2} & \multicolumn{2}{c@{\hspace{2em}}}{set 3} & \multicolumn{2}{c@{\hspace{0.5em}}}{set 4}\\\hline
$A_x^{(4)}$
& $0.464$ & $ 0.017$
& $0.457$ & $ 0.040$
& $0.459$ & $0.011$
& $0.452$ & $0.016$
& $0.462$ & $0.010$
\\
$B_x^{(4)}$
& $1.95$ & $0.15$
& $1.90$ & $0.21$
& $1.91$ & $0.05$
& $1.88$ & $0.09$
& $1.93$ & $0.06$
\\
$C_x^{(4)}$
& $-0.42$ & $ 4.4$
& $-0.68$ & $0.91$
& $-0.62$ & $0.17$
& $-0.76$ & $0.41$
& $-0.51$ & $0.25$
\\
$D_x^{(4)}$
& $1.07$ & $ 3.8$
& $1.04 $ & $ 0.03$
& $1.04 $ & $ 0.01$
& $1.04 $ & $ 0.02$
& $1.04 $ & $ 0.01$
\\
\hline\hline
\end{tabular}
\caption{The values of the $O(p^4)$ dispersive parameters (in appropriate powers of [$GeV$]) corresponding to matching to ChPT NLO amplitude.}
\label{tab:sub4}
\end{table*}
%%%%%
%%%%%
\begin{table*}
\centering
\begin{tabular}{@{\hspace{1em}}c@{\hspace{2em}}r@{$\:\pm\:$}l@{\hspace{2em}}r@{$\:\pm\:$}l@{\hspace{2em}}r@{$\:\pm\:$}l@{\hspace{2em}}r@{$\:\pm\:$}l@{\hspace{2em}}r@{$\:\pm\:$}l@{\hspace{0.5em}}}
\hline\hline
& \multicolumn{2}{c@{\hspace{2em}}}{set 0} & \multicolumn{2}{c@{\hspace{2em}}}{set 1} & \multicolumn{2}{c@{\hspace{2em}}}{set 2} & \multicolumn{2}{c@{\hspace{2em}}}{set 3} & \multicolumn{2}{c@{\hspace{0.5em}}}{set 4}\\\hline
$A_x^{(6)}$
& $0.577 $ & $ 0.013$
& $0.581$ & $ 0.003$
& $0.581$ & $ 0.023$
& $0.577$ & $ 0.002$
& $0.583$ & $ 0.011$
\\
$B_x^{(6)}$
& $2.42 $ & $ 0.22$
& $2.460 $ & $ 0.012$
& $2.47$ & $ 0.19 $
& $2.44$ & $ 0.01$
& $2.49$ & $  0.10 $
\\
$C_x^{(6)}$
& $0.24 $ & $ 3.4$
& $0.30 $ & $  0.11$
& $0.38$ & $ 1.7 $
& $0.20 $ & $ 0.09$
& $0.55$ & $  0.90 $
\\
$D_x^{(6)}$
& $1.55 $ & $ 2.4$
& $ 1.57 $ & $ 0.02$
& $1.58$ & $ 0.03$
& $1.58 $ & $ 0.02$
& $1.58$ & $  0.02$
\\
$E_x^{(6)}$
& $5 $ & $  149$
& $5.4 $ & $  0.7$
& $5.6$ & $ 4.7 $
& $5.1 $ & $ 0.6 $
& $6.1$ & $  2.6$
\\
$F_x^{(6)}$
& $-4 $ & $  84$
& $-3.6 $ & $  0.1$
& $-3.7$ & $ 0.2$
& $-3.7 $ & $ 0.1$
& $-3.7$ & $  0.1$
\\\hline\hline
\end{tabular}
\caption{The values of the $O(p^6)$ dispersive parameters (in appropriate powers of [$GeV$]) corresponding to matching to ChPT NNLO amplitude.}
\label{tab:sub6}
\end{table*}

At the moment we have in hands the dispersively constructed amplitude (i.e.\ the analytic formula) which is numerically equivalent (or very close) to NNLO ChPT amplitude. We can verify the equivalence also by computing the decay width.

Our dispersive representation was constructed in accordance with chiral
perturbation theory and we have chosen similar normalization as used in \cite{BG} (with an extra factor $1/F_\pi^2$).
We can thus compare directly a neat amplitude ${\cal M}_x(s,t,u)$ with the isospin-breaking
factor pulled out as defined in (\ref{ARM}).
The result of the integration of the amplitude square over the physical phase space is (cf. (6.7) in \cite{BG}):
\begin{equation}
 \Gamma(\eta\to\pi^+\pi^-\pi^0) = \sin^2 \epsilon \times 2.68 \text{ MeV}, \quad
\text{[ChPT]}
\end{equation}
where we have introduced
\begin{equation}
\sin\epsilon = \frac{\sqrt{3}}{4R}\,.
\end{equation}
Comparing this result with the experimental measurement for the decay rate \cite{PDG} we
arrive at the value which exactly reproduces the one of \cite{BG} (mind the typo in \cite{BG})
\begin{equation}
 R  = 41.3. \quad
\text{[ChPT]}
\end{equation}

We can use result \cite{BG} also for a numerical estimate of the error induced by a few approximations in our parametrization we have made with respect to the ChPT computation. As was discussed in Sec.~\ref{sec:conwithchpt}, we have neglected the imaginary parts of our parameters (which are connected with the contribution of the sunset diagram). In the physical region we have performed fits, in which we have allowed the $O(p^6)$ parameters to be complex. We have found that the NNLO ChPT result is very well approximated by adding a constant imaginary term $\im A_x^{(6)}=0.080\pm0.064$. By neglecting this term in the computation of $R$ we introduce an error of 0.1\%. Similarly, we have neglected higher than third order polynomial terms in the expansion of $KK$ and $\pi\eta$ contributions (in the decay region). We can estimate the corresponding error by addition of some higher-order terms into the polynomial. The symmetries dictate that the fourth order polynomial would contain terms $G_x(s-s^c)^4+H_x(s - s^c)^2 (t - s^c) (u - s^c) + I_x\left((t - s^c)^4 + (u - s^c)^4\right)$. From the dimensional considerations, the contribution of $KK$ intermediate states into these parameters should be $\sim\frac{1}{M_K^4(4\pi F_\pi)^2}\lesssim 10^2\,\mathrm{GeV}^{-6}$ (and similarly for $\pi\eta$), whereas even if all of them were $\sim10^4\,\mathrm{GeV}^{-6}$ the shift in the determined $R$ would be 0.1\%. Both of the errors are therefore negligible with respect to the other sources or error discussed in the following analyses.

\subsection{Correction to order-by-order fit: Correcting the \texorpdfstring{$C_i$s}{Ci's} in ChPT}\label{sec:corOO}
In the previous subsection we have constructed the dispersive amplitude reproducing ChPT in the region where our method is
applicable. It is no surprise that if we fitted this dispersive representation to the Dalitz parametrization (\ref{dalitz etac}) as was done in \cite{BG}, we would obtain the same values of the Dalitz parameters as \cite{BG}. In Sec.~\ref{sec:Ci} we have found an indication that the discrepancy between so obtained values and the values measured by KLOE can be (at least partially) caused by the incorrect values used for the $O(p^6)$ LECs $C_i$ of ChPT. The contribution of the $C_i$s to the $O(p^6)$ amplitude is polynomial and real and so changing them means changing the $O(p^6)$ part of our polynomial (\ref{P6}) --- shifting the parameters appearing in it. By studying the chiral amplitude obtained from our previous analysis with an unknown $O(p^6)$ polynomial added,
\begin{equation}\label{opravena amplitude}
 {\cal A}_x^{cor}(s,t,u) = {\cal A}_x(s,t,u) + \Delta {\cal A}_x(s,t,u),
\end{equation}
with $\Delta {\cal A}_x(s,t,u) = {\cal N}_\eta \Delta {\cal P}_x(s,t,u)$ and
\begin{multline}
 \Delta {\cal P}_x (s,t,u) = \Delta A_x M_\eta^2 + \Delta B_x (s-s^c) +
\Delta C_x(s-s^c)^2
+ \Delta D_x \bigl[(t-s^c)^2 + (u-s^c)^2 \bigr]\\
+ \Delta E_x (s-s^c)^3
 + \Delta F_x \bigl[(t-s^c)^3 + (u-s^c)^3 \bigr]\,,
\label{DPx}
\end{multline}
we can thus study the impact of the corrected $C_i$s on the chiral $O(p^6)$ amplitude.

Provided the dominant part of the discrepancy between the NNLO chiral result and the measured amplitude is hidden just in the incorrect determination of the $C_i$s, the chiral $O(p^6)$ amplitude with the correct set of the $C_i$s, and thereby also the corrected amplitude ${\cal A}_x^{cor}(s,t,u)$, should reproduce the physical data. Therefore by fitting the KLOE-like distribution, we should obtain the values of the dispersive parameters corresponding to the correct values of the $C_i$s. By comparison of these values with the analytic expressions of these parameters in terms of the $C_i$s, one could obtain approximate constraints that the correct values of the $C_i$s should fulfill\footnote{At the current level these constraints could be formulated in terms of reproducing the measured Dalitz plot parameters. For every such parameter by using relations of Section III.A and the observed difference between its experimental value and the value coming from \cite{BG} with all $C_i=0$, one obtains one constraint on the $C_i$s. Note that provided the information on $R$ was supplied from another source with enough accuracy, one could obtain one additional constraint on the $C_i$s. Unfortunately, such constraints are very complicated and would need to be analyzed together with additional constraints coming from other processes (similarly as was done in \cite{Bijnens:2011tb}) in order to provide any useful information on the values of $C_i$s.}.

However, in the case the change of the $C_i$s is insufficient in order to obtain the physically measured amplitude and there still remains a big difference between the physical amplitude and the one obtained from NNLO ChPT, the fit of the amplitude ${\cal A}_x^{cor}(s,t,u)$ to the physical data would mean this time that this difference was parametrized (and approximated) by the polynomial $\Delta {\cal A}_x(s,t,u)$.

Unfortunately, just from the fit of the amplitude ${\cal A}_x^{cor}(s,t,u)$ to the physical data there is no way how to distinguish between these two scenarios --- either that the $O(p^6)$ chiral amplitude with the correct values of the $C_i$ describes well the physical amplitude or that the higher-order remainder can be on the physical region approximated by the polynomial (or if both of the situations are present in a combination, there is no way how to separate these two contributions). We should be therefore careful with the interpretation of the result of such a fit and take this fit just as a starting point for the deeper analysis of the chiral amplitudes. Note that in Sec.~\ref{sec:Ci} we have listed a few criteria that would indicate the realization of the first scenario.

In any case the amplitude constructed that way should describe the data better than the ChPT parametrization of the previous subsection and the determination of $R$ from this distribution will be closer to the real one.

We have therefore fitted all $\Delta$ parameters from (\ref{DPx}) using the condition that the amplitude (\ref{opravena amplitude}) has to fulfill the distribution of data based on KLOE. This distribution is
limited only within the ellipse of the physical region (cf.\ Figure~\ref{fig:domains}).
We have performed the fit for the following two data sets. First we have created exactly the same points as done for set 0 (or equivalently in \cite{BG}) and then much more (2500) in order to study how this affects the dependence on the statistics. Using the different sets of parameters for the chiral amplitude ${\cal A}_x(s,t,u)$ as summarized in Table~\ref{tab:sub4} and \ref{tab:sub6} has very little effect on the resulting $\Delta$ parameters, so we display only their average, cf.\ Table~\ref{tab:cor}. For further applications we have also fitted the complete $O(p^6)$ polynomial
${\cal P}_x^{(6)}(s,t,u)+\Delta {\cal P}_x (s,t,u)$, whose results are presented in the first column of Table~\ref{tab:resummed}.
\begin{table}
\centering
\begin{tabular}{@{\hspace{1em}}c@{\hspace{2em}}r@{$\:\pm\:$}l@{\hspace{2em}}r@{$\:\pm\:$}l@{\hspace{1em}}}\hline\hline
& \multicolumn{2}{c@{\hspace{2em}}}{\# 174} & \multicolumn{2}{c@{\hspace{1em}}}{\# 2500} \\\hline
$\Delta A_x$
& $-0.05 $ & $ 0.3$
& $-0.029$ & $ 0.003$
\\
$\Delta B_x$
& $-0.5 $ & $ 1$
& $-0.46 $ & $ 0.01$
\\
$\Delta C_x$
& $-7 $ & $ 2$
& $-6.97 $ & $ 0.07$
\\
$\Delta D_x$
& $-0.7 $ & $ 0.8$
& $-0.64 $ & $ 0.02$
\\
$\Delta E_x$
& $-37 $ & $  18$
& $-36 $ & $  3$
\\
$\Delta F_x$
& $24 $ & $  5$
& $24 $ & $  1$
\\\hline\hline
\end{tabular}
\caption{Corrections to the chiral dispersive parameters in order to reproduce the KLOE data (in appropriate powers of $GeV$).}
\label{tab:cor}
\end{table}
Since we have no further information from KLOE concerning the efficiency
of every selected bin, the appointed errors in both tables are not very reliable.

As we have discussed above, it is not easy to interpret the result (moreover, if the fit relies just on the KLOE-like distribution we have made just from the 4+1 measured Dalitz parameters and do not have any notion of the systematic uncertainties here), but let us present some interesting observations. All the corrections of the parameters seems to reduce the original $O(p^6)$ contributions to these parameters coming from \cite{BG}, some of them even tend to change the sign of the total $O(p^6)$ contribution with respect to the original one. Taking into account relation (\ref{Eplus2F}) together with the condition (\ref{f rovna se g}) imply that in the case that all the difference between the physical amplitude and the original $O(p^6)$ chiral one is hidden in the wrong determination of the $C_i$s,
\begin{equation}
\Delta E_x+2\Delta F_x =0.
\end{equation}
From Table~\ref{tab:cor} we see this tendency.

Let us return to our main interest, the determination of the ration $R$. The integrated decay rate computed from this distribution is
\begin{equation}
 \Gamma(\eta\to\pi^+\pi^-\pi^0) = \sin^2 \epsilon \times 2.24(10) \text{ MeV}.
\end{equation}
Together with the experimental input for $\Gamma$ this leads to
\begin{equation}
 R  = 37.7(9).\quad
\text{[ChPT+disp.+KLOE]}
\end{equation}
The quoted error is based only on getting values for parameters in different regions as explained in the previous text.
It does not take into account systematic errors coming from the experimental data, which we do not know, and the theoretical error from the ChPT part. The later one can be estimated from the convergence of results coming from the chiral expansion.

The LO value coming from the current algebra is $R=19.1$ and the NLO result was $R=31.8$. Taking these values into account, our prediction for this ratio from the NNLO ChPT and data of KLOE is
\begin{equation}
 R  = 37.7\pm 2.8.\quad
\text{[ChPT+disp.+KLOE]}\label{Roo}
\end{equation}

\subsection{Overall fit: using the measured distribution}\label{sec:res}
So far we were using the dispersive representation in a very close connection with ChPT and up to the fact that we were trying to correct it by the values of the $C_i$s coming from KLOE, we just reproduced the amplitude
coming from this theoretical framework. But we can also change our strategy totally; we can use the experimentally measured distribution and employ ChPT just for the normalization. In this analysis we therefore assume that no matter what the proper description of this process leading to the correct physical amplitude would be, it would fulfill the general principles of quantum field theory together with the observed hierarchy of various contributions\footnote{This hierarchy is expressed in the construction of the representation in terms of the very basic chiral counting of the partial waves of the amplitude -- cf.~relations (2.2) and (2.3) in \cite{nas-proceeding2}.} which were used for the construction of our analytic dispersive representation. It then means that such correct physical amplitude can be to a good extent described by our parametrization and the later can be used for its analytic continuation to some region where ChPT gives a reliable result for the amplitude, and can be used there for the matching. Such procedure will substantially reduce the influence on $R$ of the error connected with the chiral expansion of the amplitude.

Let us start with the easier part --- fitting the KLOE-like distribution, which plays in this analysis a role of the experimental distribution. In contrast to our previous fits, where respecting the chiral orders of the dispersive parameters was natural (and important), in this case keeping the different chiral orders of the parameters makes no sense. The more natural approach is using our representation in the \emph{resummed} form --- the values of the parameters in the polynomial and in the unitarity part are the same.

The fit of this general representation to KLOE-like distribution was performed for the same data set as in the previous subsection (174 data points). Note that in the fit, the overall normalization is set so that the amplitude is equal to one at the center of the Dalitz plot. In order to simplify the comparison between this and the previous fits, we have decided to multiply all these data by the numerical factor that produces the same number for parameter $A_x$ as the corrected value obtained from the analysis in the previous subsection (average over all data sets), i.e.\ $A_x =
A_x^{(6)}+ \Delta A_x =  0.575$.

\begin{table}
\centering
\begin{tabular}{@{\hspace{1em}}c@{\hspace{2em}}r@{$\:\pm\:$}l@{\hspace{2em}}r@{$\:\pm\:$}l@{\hspace{1em}}}\hline\hline
& \multicolumn{2}{c@{\hspace{2em}}}{cor.set} & \multicolumn{2}{c@{\hspace{1em}}}{fit to KLOE} \\\hline
$ A_x$
& $0.575 $ & $ 0.006$
& $0.575 $ & $ 0.001$
\\
$ B_x$
& $1.99 $ & $ 0.04$
& $2.15 $ & $ 0.02$
\\
$ C_x$
& $-6.8 $ & $ 0.3$
& $-5.8 $ & $ 0.2$
\\
$ D_x$
& $0.94 $ & $ 0.03$
& $0.87 $ & $ 0.08$
\\
$ E_x$
& $-31 $ & $ 3$
& $-19 $ & $ 9$
\\
$ F_x$
& $20 $ & $ 1$
& $21 $ & $  5$
\\\hline\hline
\end{tabular}
\caption{Dispersive parameters as free parameters fitted to KLOE (ultimate column; in
appropriate powers of $GeV$). For a comparison in the first column we have also summarized the corrected
parameters of the previous subsection.}
\label{tab:resummed}
\end{table}

The values of so normalized dispersive parameters coming from the overall fit of KLOE are presented in the ultimate column of Table~\ref{tab:resummed}. For the comparison there are displayed also the corrected values of these parameters from the previous subsection. Note that these two sets of values obtained from fit to KLOE correspond to two amplitudes with different unitarity parts (the first one contains the parameters of ChPT respecting their chiral orders, whereas the later contains in the unitarity part exactly those values of the parameters appearing in the polynomial part) and that in both cases the unitarity part forms an important part of the amplitude. The agreement between these two sets that is seen in this table is therefore quite interesting (the small disagreement is seen only for parameter $E_x$ which is, however, given with the biggest error in both approaches). Let us stress once more that the particular overall normalization was taken only to simplify this comparison and we still have to remember that values in Table~\ref{tab:resummed} are multiplied by, for the moment, unknown constant.

To set this overall normalization is, in our opinion, the main issue of the dispersive study for
$\eta\to3\pi$, so we try to be as cautious as possible. First of all, we will rely on a set of points rather than only on one point even though it would be sufficient for setting the normalization. Thus, we need to select the region of the points where we believe ChPT result. In order to achieve this task, we discuss the following articles which could be important for its selection:
\begin{itemize}
\item[\rim1)] Adler zero condition
\item[\rim2)] correspondence with order-by-order fit and convergence of chiral orders
\item[\rim3)] plateau argument
\end{itemize}
Let us explain them in detail.  The point \rim1) is connected with the $SU(2)\times SU(2)$ theorem
which was summarized in Sec.~\ref{sec:adler}. According to this
theorem the values of the charged amplitude at the points (\ref{adlerxi}), in
a small vicinity of $SU(2)\times SU(2)$ Adler zero, are $O(m_\pi^2)$ for $m_\pi\to 0$ and therefore protected
from being large. At $O(p^4)$ the zero of the real part of the
amplitude, namely, the point $s=u=1.4 m_{\pi}^2$ belonging to
the above set of protected points, possesses the following additional convenient
property. The slope of the amplitude (which is generally not
protected by that theorem) develops $O(p^4)$ corrections that are accidentally small. In the previous dispersive analysis \cite{Anisovich:1996tx}, this fact was the main motivation for matching of the dispersively constructed amplitudes exactly at this point. However, as we have learned from Sec.~\ref{sec:adler}, there is no guarantee that at the points near the Adler zero the chiral corrections to any calculated order are small.
Nevertheless, the points where the real part of the amplitude vanishes (which are often  called ``Adler zeros`` in this context too) can serve as good reference points, or benchmarks, of the individual chiral orders.
In such a way these points were used in the analysis of the NNLO ChPT calculations \cite{BG} with a result that the best convergence of their positions is observed on the line\footnote{Note that these points do
not belong to the set (\ref{adlerxi}), i.e.\ they are not close to the Adler zero in the strict sense of the $SU(2)\times SU(2)$ theorem.}
$t=u$. On the other hand, according to the same analysis, the point $s=u=1.4 m_{\pi}^2$ does not seem to be particularly stable with respect to the $O(p^6)$ corrections. The matching of the overall normalization at the
``Adler zeroes`` has however the advantage that at these points just the imaginary part of the amplitude is matched and therefore the uncertainty corresponding to the not well known $O(p^6)$ LECs is eliminated (or suppressed when we match in their vicinity).

The second article \rim2) can help us to reformulate the previous conclusion in different words.
The order-by-order fit should be by construction very similar to the chiral expansion. The
convergence of this expansion was crucial in \cite{Anisovich:1996tx} for setting
the matching point in $s=u$ around the $O(p^4)$ Adler zero ($s=1.4 m_\pi^2$). However, following the
detailed analysis of \cite{BG} we have to conclude that this choice becomes to be very
dangerous for matching at $O(p^6)$ (for $s=u$ the imaginary part at NNLO is even of the opposite sign when compared with NLO). Much better convergence when
coming from NLO to NNLO is seen for $t=u$, which we use in the following.

Finally \rim3) reflects the stability of the points within the given region or
cut. Studying some physical observable (for example the decay width) as a function
of the matching point one would expect a plateau behavior in the correct region.

Using the previous arguments we fit the normalization of subthreshold
parameters for $t=u$ cut (below the physical threshold) matching only the
imaginary part of ChPT amplitude where we interpolate between both variants,
the standard and the resummed one (cf.\ also Fig.~\ref{figtu} below).
Within the straightforward analysis one obtains
\begin{equation}
 \Gamma(\eta\to\pi^+\pi^-\pi^0) = \sin^2 \epsilon \times 2.25(40) \text{ MeV},
\end{equation}
which corresponds to the value
\begin{equation}\label{R disp}
 R  = 37.8 \pm 3.3.\quad
\text{[disp.+KLOE]}
\end{equation}

This number depends more strongly on the data than it was in the case of the order-by-order fit of the previous subsection. This is the reason why we have dropped ``ChPT'' in
its description even though one should remember that it enters the determination of this number through the normalization as explained above. The sources of errors are thus of the two types: the uncertainties connected with the experiment --- the uncertainties of the experimental data we fitted and the accuracy of their parametrization by our dispersive representation; and the uncertainties connected with the normalization procedure --- the error that is induced by the analytic continuation of the parametrization from the region where we have fitted the data to the region where we have matched with ChPT and finally, the error of the the determination of the values of the amplitudes in this region from ChPT, the error of the chiral expansion.

\begin{figure}
\begin{center}
\includegraphics{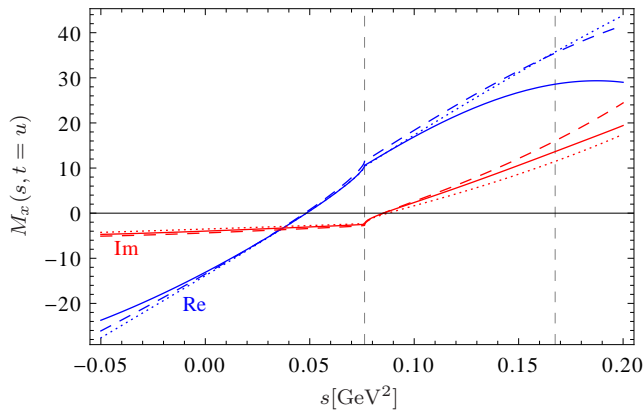}
\end{center}
\caption{The real (blue) and the imaginary (red) part of the amplitude for $t=u$ line.
The dotted lines represent order-by-order fit (Set~3), the dashed ones stand for resummed fit constructed from it (with the values of all parameters equal to their $O(p^6)$ values from Set~3), and finally the solid lines reflect an overall fit corresponding to KLOE values, with normalization set to interpolate between the dotted and the dashed lines for the imaginary part. The vertical lines indicate the physical region.}
\label{figtu}
\end{figure}

From the way our parametrization was constructed and from the normalization procedure described above, it should be obvious that we have concentrated mainly on the reduction of the errors of the second type. Moreover, since we used only a distribution coming from 4+1 Dalitz parameters measured by KLOE without any information of the systematic uncertainties in different regions of the Dalitz plot, it should be obvious that these uncertainties prevail and the error quoted in (\ref{R disp}) corresponds solely to them. Provided we have more precise measurement of the distribution of the physical amplitude, we could quantify the uncertainty coming from fitting these data to our parametrization and a deeper analysis of the ones coming from the normalization procedure would be required.

To conclude the study on the charged decay $\eta\to \pi^+\pi^-\pi^0$ let us
summarize all the analyses performed in this and the previous
Sec.~\ref{sec:ObOC} and~\ref{sec:corOO} in one plot (Fig.~\ref{figtu})
focusing on the $t=u$ line.

%%%%%%%%%%%%%%%%%%%%%%%%%%%%%%%%%%%%%%%%%%%%%%%%%%%%%%%%%%%%%%%%%%%%%%%%%%%%
\section{Neutral decay: \texorpdfstring{$\eta\to3\pi^0$}{eta -> 3 pi(0)}}\label{sec:neutral}
Before we present our results for the $\eta\to3\pi^0$ amplitude, let us shortly discuss the number of free parameters
of our dispersive parametrization for the $\eta\to3\pi$ decays that has to be determined from a fit.

The polynomial part of the charged amplitude was given in (\ref{polstu}). For the $\eta\to3\pi^0$ decay the situation is simpler since
{\allowdisplaybreaks[1]
\begin{equation}
  {\cal P}_0 (s,t,u) =  A_0 \met^2 +
 C_0 \bigl[ (s-s^c)^2 + (t-s^c)^2 + (u-s^c)^2 \bigr]
 +   E_0
\bigl[ (s-s^c)^3 + (t-s^c)^3 + (u-s^c)^3 \bigr].
\end{equation}
}
These two decays are related by the 2-particle unitarity and so in the unitarity part of the $\eta\to\pi^+\pi^-\pi^0$ decay there appear 2 parameters from the neutral decay, whereas in the unitarity part of the $\eta\to3\pi^0$ decay there appear 4 parameters from the charged one. When one takes the full isospin breaking into account there is no further connection between these two decays and one needs to determine all these parameters appearing in the considered amplitude.
However, in the case we work in the leading order of the isospin breaking, relation (\ref{M0Mxiso}) bounds these two amplitudes together and all the parameters of the neutral decay can be expressed in terms of the charged parameters. The number of the parameters needed to be determined in the various studies in these two cases are given in Table~\ref{tab:pocty}.
\begin{table}
\centering
\begin{ruledtabular}
\begin{tabular}{lcc}
& $\mpp=\mpn$ & $\mpp\neq\mpn$
\\\hline
$\eta\to\pi^+\pi^-\pi^0$ & 6 & 8\\
$\eta\to3\pi^0$ & 5 & 7\\
$\eta\to3\pi$ & 6 & 9
\end{tabular}
\end{ruledtabular}
\caption{Numbers of free parameters of the dispersive parametrization that are needed to be determined in the various analysis. The first two lines describe the individual studies of the charged and the neutral $\eta\to3\pi$ decays, while the ultimate corresponds to the combined fit of both of them.}
\label{tab:pocty}
\end{table}

From the table and the present status of information we have from experiment on these amplitudes (as summarized in
Sec.~\ref{Dalitz plot parametrization}), it is obvious why we were focusing just on the charged decay modes of $\eta$.
In that case we have six unknown parameters in our dispersive formula (in the case $\mpp=\mpn$) which could be saturated by five known Dalitz parameters of this decay. On the other hand, although the neutral decay is theoretically much simpler (having less parameters and there is no P-wave contribution to the unitarity part), so far only one Dalitz parameter ($\alpha$) was measured for $\eta\to3\pi^0$. The procedure elaborated in the previous section will not be
thus very reliable in this case.

Working in the $\mpp=\mpn$ we can obtain the values of the neutral dispersive parameters from the values of the charged ones (using (\ref{M0Mxiso})), from which we can compute the neutral Dalitz parameters.

Doing so we get $\alpha = -0.042$ for the order-by-order correspondence (ChPT with an additional $O(p^6)$ polynomial) and $\alpha = -0.047$ for the overall fit (fit to KLOE with the normalization from ChPT). This two predictions can be put together to:
\begin{equation}
\alpha = -0.044 \pm 0.004.
\label{falpha}
\end{equation}
In these error bars the systematic uncertainties from KLOE are not included. This error also does not take into account the isospin corrections to the relation (\ref{M0Mxiso}).

The corresponding study of KLOE \cite{KLOE-etac} based on the charged decay mode leads to
the value $\alpha= -0.038(3)(12)$, which is also higher (in absolute value) than are the most precise values on this quantity (see Table~\ref{tabulka Dalitz eta0}, e.g. MAMI-C: $\alpha = -0.032(3)$). Note that the similar study done recently in~\cite{Schneider:2010hs} obtained the same prediction at the leading order in $\pi\pi$ rescattering taking using KLOE data: $\alpha^{(1)}_\text{NREFT} = -0.042(2)\tbinom{3}{5}$. Employing higher orders to this prediction leads to $\alpha_\text{NREFT}\approx -0.06$, the value even more off the current most precise experimental determination. However, these two values were really based on the KLOE data. The independent determination of \cite{Schneider:2010hs} for the parameter $\alpha$ is in much better agreement with the present most precise experimental value (see e.g.\ Table~\ref{tab:inputdalitz}).

Therefore, we have verified a deviation in the neutral Dalitz parameter $\alpha$ obtained from the measurement of the charged Dalitz parameters (using isospin relation (\ref{M0Mxiso})) and the one obtained from the direct
measurement of the neutral decay. This discrepancy can be attributed to the fact that we
do not have at disposal the exact KLOE distribution but only the simplified one. The second
reason is connected with the following disadvantage of the dispersive representation when connected to the isospin simplification $\mpp=\mpn$.
This limit forces us to use only one value for the two masses of pions. In the case of $\eta\to \pi^+\pi^-\pi^0$ we use an average value of the charged and the neutral one, whereas in the neutral case there is used naturally the neutral mass itself. In ChPT we are free to make this change without changing the other parameters of the amplitude (at least in principle LECs do not depend on the light quark masses).
This is, however, not true for the polynomial parameters used in the dispersive approach, which do depend on the masses.
Indeed, repeating the calculation of $\alpha$ done for the ``average'' pion mass we obtain the value of $\alpha$ that is reduced by approximately $10\%$ with respect to our result~(\ref{falpha}), i.e.\ it goes in a good direction to the independent measurements.

\begin{figure}
\begin{center}
\includegraphics{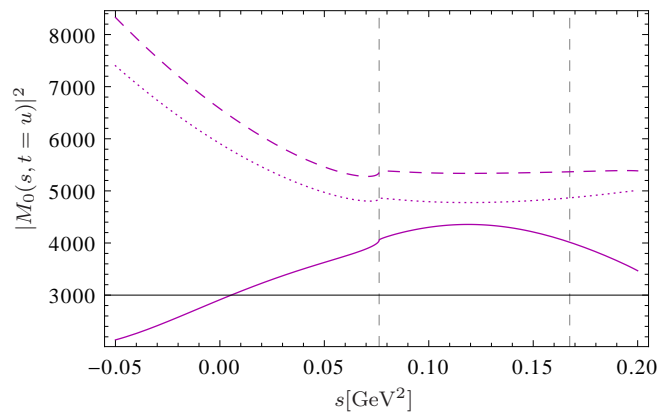}
\end{center}
\caption{The absolute value (squared) of the amplitude $\eta\to3\pi^0$ for $t=u$
line.
The notation is the same as in the previous graph: the dotted line represents order-by-order
fit (set 3), the dashed line denotes a resummed
fit (with $O(p^6)$ set~3 values), and the solid line stands for an overall fit
corresponding to KLOE value. With the vertical lines we have demarcated the physical region.}
\label{figtuz}
\end{figure}

Similarly as in the previous section for the charged decay, we summarize our analyses for the neutral decay in one
plot (Fig.~\ref{figtuz}) again on the line $t=u$, now for absolute value squared of the amplitude.
Note the only slight change of the slope ($\alpha$) when passing from the order-by-order fit (or equivalently NNLO ChPT) to the resummed one (which is equivalent to a different resummation of the $O(p^8)$ chiral orders in ChPT). A bigger change in $\alpha$, connected even with a flip of the sign, is seen only after inclusion of the KLOE data.

Before closing our study of the neutral decay mode let us return back to the resummed correspondence introduced in Sec.~\ref{sec:res}. The major problem addressed there, the overall normalization, can be totally ignored for a
special quantity, the ratio of decay rates
\begin{equation}
r_\eta = \frac{\Gamma(\eta\to \pi^0\pi^0\pi^0)}{\Gamma(\eta\to \pi^+\pi^-\pi^0)},
\end{equation}
where such normalization simply drops out. Using the numbers from fit to KLOE in Table~\ref{tab:resummed} one obtains
\begin{equation}
r_\eta = 1.475\pm 0.015,
\end{equation}
the value that is in an excellent agreement with the most precise measurement of \cite{Achasov:2006dv}:
$r_\eta= 1.46(3)(9)$. We have verified that our result is stable against isospin corrections and is very close to the original ChPT value \cite{BG} $r_\eta^{NNLO} = 1.47$. Note that the prediction of NREFT \cite{Schneider:2010hs} is somewhat smaller $r_\eta = 1.40(1)(4)$.

%%%%%%%%%%%%%%%%%%%%%%%%%%%%%%%%%%%%%%%%%%%%%%%%%%%%%%%%%%%%%%%%%%%%%%%%%%%%
\section{Conclusion}\label{sec:conclusions}

The main purpose of this paper was a demonstration of the possibilities of using the dispersive approach in the case of the $\eta \to 3 \pi$ decays. In the advent of the new precise measurements which could be sensitive to higher-order isospin-breaking effects it is a reliable theoretical tool, based only on general assumptions, as relativistic invariance,
unitarity, analyticity and crossing symmetry together with chiral counting, which ensures that our
amplitude is valid up to and including $O(p^6)$. It easily reproduces the ChPT amplitudes, providing them in a simple and compact analytic form. It is therefore useful for studies of the structure of the amplitudes and enables us to identify the basic independent combinations of the LECs appearing in them. In addition, it is more general and can be analytically continued outside the kinematic decay region. Further, it incorporates naturally the isospin corrections induced by the $\mpp-\mpn$ difference.

However, in the present situation when experiments are limited to 4+1 (one parameter is
zero or close to zero) Dalitz parameters for $\eta\to\pi^+\pi^-\pi^0$ and 1
parameter for $\eta\to3\pi^0$ decay it seems reasonable to use only leading
order of the isospin breaking. This leading order is hidden in the parameter $R$
(see (\ref{ARM})) and $\M(s,t,u)$ is thus effectively taken in the isospin
limit. Even such simplified analysis has an important theoretical outcome since by comparison of the experimentally measured decay rates of these processes with the predictions coming from theory, we can determine\footnote{Note that instead of $R$ we could use in our normalization of the amplitudes the parameter $Q$ as well. Such choice is favored in the analyses based on ChPT at NLO since at this order $Q$ depends only on QCD meson masses \cite{Gasser:1984gg} and is reasonably stable with respect to the Kaplan-Manohar transformation \cite{Kaplan:1986ru} of quark masses of ChPT. However, both these advantages of $Q$ are lost when one includes chiral two-loop effects because the relation between $Q$ and the meson masses gains noticeable $r=m_s/\hat{m}$ dependent chiral corrections at NNLO (cf.\ \cite{Amoros:2001cp}). When matching the amplitudes with the results of NNLO ChPT \cite{BG}, it is more natural to employ the normalization containing $R$ and assume that the Kaplan-Manohar ambiguity is fixed by the values of LECs used in that computation (for instance the value of $L_6^r=0$ stemming from large $N_c$ considerations).} the parameter $R$, thereby obtaining information on the individual masses of $m_u$ and $m_d$.

Such analysis was performed within NNLO ChPT, but the necessity of determination of the large amount of NNLO LECs of ChPT, together with the observed discrepancies between the values of the Dalitz parameters describing the energy dependence of the amplitude predicted by ChPT and those measured by experiments question also the accuracy of the result obtained for $R$.

On the analysis of the charged data from KLOE we have presented two methods how to make use both the information we have from the NNLO ChPT and the one from experiment together in order to determine the corrected value of this parameter. The first one is motivated by the possibility that all the discrepancy between the NNLO ChPT predictions and the KLOE measurement comes from the incorrect determination of the $O(p^6)$ LECs of ChPT (or more generally that all the discrepancy can be approximated well by a polynomial of the third order). Using it we have obtained the value $R = 37.7\pm 2.9$. The second method is based on the fit to the experimental data, in which the error of a slower convergence of the chiral amplitude is reduced by using ChPT just for the normalization of the parametrization in the region where the chiral expansion seems to converge well. Here the main source of uncertainties was due to the experimental error and we obtained $R=37.8\pm 3.3$. Even though both of these results depend on KLOE, the nature of the dominant error in each of the methods is different and so we can combine them to obtain our final prediction\footnote{Let us emphasize once more that this number stands and falls with the assumption that the genuine physical amplitude is described well by the KLOE-like distribution constructed from the values of 4+1 Dalitz plot parameters presented by KLOE \cite{KLOE-etac}. It is therefore desirable to remeasure the charged $\eta\to3\pi$ decay and repeat the performed analysis with the access to the real data.}
\begin{equation}
R=37.7\pm 2.2.
\end{equation}

This value agrees very well with the lattice average prediction~\cite{Colangelo:2010et} and is compatible with the Dashen's limit at NNLO order. (However, it would correspond to large Dashen violation if one worked at NLO as studied in \cite{Amoros:2001cp} --- cf.\ also Table 6 in \cite{BG}).
Using relation (\ref{RQ relace}) and the value of $r\equiv m_s/\hat m=27.4\pm0.4$ provided by lattice simulation~\cite{Colangelo:2010et}, we obtain
\begin{equation}
Q= 23.1\pm0.7.
\end{equation}
Note that \cite{Oller:2006xb} and several non-lattice determinations (for the recent overview see \cite{Leutwyler:2009jg,Zdrahal:2011ta}) point towards a smaller value for $r$, which will mean also a smaller value of $Q$, e.g.~the value of $r\sim24$ leads to $Q\sim21.5$. For the sake of completeness, let us also mention the value $Q=20.7\pm 1.2$ obtained from large deviation from Dashen's limit in \cite{Ananthanarayan:2004qk} (cf.\ also \cite{Kastner:2008ch}).

In the case we want to fully complete the task set in the Introduction and determine the current masses of $m_u$ and $m_d$ at some scale $\mu$, we need as inputs the values of $\hat{m}$ and $m_s$ at the same scale obtained from somewhere else. Introducing the most recent averaged values of PDG \cite{PDG}, in $\overline{\mathrm{MS}}$ scheme at the running scale $\mu=2\,\mathrm{GeV}$, $m_s=(100.6\pm2.1)\,\mathrm{MeV}$ and $\hat{m}=(3.8\pm0.1)\,\mathrm{MeV}$ into the definitions (\ref{hatm}) and (\ref{RQ}) as inputs, one obtains the current masses at this scale
\begin{equation}\begin{split}
m_u(\overline{\mathrm{MS}},\mu=2\,\mathrm{GeV})&=(2.52\pm0.13)\,\mathrm{MeV}, \\
m_d(\overline{\mathrm{MS}},\mu=2\,\mathrm{GeV})&=(5.08\pm0.13)\,\mathrm{MeV}.
\end{split}\end{equation}
Employing instead the averaged lattice results \cite{Colangelo:2010et} $m_s=(94\pm3)\,\mathrm{MeV}$ and $\hat{m}=(3.43\pm0.11)\,\mathrm{MeV}$ leads to
\begin{equation}\begin{split}
m_u(\overline{\mathrm{MS}},\mu&=2\,\mathrm{GeV})=(2.23\pm0.14)\,\mathrm{MeV}, \\
m_d(\overline{\mathrm{MS}},\mu&=2\,\mathrm{GeV})=(4.63\pm0.14)\,\mathrm{MeV}.
\end{split}\end{equation}
Our results are compatible both with the averaged values of these individual masses quoted in PDG \cite{PDG} and with the averaged lattice results of \cite{Colangelo:2010et}. Note that these two sets for each mass are correlated since the PDG values contains among others also the lattice results.

We have also made a quick look onto neutral decay mode $\eta\to3\pi^0$.
Assuming that the original KLOE measurement of charged Dalitz parameters
$a,b,d,f,g$ is correct we have verified some deviation in neutral Dalitz
parameter $\alpha$ obtained from our representation using them (and the isospin relation (\ref{M0Mxiso})) and from its direct measurements. In the previous section we have discussed the possible explanations of this discrepancy.
Note that the neutral decay $\eta\to3\pi^0$ hides more --- in the physical region there occurs an intrinsic cusp, however, its position is naturally very close to the edge of the phase-space. Its shape is
thus very suppressed and the techniques similar to the extraction of $K_L\to3\pi^0$
have to be employed. An unambiguous description of the amplitude including corrections
induced by $\mpp-\mpn$ difference can thus naturally help in this extraction.

At this point let us mention also the second neutral Dalitz parameter $\beta$, whose
measurement is planned also in the forthcoming experiments (either WASA or KLOE-II).
Its theoretical determination in ChPT is not influenced by $O(p^6)$ LECs and is
thus important for its consistency check. In Sec.~\ref{sec:Ci} we also list the other combinations of the Dalitz parameters that are suitable for such a check and their evaluation can shed light on the discrepancies in the determinations of the Dalitz parameters --- not only the discrepancy in $\alpha$ whose solution was proposed in \cite{Schneider:2010hs} but also the discrepancy in $b$ that even in \cite{Schneider:2010hs} remains.

The last physical quantity we have discussed was $r_\eta$, the ratio of the
neutral and charge decay width, with the result
\begin{equation}
r_\eta=1.475\pm 0.015.
\end{equation}
Our prediction, or more precisely the prediction of KLOE based on our dispersive parametrization, agrees not
only with the most recent experiment but also with the NNLO ChPT calculation.
Naively, seeing Figures \ref{figtu} and \ref{figtuz} and the change in the absolute value for both amplitudes, one would expect some impact, but apparently these two shifts canceled out in the ratio.

\acknowledgments{We would like to thank Hans Bijnens, Andrzej Kup\'s\'c and Bachir Moussallam for stimulating discussions and/or remarks on the text.
We are thankful to Hans Bijnens for providing us with \textsf{FORTRAN} programs which enable two-loop ChPT calculations.
Last but not least we have benefited from discussions with our late colleague Jan Stern.
This work is supported in part by the European Community-Research
Infrastructure Integrating Activity ``Study of Strongly Interacting
Matter'' (HadronPhysics2, Grant Agreement n.\ 227431) and the Center for
Particle Physics (project no.\ LC 527) of the Ministry of Education of the
Czech Republic.}

%%%%%%%%%%%%%%%%%%%%%%%%%%%%%%%%%%%%%%%%%%%%%%%%%%%%%%%%%%%%%%%%%%%%%%%%%%%%
%%%%%%%%%%%%%%%%%%%%%%%%%%%%%%%%%%%%%%%%%%%%%%%%%%%%%%%%%%%%%%%%%%%%%%%%%%%%
%%%%%%%%%%%%%%%%%%%%%%%%%%%%%%%%%%%%%%%%%%%%%%%%%%%%%%%%%%%%%%%%%%%%%%%%%%%%
\appendix
\section{Kinematic functions appearing in the dispersive representation}

\bigskip Five basic functions $\mathcal{G}_{i}(s)$, $\ i=1,\ldots 5$
entering general formulae (\ref{WS0})--(\ref{derGi-last}) for $W_{S,T}^{0}(s)$ and
$W_{T}^{1}(s)$ are given for complex $s$ by the dispersion integrals
with an appropriate number of subtractions $k_{i}$
\begin{equation}
\mathcal{G}_{i}(s)=\frac{s^{k_{i}}}{\pi }\int_{4m_{\pi }^{2}}^{\infty }\frac{\mathrm{d}x}{x^{k_{i}}}\frac{\mathcal{F}_{i}(x)}{x-s}\,,  \label{disp_def}
\end{equation}
where $\mathcal{F}_{i}$ is the set of generally complex functions describing the discontinuities of the amplitudes listed below.
Therefore the functions ${\cal G}_i$ are analytic in the cut complex plane with cuts along $(4\mpi^2,\infty)$ and with the discontinuities across these cuts
\begin{equation}
\disc \mathcal{G}_{i}(s)=\frac{1}{2\mathrm{i}}\left( \mathcal{G}_{i}(s+\mathrm{i}0)-\mathcal{G}_{i}(s-\mathrm{i}0)\right) =\mathcal{F}_{i}(x).
\label{disc_def}
\end{equation}
For real $s\in(4\mpi^2,\infty)$ one has
\begin{equation}
\mathcal{G}_{i}(s\pm \mathrm{i}0)=\frac{s^{k_{i}}}{\pi }\text{v.p.}\int_{4m_{\pi
}^{2}}^{\infty }\frac{\mathrm{d}x}{x^{k_{i}}}\frac{\mathcal{F}_{i}(x)}{x-s}
\pm \mathrm{i}\mathcal{F}_{i}(s).  \label{hilb_def}
\end{equation}
The physical value of these functions corresponds then to the $+\mathrm{i}0$ prescription. In
mathematical language, each function $\mathcal{G}_{i}$ is Hilbert transform of the corresponding function $\mathcal{F}_{i}$.

The discontinuities $\mathcal{F}_{i}(x)$ needed for the construction of Sec.~\ref{sec:dispersive construction} are expressed in terms of the
following two logarithmic functions
\begin{align}\label{funkce L}
L(s)& =\log \frac{1-\sigma (s)}{1+\sigma (s)}\,, \\
M(s)& =-2\log \left( 1-\frac{m_{+}m_{-}}{s}+\frac{\lambda ^{1/2}(s)}{s}\right) +\log \frac{4\mpi^{2}}{s}\,, \label{funkce M}
\end{align}
where $\lambda(s)\equiv\lambda_{\eta\pi}(s)$ and $\sigma(s)\equiv\sigma_\pi(s)$ were defined in (\ref{lambda}) and (\ref{sigma}), respectively; and
\begin{equation}
m_{\pm }=m_{\eta }\pm m_{\pi }.
\end{equation}
For the logarithms we place the
branch cut along the negative real axis and $\im\log z\in (-\pi ,\pi \rangle $.
The function $L(s)$ is then real on the physical region. This is, however, not
true for $M(s)$, which is real only for $s>m_{+}^{2}$ (this corresponds to
the $\eta \pi \rightarrow \pi \pi $ scattering region).  We have
\begin{align}
\mathcal{F}_{1}(x)& =\sigma (x), \\
\mathcal{F}_{2}(x)& =L(x), \\
\mathcal{F}_{3}(x)& =\frac{L^{2}(x)}{x\sigma (x)}\,, \\
\mathcal{F}_{4}(x)& =\sigma (x)\frac{M(x)}{\lambda ^{1/2}(x)}\,, \label{funkce F4}\\
\mathcal{F}_{5}(x)& =L(x)\frac{M(x)}{\lambda ^{1/2}(x)}\,.\label{funkce F5}
\end{align}
and the numbers of subtraction taken for them in (\ref{disp_def}) read $k_{1}=k_{2}=1$ and $k_{3}=k_{4}=k_{5}=0$.
Let us note that in these expressions the branch of the square root $\lambda ^{1/2}(s)$ is inessential.

To find an analytic form of the Hilbert transform from its integral definition
is a non-trivial task. For some functions we can use the roundabout way
using the formula (\ref{disc_def}) trying to find a function analytic in the
complex plane except the branch cut on the interval $(4\mpi^{2},\infty )$
where it has the discontinuity equal  to the value of the function $\mathcal{F}_{i}(s)$. All the functions satisfying this requirement differ
just by polynomials that can be restricted by the UV and IR asymptotics of
the integrals depending on the number of subtractions.

It is easy to find that
\begin{equation}
\disc\left( \sigma (s)\log \frac{\sigma (s)-1}{\sigma (s)+1}\right)
=\theta (s-4\mpi^{2})\pi \sigma (s)=\theta (s-4\mpi^{2})\pi \mathcal{F}_{1}(s).
\end{equation}
Since we have defined  $\mathcal{G}_{1}$ with one subtraction, we conclude
that
\begin{equation}\label{funkce G1}
\mathcal{G}_{1}(s)=\frac{1}{\pi }\Bigl(2+\sigma (s)\log \frac{\sigma (s)-1}{\sigma (s)+1}\Bigr)=16\pi \bar{J}(s).
\end{equation}
As we have seen,  $\mathcal{G}_{1}(s)$ is connected with the
$O(p^{4})$ unitarity part of the  amplitude corresponding to
the single two-pion rescattering in the final state, it is
therefore no surprise that this result restores the one-loop
function $\bar{J}(s)$ which is a once subtracted scalar bubble
with mass $m_{\pi }$.

Similarly, one arrives at
\begin{align}\label{funkce G2}
\mathcal{G}_{2}(s)& =\frac{1}{2\pi }\log ^{2}\frac{\sigma (s)-1}{\sigma (s)+1}\,, \\
\mathcal{G}_{3}(s)& =\frac{1}{3\pi s\sigma (s)}\log \frac{\sigma (s)-1}{\sigma (s)+1}\left( \log ^{2}\frac{\sigma (s)-1}{\sigma (s)+1}+\pi
^{2}\right).
\end{align}
Note that these three functions have appeared also in the two-loop pion
scattering computation \cite{Knecht:1995tr}.

For functions $\mathcal{G}_{4}(s)$ and $\mathcal{G}_{5}(s)$, this roundabout
way does not work and we have to employ their integral representations (\ref{disp_def}) and (\ref{hilb_def}).  We have two possibilities how to compute
them: either by integrating these expressions numerically or by means of the
construction of analytic approximations which is described in Appendix~B.
One should note that having these two functions only in either of these approximate forms does not mean
that they are worse than the others --- since they depend only on two masses $\mpi$ and $\met$ and on the variable $s$, once we fix the masses we can tabulate them.

From these five functions we can obtain all the other functions appearing in
the dispersive representation by using the following properties of the
general  Hilbert transform with $n$ subtractions
\begin{equation}
\Han\!(s)=\frac{s^{n}}{\pi }\int \frac{\mathrm{d}x}{x^{n}}\frac{\mathcal{K}(x)}{x-s}.
\end{equation}

In order to simplify the following relations, we define
\begin{equation}
\han(s,x,l)=\frac{\Han\!(s)-\left(\frac{s}{x}\right)^{l}\Han\!(x)}{s-x}\,.
\end{equation}

\begin{enumerate}
\item The formula for raising the number of subtraction is
\begin{equation}
\Ha{n+1} \!(s)
=\Han\!(s)-s^{n}\lim_{s\rightarrow 0}\left(
\frac{\Han\!(s)}{s^{n}}\right) .  \label{zmena poctu subtrakci}
\end{equation}
In terms of $n$-th derivation of $\Han (s)$ with respect to
$s$, we can write this expression also as
\begin{equation}
\Ha{n+1} \!(s)=\Han\!(s)-\frac{s^{n}}{n!}\frac{\partial
^{n}}{\partial s^{n}}\Han\!(0).
\end{equation}

\item Hilbert transform $\Han_{(\lambda )}\!(s)$ of function
\begin{equation}
\mathcal{K}_{(\lambda)}(s)=\frac{\mathcal{K}(s)}{\lambda (s)}
\end{equation}
(where $\lambda (s)$ is from (\ref{lambda}) equal to $(s-m_+^2)(s-m_-^2)$) using the
knowledge of the Hilbert transform $\mathcal{H}^{n}(s)$ of
function $\mathcal{K}(s)$ reads
\begin{equation}
\Han_{(\lambda)}\!(s)=\frac{\han(s,m_+^2,n)-\han(s,m_-^2,n)}{m_{+}^{2}-m_{-}^{2}}\,.
\end{equation}

\item Obviously, we can lower the number of subtraction for $\mathcal{K}_{(\lambda )}(s)$,
\begin{equation}
\Ha{n-1}\!\!\! _{(\lambda)}(s) =\frac{m_{+}^{2}\han(s,m_+^2,n)
-m_{-}^{2}\han(s,m_-^2,n)}{s(m_{+}^{2}-m_{-}^{2})}
\end{equation}
and
\begin{equation}
\Ha{n-2}\!\!\! _{(\lambda)}(s) =\frac{1}{s^{2}}\Bigg(\Han\!(s)
+\frac{m_{+}^{4}\han(s,m_+^2,n)-m_{-}^{4}\han(s,m_-^2,n)}{m_{+}^{2}-m_{-}^{2}}\Bigg).
\end{equation}

\item In the case of same-mass particles, there appear functions
\begin{equation}
\mathcal{K}_{(\sigma )}(s)=\frac{1}{s^{2}\sigma ^{2}(s)}\,\mathcal{K}(s)=
\frac{1}{s(s-4m_{\pi }^{2})}\,\mathcal{K}(s).
\end{equation}
Their Hilbert transform is
\begin{equation}
\Han\!\!\!_{(\sigma)}(s)=\frac{\han(s,4m_\pi^2,n+1)}{s}
+\frac{s^{n-1}}{4m_{\pi }^{2}}\left( \frac{\Han\!(s)}{s}\right) _{\!\!s=0}\!.
\end{equation}

\item Again by lowering the number of subtractions we arrive at
\begin{align}
\Ha{n-1}\!\!\! _{(\sigma)}(s)& =\frac{\han(s,4m_\pi^2,n)}{s}\,, \\
\begin{split}
\Ha{n-2}\!\!\! _{(\sigma)}(s)& =\frac{\han(s,4m_\pi^2,n-1)}{s}\,.
\end{split}
\end{align}
\end{enumerate}

%%%%%%%%%%%%%%%%%%%%%%%%%%%%%%%%%%%%%%%%%%%%%%%%%%%%%%%%%%%%%%%%%%%%%%%%%%%%
%%%%%%%%%%%%%%%%%%%%%%%%%%%%%%%%%%%%%%%%%%%%%%%%%%%%%%%%%%%%%%%%%%%%%%%%%%%%
%%%%%%%%%%%%%%%%%%%%%%%%%%%%%%%%%%%%%%%%%%%%%%%%%%%%%%%%%%%%%%%%%%%%%%%%%%%%
\section{Functions \texorpdfstring{$\mathcal{G}_{4}$}{G4} and \texorpdfstring{$\mathcal{G}_{5}$}{G5}, analytical
approximations}

In this appendix we discuss in more detail the functions
$\mathcal{G}_{4}(s)$ and $\mathcal{G}_{5}(s)$ from the previous appendix, which we do not know
analytically. We first find their relation to some explicitly known
analytic functions, which allows us to express them through more
simple dispersive integrals, and then we construct an analytical
approximation to the latter.

We start by defining the following functions for complex $z$
\begin{align}
\sigma _{\eta \pi }(z) &=\frac{\sqrt{m_{+}^{2}-z}}{\sqrt{m_{-}^{2}-z}}\,, \\
\lambda _{1/2}(z) &=(m_{-}^{2}-z)\sigma _{\eta \pi }(z)=\sqrt{m_{-}^{2}-z}\sqrt{m_{+}^{2}-z}\,, \\
\sigma (z) &=\lim_{m_{\eta }\rightarrow m_{\pi }}\sigma _{\eta \pi }(z)=\sqrt{1-\frac{4m_{\pi }^{2}}{z}}\,,
\end{align}
where the square roots are defined as $\sqrt{1}=1$ with the cut along the negative real axis.
Therefore, $\sigma _{\eta \pi }(z)$ has a cut along $(m_{-}^{2},m_{+}^{2})$ with
the boundary values on the real axis
\begin{equation}
\sigma _{\eta \pi }(x+\mathrm{i}0)=\left\{
\begin{array}{c@{\qquad}l}
|\sigma _{\eta \pi }(x)|^{1/2} & x<m_{-}^{2} \\
\mathrm{i}|\sigma _{\eta \pi }(x)|^{1/2} & m_{-}^{2}<x<m_{+}^{2} \\
|\sigma _{\eta \pi }(x)|^{1/2} & x>m_{+}^{2}
\end{array}
\right.
\end{equation}
and the discontinuity equal to
\begin{equation}
\disc \sigma _{\eta \pi }(x)=\theta (x-m_{-}^{2})\theta
(m_{+}^{2}-x)|\sigma _{\eta \pi }(x)|^{1/2}.
\end{equation}
The function $\sigma (z)$ has the similar properties obtained by replacing $m_{-}\rightarrow 0$ and $m_{+}\rightarrow 2m_{\pi }$ in the previous formulae. Analogously,
\begin{align}
\lambda _{1/2}(x+\mathrm{i}0) &=\left\{
\begin{array}{c@{\qquad}l}
|\lambda _{1/2}(x)|^{1/2} & x<m_{-}^{2} \\
-\mathrm{i}|\lambda _{1/2}(x)|^{1/2} & m_{-}^{2}<x<m_{+}^{2} \\
-|\lambda _{1/2}(x)|^{1/2} & x>m_{+}^{2}
\end{array}
\right. , \\
\disc \lambda _{1/2}(x) &=-\theta (x-m_{-}^{2})\theta
(m_{+}^{2}-x)|\lambda _{1/2}(x)|^{1/2}.
\end{align}
With the help of these elementary functions we can construct\footnote{Note the relation of the functions $F_4(z)$ and $F_5(z)$ to the functions $\mathcal{G}_1(s)$ and $\mathcal{G}_2(s)$ from (\ref{funkce G1}) and (\ref{funkce G2}), respectively.}
\begin{align}
F_{4}(z) &=\sigma (z)\log \left( \frac{\sigma (z)-1}{\sigma (z)+1}\right),  \\
F_{5}(z) &=\frac{1}{2}\frac{F_4(z)^{2}}{\sigma (z)^{2}}=\frac{1}{2}\log
^{2}\left( \frac{\sigma (z)-1}{\sigma (z)+1}\right),  \\
I(z) &=\frac{1}{\lambda _{1/2}(z)}\Bigg( \log \left( \frac{\sigma _{\eta \pi}(z)+1}{\sigma _{\eta \pi }(z)-1}\right)
+\log \left( \frac{1+\frac{m_{-}}{m_{+}}\sigma _{\eta \pi }(z)}{1-\frac{m_{-}}{m_{+}}\sigma _{\eta \pi }(z)}\right) -\mathrm{i}\pi \Bigg).
\end{align}
Taking the principal branch of the logarithm with the cut along $(-\infty ,0)$
and $-\pi <\im \log z\leq \pi $, their discontinuities are equal to
\begin{align}\label{funkce pomocna F4}
\frac{1}{\pi }\disc F_4(x) &=\theta (x-4m_{\pi }^{2})\sigma (x), \\
\frac{1}{\pi }\disc F_5(x) &=\theta (x-4m_{\pi }^{2})L(x),\label{funkce pomocna F5}\\
\frac{1}{\pi }\disc I(x)&=\theta (x)\,\frac{1}{\widetilde{\lambda }
_{1/2}(x)}\,,
\end{align}
where $L(s)$ was defined in (\ref{funkce L}) and\footnote{
Note that contrary to $\lambda _{1/2}(x)$, the values of $\widetilde{\lambda }_{1/2}(x)$ are not $x-\mathrm{i}0$ boundary values of any analytic function in the cut complex plane. We can however write
\begin{equation}
\widetilde{\lambda }_{1/2}(x)=\hat{\lambda }_{1/2}\left(x+\sign (x-m_{-}^{2})\mathrm{i}0\right)
=\hat{\lambda }_{1/2}\left(x+\sign(x-m_{+}^{2})\mathrm{i}0\right),
\end{equation}
where with the above choice of the cut of the square root, the function
\begin{equation}
\hat{\lambda }_{1/2}(z)=\mathrm{i}\sqrt{z-m_{-}^{2}}\sqrt{m_{+}^{2}-z}
\end{equation}
is analytic with a cut along $(-\infty ,m_{-}^{2})\cup (m_{+}^{2},\infty )$.}
for $x>0$
\begin{equation}
\widetilde{\lambda }_{1/2}(x)=\left\{
\begin{array}{c@{\qquad}l}
|\lambda (x)|^{1/2} & x<m_{-}^{2} \\
\mathrm{i}|\lambda (x)|^{1/2} & m_{-}^{2}<x<m_{+}^{2} \\
|\lambda (x)|^{1/2} & x>m_{+}^{2}
\end{array}
\right. .  \label{lambdatilde}
\end{equation}

The functions $\mathcal{F}_{4,5}(x)$ from (\ref{funkce F4})--(\ref{funkce F5}) for $x>4m_{\pi }^{2}$ can now be expressed in a convenient form as
\begin{equation}
\mathcal{F}_{j}(x)=\frac{1}{\pi }I(x-\mathrm{i}0)\disc F_{j}(x) \qquad j=4,5.
\end{equation}
Using the formula
\begin{equation}
\disc f(x)g(x)=f(x-\mathrm{i}0)\disc g(x)+g(x+\mathrm{i}0)\disc f(x),
\end{equation}
we get then for $j=4,5$
\begin{equation}
\label{dibp}
\frac{1}{\pi }\disc \left[ F_{j}(x)I(x)\right] =\theta (x-4m_{\pi
}^{2})\mathcal{F}_{j}(x)
+\theta (x)F_{j}(x+\mathrm{i}0)\,\frac{1}{\widetilde{\lambda }^{1/2}(x)}.
\end{equation}
To find the functions with the desired discontinuities $\mathcal{F}_{4,5}(x)$, it
remains therefore to evaluate the dispersion integrals
\begin{equation}
K_{j}(z)=\int_{0}^{\infty }\frac{dx}{x-z}\frac{1}{\widetilde{\lambda }^{1/2}(x)}F_{j}(x+\mathrm{i}0).
\end{equation}
By using (\ref{dibp}) and ``dispersive integrations by parts'', there can be easily proved that\footnote{
It may seem that as a consequence of (\ref{dibp}) this relation holds only up to a polynomial which does not contribute
to the discontinuity. However, taking carefully into account the necessary number of subtractions for each term on both sides of the relations, there can be proved that such a polynomial is in fact absent.}
\begin{equation}
\mathcal{G}_{j}(z)=\frac{1}{\pi }\left( F_{j}(z)I(z)-K_{j}(z)\right) .
\end{equation}
This representation of the functions $\mathcal{G}_{j}(z)$ for $j=4,5$ allows to construct a systematic analytical approximation of them based on the approximations of much simpler integrals $K_{j}(z)$.

The integrals $K_{j}(z)$  have the following structure
\begin{equation}
K_{j}(z)=\int_{4m_{\pi }^{2}}^{\infty }\frac{dx}{x-z}\frac{1}{\widetilde{\lambda }_{1/2}(x)}f_{j}(\sigma (x))
+\int_{0}^{4m_{\pi }^{2}}\frac{dx}{x-z}\frac{1}{\widetilde{\lambda }_{1/2}(x)}g_{j}(|\sigma (x)|),
\end{equation}
where the functions $f_j(\sigma)$ are given by (\ref{funkce pomocna F4}) and (\ref{funkce pomocna F5}) such that
\begin{equation}
F_j(x)=f_j(\sigma(x))
\end{equation}
and the functions $g_{j}(\sigma )$ for real $\sigma >0$ are defined as
\begin{equation}
g_{j}(\sigma )=f_{j}(\mathrm{i}\sigma ).
\end{equation}
The key idea is now to find appropriate series expansions of the functions $f_{j}(\sigma )$ and $g_{j}(\sigma )$ in the variable $\sigma $ and integrate then term by term. This can be done analytically and as we will see in what follows and
the results can be partially summed.

In order to perform this task, we have to distinguish three different regions, namely, \Rim1: $x>4m_{\pi }^{2}$, where we have $0<\sigma <1$ and $f_{j}(\sigma )$ have a convergent Taylor expansion in the domain $|\sigma |<1$; \Rim2: $2m_{\pi}^{2}<x<$ $4m_{\pi }^{2}$, where the function $\sigma (x+\mathrm{i}0)=\mathrm{i}|\sigma (x+\mathrm{i}0)|$ with $|\sigma |<1$ and $g_{j}(|\sigma |)$ have a convergent Taylor expansion in powers of $|\sigma|$ and finally \Rim3: $0<x<2m_{\pi }^{2}$, where again $\sigma (x+\mathrm{i}0)=\mathrm{i}|\sigma (x+\mathrm{i}0)|$ but now with $|\sigma |>1$ and $g_{j}(|\sigma |)$ can be expanded in a convergent Taylor expansion in the variable $1/|\sigma |$.

Furthermore, we split the region \Rim1 into three subregions \Rim1a,\Rim1b,\Rim1c corresponding to the different explicit form of the function $\widetilde{\lambda }_{1/2}(x)$ for $x<m_{-}^{2}$, $x\in (m_{-}^{2},m_{+}^{2})$ and $x>m_{+}^{2}$ respectively
(see (\ref{lambdatilde})). In what follows we therefore write $K_{j}(z)$ as a sum of the contributions of five regions
\begin{equation}
K_{j}(z)=K_{j}^{\mRim3}(z)+K_{j}^{\mRim2}(z)+K_{j}^{\mRima1a}(z)+K_{j}^{\mRima1b}(z)+K_{j}^{\mRima1c}(z),
\end{equation}
which are ordered according to increasing $x$.

Let us illustrate the the general recipe using $K_{j}^{\mRim3}(z)$ on the
region \Rim3. Here
\begin{align}
g_{4}(|\sigma |) &=-2|\sigma |\arctan \left( \frac{1}{|\sigma |}\right),  \\
g_{5}(|\sigma |) &=-2\arctan ^{2}\left( \frac{1}{|\sigma |}\right)
\end{align}
and after the substitution $u=1/|\sigma (x+\mathrm{i}0)|$ we have in this
region (we use the shortcuts $\sigma _{\pm }=\sigma (m_{\pm }^{2})$, $\Delta =m_{+}m_{-}$)
\begin{equation}
K_{j}^{\mRim3}(z) =\left( \frac{4m_{\pi }^{2}}{4m_{\pi }^{2}-z}\right) \frac{1}{\sqrt{(m_{+}^{2}-4m_{\pi }^{2})(m_{-}^{2}-4m_{\pi }^{2})}} 
\int_{0}^{1}\frac{2u\mathrm{d}u}{(u^{2}+\sigma (z)^{-2})}
\frac{g_{j}\left( \frac{1}{u}\right) }{\sqrt{(u^{2}+\sigma _{+}^{-2})(u^{2}+\sigma
_{-}^{-2})}}\,.
\end{equation}
Using the expansions
\begin{equation}
g_{j}\left( \frac{1}{u}\right) =\sum_{k=0}^{\infty }a_{k}^{(j)}u^{2k},
\end{equation}
we can write
\begin{equation}
K_{j}^{\mRim3}(z)=\frac{1-\sigma (z)^{-2}}{\Delta \sigma _{+}\sigma _{-}}
\sum_{k=0}^{\infty }a_{k}^{(j)}
\int_{0}^{1}\frac{\mathrm{d}t}{(t+\sigma (z)^{-2})}\frac{t^{k}}{\sqrt{(t+\sigma _{+}^{-2})(t+\sigma _{-}^{-2})}}\,.
\end{equation}
Note that the series for $g_{j}\left(\frac{1}{u}\right) $ converge absolutely, an in addition the partial sums of them have the integrable majorants $1/2u\log [(1+u)/(1-u)]$ and $1/2\log^{2}[(1+u)/(1-u)]$, respectively. Therefore, the
sum and the integral are interchangeable.

Apparently we have to calculate one extra integral for each $k$, however, in
fact all the integrals can be obtained easily from one such integrals. Indeed, let us define for complex $w$
\begin{equation}
M_{\mRim3}^{(k)}(w)=\int_{0}^{1}\frac{\mathrm{d}t}{(t-w)}\frac{t^{k}}{\sqrt{(t+\sigma _{+}^{-2})(t+\sigma _{-}^{-2})}}
\end{equation}
so that
\begin{equation}
K_{j}^{\mRim3}(z)=\frac{1-\sigma (z)^{-2}}{\Delta \sigma _{+}\sigma _{-}}
\sum_{k=0}^{\infty }a_{k}^{(j)}M_{\mRim3}^{(k)}(-\sigma (z)^{-2}).
\end{equation}
Then we can write
\begin{equation}
M_{\mRim3}^{(k)}(w)=\frac{1}{k!}\frac{\partial ^{k}}{\partial \alpha ^{k}}
M_{\mRim3}(w;\alpha )\Big|_{\alpha =0}\,,
\end{equation}
using the ``generating integral'' $M_{\mRim3}(w;\alpha )$ that is for complex $\alpha $, $|\alpha |<1$, equal to
\begin{equation}\begin{split}
M_{\mRim3}(w;\alpha )&=\sum_{k=0}^{\infty }\alpha
^{k}M_{\mRim3}^{(k)}(w)
=\int_{0}^{1}\frac{\mathrm{d}u}{(t-w)}\frac{1}{\sqrt{(t+\sigma _{+}^{-2})(t+\sigma _{-}^{-2})}}\frac{1}{1-\alpha t}
=\frac{1}{1-\alpha w}\left( M_{\mRim3}^{(0)}(w)-M_{\mRim3}^{(0)}\Big(\frac{1}{\alpha }\Big)\right).
\end{split}\end{equation}
Now, as a consequence,
\begin{equation}
M_{\mRim3}^{(k)}(w) =\frac{1}{k!}\frac{\partial ^{k}}{\partial \alpha ^{k}}
\frac{1}{1-\alpha w}\left( M_{\mRim3}^{(0)}(w)-M_{\mRim3}^{(0)}\Big(\frac{1}{\alpha }\Big)\right) \bigg|_{\alpha =0} 
=w^{k}M_{\mRim3}^{(0)}(w)-w^{k}\sum_{i=0}^{k}\frac{w^{-i}}{i!}\frac{\partial
^{i}}{\partial \alpha ^{i}}M_{\mRim3}^{(0)}\Big(\frac{1}{\alpha }\Big)\Big|_{\alpha =0}\,.
\end{equation}
This formula can be easily understood. $M_{\mRim3}^{(k)}(w)$ is by definition an
analytic function with a cut along $(0,1)$ and the discontinuity
\begin{equation}
\disc M_{\mRim3}^{(k)}(t)=\theta (t)\theta (1-t)\,\frac{t^{k}}{\sqrt{(t+\sigma _{+}^{-2})(t+\sigma _{-}^{-2})}}\,.
\end{equation}
It is therefore determined uniquely up to a polynomial. Because
\begin{equation}
\disc M_{\mRim3}^{(0)}(t)=\theta (t)\theta (1-t)\,\frac{1}{\sqrt{(t+\sigma
_{+}^{-2})(t+\sigma _{-}^{-2})}}\,,
\end{equation}
we can identify $M_{\mRim3}^{(k)}(w)$ up to some polynomial $P_{\mRim3}^{(k)}(w)$ with $w^{k}M_{\mRim3}^{(0)}(w)$,
\begin{equation}
M_{\mRim3}^{(k)}(w)=w^{k}M_{\mRim3}^{(0)}(w)+P_{\mRim3}^{(k)}(w).
\end{equation}
The unknown polynomial $P_{\mRim3}^{(k)}(w)$ can be fixed by imposing the
requirement of the appropriate asymptotics for $w\rightarrow \infty $, where $M_{\mRim3}^{(k)}(w)\rightarrow 0$. Choosing
\begin{equation}
P_{\mRim3}^{(k)}(w)=-w^{k}\sum_{i=0}^{k}\frac{1}{i!}w^{-i}\frac{\partial ^{i}}{\partial \alpha ^{i}}M_{\mRim3}^{(0)}\Big(\frac{1}{\alpha }\Big)\Big|_{\alpha =0}\,,
\end{equation}
we subtract from $w^{k}M_{\mRim3}^{(0)}(w)$ just the singular (and finite) part at $w\rightarrow \infty $ which ensures the right asymptotics.

By explicit integration we obtain the result
\begin{widetext}\begin{equation}
M_{\mRim3}^{(0)}(w)=2\sigma _{+}\sigma _{-}w^{-1}\frac{\tanh ^{-1}\left( \frac{\sqrt{w^{-1}+\sigma _{-}^{2}}}{\sqrt{w^{-1}+\sigma _{+}^{2}}}\right) -\tanh
^{-1}\left( \frac{\sqrt{1+\sigma _{+}^{2}}}{\sqrt{1+\sigma _{-}^{2}}}\frac{\sqrt{w^{-1}+\sigma _{-}^{2}}}{\sqrt{w^{-1}+\sigma _{+}^{2}}}\right) }{\sqrt{w^{-1}+\sigma _{+}^{2}}\sqrt{w^{-1}+\sigma _{-}^{2}}}\,.
\end{equation}

In conclusion, the result for this region is given by the formula
\begin{equation}
K_{j}^{\mRim3}(z)=\frac{1-\sigma (z)^{-2}}{\Delta \sigma _{+}\sigma _{-}}
\sum_{k=0}^{\infty }a_{k}^{(j)}\left( w^{k}M_{\mRim3}^{(0)}(w)-w^{k}\sum_{i=0}^{k}
\frac{1}{i!}w^{-i}\frac{\partial ^{i}}{\partial \alpha ^{i}}M_{\mRim3}^{(0)}
\Big(\frac{1}{\alpha }\Big)\Big|_{\alpha =0}\right)\Biggr\rvert_{w=-\sigma (z)^{-2}},
\end{equation}\end{widetext}
which allows for systematic analytic calculation of successive
approximations.

The remaining regions can be treated in the same way as was described above,
with the only difference that the series expansion of the integrand contains
now also the odd powers of $\sigma $ and therefore two ``generating integrals'' $M_{A}^{(0)}$ and $N_{A}^{(0)}$,
$A=\mRim2,\mRima1a, \mRima1b, \mRima1c$ are needed instead of one.

The above calculation of the integrals $K_{j}^{A}(z)$, $A=\mRim3,\mRim2,\mRima1a, \mRima1b, \mRima1c$
gives the result in the form of the series expansion in the variables $w^{A}(z)$,
where
\begin{align}
w^{\mRim3}(z) &=-\sigma (z)^{-2}=\frac{z}{4m_{\pi }^{2}-z}\,, \\
w^{\mRim2}(z) &=-\sigma (z)^{2}=\frac{4m_{\pi }^{2}-z}{z}\,, \\
w^{\mRima1a}(z) &=w^{\mRima1b}(z)=w^{\mRima1c}(z)=\sigma (z)^{2}=\frac{z-4m_{\pi
}^{2}}{z}\,,
\end{align}
in the general form
\begin{equation}
K_j^{A}(z)=P^{A}(w^{A}(z))\sum_{k=0}^{\infty }\Big(
a_{k;A}^{(j)}M_{A}^{(k)}(w^{A}(z))
+b_{k;A}^{(j)}N_{A}^{(k)}(w^{A}(z))\Big),
\end{equation}
where the first order polynomials $P^{A}(w)$ are
\begin{align}
P^{\mRim3}(w) &=\frac{1+w}{\Delta \sigma _{+}\sigma _{-}}\,, \\
P^{\mRim2}(w) &=-\frac{1+w}{\Delta }\,, \\
P^{\mRima1a}(w) &=P^{\mRima1b}(w)=P^{\mRima1c}(w)=\frac{1-w}{\Delta }
\end{align}
and the coefficients $a_{k;A}^{(j)}$ and $b_{k;A}^{(j)}$ correspond to the expansions of
the functions $f_{j}(\sigma)$ and $g_{j}(1/\sigma )$, $j=4,5$, in the integrands within the region considered. Let us remind that $M_{A}^{(k)}(w)$ can be obtained from the ``generating function'' in the general form
\begin{equation}
M_{A}^{(k)}(w) =w^{k}M_{A}^{(0)}(w)-w^{k}\sum_{i=0}^{k}\frac{w^{-i}}{i!}
\frac{\partial ^{i}}{\partial \alpha ^{i}}M_{A}^{(0)}\big(\alpha ^{-1}\big)\Big|_{\alpha
=0} 
=w^{k}M_{A}^{(0)}(w)-Q_{A}^{(k)}(w),
\end{equation}
where $Q_{A}^{(k)}(w)$ is a polynomial of order $k$, and similarly for the
functions $N_{A}^{(k)}(w)$. In their convergence region the series $\sum_{k=0}^{\infty }a_{k;A}^{(j)}w^{k}$ and $\sum_{k=0}^{\infty}b_{k;A}^{(j)}w^{k}$ can be summed up reproducing the even and the odd parts of the
functions $f_{j}(\sigma )$ and $g_{j}(1/\sigma )$ or $g_{j}(\sigma )$ in the
integrands within the region A, e.g.\ for $K_{4}^{\mRim3}(z)$ and
for $|w|<1$ we get
\begin{equation}
\sum_{k=0}^{\infty }a_{k,A}^{(4)}w^{k}=-2\frac{1}{\sqrt{w}}\arctan \sqrt{w}\,.
\end{equation}
\begin{figure}
\begin{center}
\includegraphics{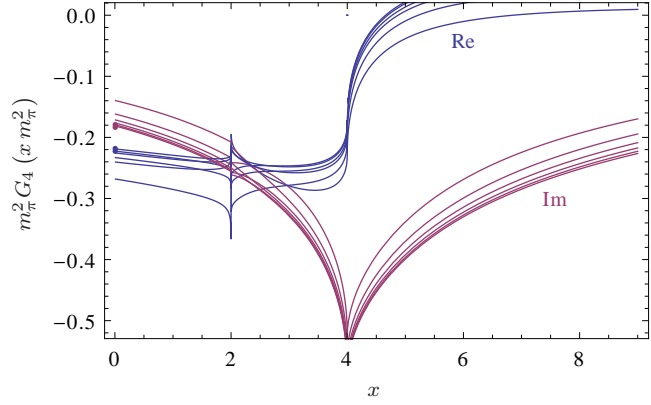}
\end{center}
\caption{Six successive (resummed) approximations (obtained by truncation
of the infinite series at $i=1,\dots ,6$) of the functions $G_4(s)$. The
spurious cusp at $s=2\mpi^2$ is an artifact of the approximation, caused by a slower convergence at this
point.}
\label{fig}
\end{figure}
Note, however, that these convergence regions do not generally coincide, so
that this partial summations can not be made simultaneously for all $n$. Nevertheless, for real $z=x+\mathrm{i}0$, such a summation reproduces
exactly either the imaginary or the real part of the corresponding integral $K_j^{A}(z)$. For instance, in the region \Rim3 we have
\begin{equation}
K_{j}^{\mRim3}(x+\mathrm{i}0)=\int_{0}^{2m_{\pi }^{2}}\frac{dy}{y-x-\mathrm{i}0}
\frac{1}{\widetilde{\lambda }_{1/2}(y)}g_{j}(|\sigma (y+\mathrm{i}0)|)\,,
\end{equation}
which gives\footnote{Note that $\widetilde{\lambda }^{1/2}(x)=\lambda ^{1/2}(x)$ for $0<x<2M_{\pi }^{2}$.}
\begin{equation}
\im K_{4}^{(1)}(x+\mathrm{i}0)=\pi \theta (x)\theta (2m_{\pi }^{2}-x)
\,\frac{-2\left\vert \sigma (x)\right\vert \arctan \left\vert \sigma
(x)\right\vert ^{-1}}{\lambda ^{1/2}(x)}
\end{equation}
The above summation of the terms is possible for $\left\vert \sigma (x)\right\vert >1$ and gives the following
contribution to the whole integral (note that the remaining part which
cannot be summed explicitly is real)
\begin{equation}
\frac{1-\sigma (x)^{-2}}{\Delta \sigma _{+}\sigma _{-}}\sum_{k=0}^{\infty
}a_{k;\mRim3}^{(4)}(-\sigma (x)^{-2})^{k}M_{\mRim3}^{(0)}\big(-\sigma (x)^{-2}\big)
=2\sqrt{-\sigma (x)^{2}}\arctan \sqrt{-\sigma (x)^{-2}}
\frac{1-\sigma
(x)^{2}}{\sigma (x)^{2}\Delta \sigma _{+}\sigma _{-}}\,M_{\mRim3}^{(0)}\big(-\sigma
(x)^{-2}\big).
\end{equation}
Thus, since
\begin{equation}
M_{\mRim3}^{(0)}\big(-\sigma (z)^{-2}\big)=\Delta \sigma _{+}\sigma _{-}\frac{\sigma
(z)^{2}}{\sigma (z)^{2}-1}
\frac{\log \left( \frac{\sigma _{\pi \eta
}(z)+\sigma _{\pi \eta }(0)}{\sigma _{\pi \eta }(z)-\sigma _{\pi \eta }(0)}\right)
 -\log \left( \frac{\sigma _{\pi \eta }(z)+\sigma _{\pi \eta }(2m_{\pi
}^{2})}{\sigma _{\pi \eta }(z)-\sigma _{\pi \eta }(2m_{\pi }^{2})}\right) }{\lambda _{1/2}(z)}\,,
\end{equation}
the imaginary part of this ``generating integral'' is equal to
\begin{equation}
\im M_{\mRim3}^{(0)}\big(-\sigma (x+\mathrm{i}0)^{-2}\big)
=\Delta \sigma
_{+}\sigma _{-}\frac{\sigma (x)^{2}}{\sigma (x)^{2}-1}\pi\, \theta
(x)\theta (2m_{\pi }^{2}-x)\,\frac{1}{\lambda _{1/2}(x)}\,.
\end{equation}
Here $\sigma (x)^{2}<0$ and the imaginary part is fully reproduced. The same
can be done for the other regions, where we need, however, $\left\vert \sigma
(x)\right\vert <1$. One finds that there again the imaginary parts (or the real parts) are completely
reproduced.

In the physical region we have $0\leq \sigma (x)<1$; therefore the summation
for the regions \Rim1 and \Rim2 can be performed. The convergence of the resummed
approximations is illustrated in Fig.~\ref{fig}.

\bibliography{bibliografie}

\begin{thebibliography}{81}%
\makeatletter
\providecommand \@ifxundefined [1]{%
 \@ifx{#1\undefined}
}%
\providecommand \@ifnum [1]{%
 \ifnum #1\expandafter \@firstoftwo
 \else \expandafter \@secondoftwo
 \fi
}%
\providecommand \@ifx [1]{%
 \ifx #1\expandafter \@firstoftwo
 \else \expandafter \@secondoftwo
 \fi
}%
\providecommand \natexlab [1]{#1}%
\providecommand \enquote  [1]{``#1''}%
\providecommand \bibnamefont  [1]{#1}%
\providecommand \bibfnamefont [1]{#1}%
\providecommand \citenamefont [1]{#1}%
\providecommand \href@noop [0]{\@secondoftwo}%
\providecommand \href [0]{\begingroup \@sanitize@url \@href}%
\providecommand \@href[1]{\@@startlink{#1}\@@href}%
\providecommand \@@href[1]{\endgroup#1\@@endlink}%
\providecommand \@sanitize@url [0]{\catcode `\\12\catcode `\$12\catcode
  `\&12\catcode `\#12\catcode `\^12\catcode `\_12\catcode `\%12\relax}%
\providecommand \@@startlink[1]{}%
\providecommand \@@endlink[0]{}%
\providecommand \url  [0]{\begingroup\@sanitize@url \@url }%
\providecommand \@url [1]{\endgroup\@href {#1}{\urlprefix }}%
\providecommand \urlprefix  [0]{URL }%
\providecommand \Eprint [0]{\href }%
\providecommand \doibase [0]{http://dx.doi.org/}%
\providecommand \selectlanguage [0]{\@gobble}%
\providecommand \bibinfo  [0]{\@secondoftwo}%
\providecommand \bibfield  [0]{\@secondoftwo}%
\providecommand \translation [1]{[#1]}%
\providecommand \BibitemOpen [0]{}%
\providecommand \bibitemStop [0]{}%
\providecommand \bibitemNoStop [0]{.\EOS\space}%
\providecommand \EOS [0]{\spacefactor3000\relax}%
\providecommand \BibitemShut  [1]{\csname bibitem#1\endcsname}%
\let\auto@bib@innerbib\@empty
%</preamble>
\bibitem [{\citenamefont {Wilson}(1974)}]{Wilson:1974sk}%
  \BibitemOpen
  \bibfield  {author} {\bibinfo {author} {\bibfnamefont {K.~G.}\ \bibnamefont
  {Wilson}},\ }\bibfield  {title} {\enquote {\bibinfo {title} {{Confinement of
  Quarks}},}\ }\href {\doibase 10.1103/PhysRevD.10.2445} {\bibfield  {journal}
  {\bibinfo  {journal} {Phys.\ Rev.}\ }\textbf {\bibinfo {volume} {D10}},\
  \bibinfo {pages} {2445--2459} (\bibinfo {year} {1974})}\BibitemShut {NoStop}%
\bibitem [{\citenamefont {Weinberg}(1979)}]{Weinberg:1978kz}%
  \BibitemOpen
  \bibfield  {author} {\bibinfo {author} {\bibfnamefont {S.}~\bibnamefont
  {Weinberg}},\ }\bibfield  {title} {\enquote {\bibinfo {title}
  {{Phenomenological Lagrangians}},}\ }\href@noop {} {\bibfield  {journal}
  {\bibinfo  {journal} {Physica}\ }\textbf {\bibinfo {volume} {A96}},\ \bibinfo
  {pages} {327} (\bibinfo {year} {1979})}\BibitemShut {NoStop}%
%%CITATION = PHYSA,A96,327;%%
\bibitem [{\citenamefont {Gasser}\ and\ \citenamefont
  {Leutwyler}(1984)}]{GL-scatlen}%
  \BibitemOpen
  \bibfield  {author} {\bibinfo {author} {\bibfnamefont {J.}~\bibnamefont
  {Gasser}}\ and\ \bibinfo {author} {\bibfnamefont {H.}~\bibnamefont
  {Leutwyler}},\ }\bibfield  {title} {\enquote {\bibinfo {title} {{Chiral
  Perturbation Theory To One Loop}},}\ }\href@noop {} {\bibfield  {journal}
  {\bibinfo  {journal} {Annals Phys.}\ }\textbf {\bibinfo {volume} {158}},\
  \bibinfo {pages} {142} (\bibinfo {year} {1984})}\BibitemShut {NoStop}%
%%CITATION = APNYA,158,142;%%
\bibitem [{\citenamefont {Gasser}\ and\ \citenamefont
  {Leutwyler}(1985{\natexlab{a}})}]{Gasser:1984gg}%
  \BibitemOpen
  \bibfield  {author} {\bibinfo {author} {\bibfnamefont {J.}~\bibnamefont
  {Gasser}}\ and\ \bibinfo {author} {\bibfnamefont {H.}~\bibnamefont
  {Leutwyler}},\ }\bibfield  {title} {\enquote {\bibinfo {title} {{Chiral
  Perturbation Theory: Expansions in the Mass of the Strange Quark}},}\ }\href
  {\doibase 10.1016/0550-3213(85)90492-4} {\bibfield  {journal} {\bibinfo
  {journal} {Nucl.\ Phys.}\ }\textbf {\bibinfo {volume} {B250}},\ \bibinfo
  {pages} {465} (\bibinfo {year} {1985}{\natexlab{a}})}\BibitemShut {NoStop}%
%%CITATION = NUPHA,B250,465;%%
\bibitem [{\citenamefont {Allton}\ \emph {et~al.}(2008)\citenamefont {Allton}
  \emph {et~al.}}]{RBC-UKQCD}%
  \BibitemOpen
  \bibfield  {author} {\bibinfo {author} {\bibfnamefont {C.}~\bibnamefont
  {Allton}} \emph {et~al.} (\bibinfo {collaboration} {RBC-UKQCD}),\ }\bibfield
  {title} {\enquote {\bibinfo {title} {{Physical Results from $2+1$ Flavor
  Domain Wall QCD and SU(2) Chiral Perturbation Theory}},}\ }\href {\doibase
  10.1103/PhysRevD.78.114509} {\bibfield  {journal} {\bibinfo  {journal}
  {Phys.\ Rev.}\ }\textbf {\bibinfo {volume} {D78}},\ \bibinfo {pages} {114509}
  (\bibinfo {year} {2008})},\ \Eprint {http://arxiv.org/abs/0804.0473}
  {arXiv:0804.0473 [hep-lat]} \BibitemShut {NoStop}%
\bibitem [{\citenamefont {Bazavov}\ \emph {et~al.}(2010)\citenamefont
  {Bazavov}, \citenamefont {Toussaint}, \citenamefont {Bernard} \emph
  {et~al.}}]{MILC}%
  \BibitemOpen
  \bibfield  {author} {\bibinfo {author} {\bibfnamefont {A.}~\bibnamefont
  {Bazavov}}, \bibinfo {author} {\bibfnamefont {D.}~\bibnamefont {Toussaint}},
  \bibinfo {author} {\bibfnamefont {C.}~\bibnamefont {Bernard}},  \emph
  {et~al.},\ }\bibfield  {title} {\enquote {\bibinfo {title} {{Nonperturbative
  QCD simulations with $2+1$ flavors of improved staggered quarks}},}\ }\href
  {\doibase 10.1103/RevModPhys.82.1349} {\bibfield  {journal} {\bibinfo
  {journal} {Rev.\ Mod.\ Phys.}\ }\textbf {\bibinfo {volume} {82}},\ \bibinfo
  {pages} {1349--1417} (\bibinfo {year} {2010})},\ \Eprint
  {http://arxiv.org/abs/0903.3598} {arXiv:0903.3598 [hep-lat]} \BibitemShut
  {NoStop}%
\bibitem [{\citenamefont {Davies}\ \emph {et~al.}(2010)\citenamefont {Davies},
  \citenamefont {McNeile}, \citenamefont {Wong} \emph {et~al.}}]{HPQCD}%
  \BibitemOpen
  \bibfield  {author} {\bibinfo {author} {\bibfnamefont {C.~T.~H.}\
  \bibnamefont {Davies}}, \bibinfo {author} {\bibfnamefont {C.}~\bibnamefont
  {McNeile}}, \bibinfo {author} {\bibfnamefont {K.~Y.}\ \bibnamefont {Wong}},
  \emph {et~al.},\ }\bibfield  {title} {\enquote {\bibinfo {title} {{Precise
  Charm to Strange Mass Ratio and Light Quark Masses from Full Lattice QCD}},}\
  }\href {\doibase 10.1103/PhysRevLett.104.132003} {\bibfield  {journal}
  {\bibinfo  {journal} {Phys.\ Rev.\ Lett.}\ }\textbf {\bibinfo {volume}
  {104}},\ \bibinfo {pages} {132003} (\bibinfo {year} {2010})},\ \Eprint
  {http://arxiv.org/abs/0910.3102} {arXiv:0910.3102 [hep-ph]} \BibitemShut
  {NoStop}%
\bibitem [{\citenamefont {Aoki}\ \emph {et~al.}(2010)\citenamefont {Aoki} \emph
  {et~al.}}]{PACS-CS}%
  \BibitemOpen
  \bibfield  {author} {\bibinfo {author} {\bibfnamefont {S.}~\bibnamefont
  {Aoki}} \emph {et~al.} (\bibinfo {collaboration} {PACS-CS}),\ }\bibfield
  {title} {\enquote {\bibinfo {title} {{Physical Point Simulation in $2+1$
  Flavor Lattice QCD}},}\ }\href {\doibase 10.1103/PhysRevD.81.074503}
  {\bibfield  {journal} {\bibinfo  {journal} {Phys.\ Rev.}\ }\textbf {\bibinfo
  {volume} {D81}},\ \bibinfo {pages} {074503} (\bibinfo {year} {2010})},\
  \Eprint {http://arxiv.org/abs/0911.2561} {arXiv:0911.2561 [hep-lat]}
  \BibitemShut {NoStop}%
\bibitem [{\citenamefont {Maltman}\ and\ \citenamefont
  {Kambor}(2001)}]{Maltman:2001nx}%
  \BibitemOpen
  \bibfield  {author} {\bibinfo {author} {\bibfnamefont {K.}~\bibnamefont
  {Maltman}}\ and\ \bibinfo {author} {\bibfnamefont {J.}~\bibnamefont
  {Kambor}},\ }\bibfield  {title} {\enquote {\bibinfo {title} {{$m_u + m_d$
  from isovector pseudoscalar sum rules}},}\ }\href {\doibase
  10.1016/S0370-2693(01)00987-X} {\bibfield  {journal} {\bibinfo  {journal}
  {Phys.\ Lett.}\ }\textbf {\bibinfo {volume} {B517}},\ \bibinfo {pages}
  {332--338} (\bibinfo {year} {2001})},\ \Eprint
  {http://arxiv.org/abs/hep-ph/0107060} {arXiv:hep-ph/0107060} \BibitemShut
  {NoStop}%
\bibitem [{\citenamefont {Narison}(2006)}]{Narison:2005ny}%
  \BibitemOpen
  \bibfield  {author} {\bibinfo {author} {\bibfnamefont {S.}~\bibnamefont
  {Narison}},\ }\bibfield  {title} {\enquote {\bibinfo {title} {{Strange quark
  mass from $e^+ e^-$ revisited and present status of light quark masses}},}\
  }\href {\doibase 10.1103/PhysRevD.74.034013} {\bibfield  {journal} {\bibinfo
  {journal} {Phys.\ Rev.}\ }\textbf {\bibinfo {volume} {D74}},\ \bibinfo
  {pages} {034013} (\bibinfo {year} {2006})},\ \Eprint
  {http://arxiv.org/abs/hep-ph/0510108} {arXiv:hep-ph/0510108} \BibitemShut
  {NoStop}%
\bibitem [{\citenamefont {Dominguez}\ \emph {et~al.}(2009)\citenamefont
  {Dominguez}, \citenamefont {Nasrallah}, \citenamefont {Rontsch},\ and\
  \citenamefont {Schilcher}}]{Dominguez:2008jz}%
  \BibitemOpen
  \bibfield  {author} {\bibinfo {author} {\bibfnamefont {C.~A.}\ \bibnamefont
  {Dominguez}}, \bibinfo {author} {\bibfnamefont {N.~F.}\ \bibnamefont
  {Nasrallah}}, \bibinfo {author} {\bibfnamefont {R.}~\bibnamefont {Rontsch}},
  \ and\ \bibinfo {author} {\bibfnamefont {K.}~\bibnamefont {Schilcher}},\
  }\bibfield  {title} {\enquote {\bibinfo {title} {{Light quark masses from QCD
  sum rules with minimal hadronic bias}},}\ }\href {\doibase
  10.1016/j.nuclphysbps.2008.12.031} {\bibfield  {journal} {\bibinfo  {journal}
  {Nucl.\ Phys.\ Proc.\ Suppl.}\ }\textbf {\bibinfo {volume} {186}},\ \bibinfo
  {pages} {133--136} (\bibinfo {year} {2009})},\ \Eprint
  {http://arxiv.org/abs/0808.3909} {arXiv:0808.3909 [hep-ph]} \BibitemShut
  {NoStop}%
\bibitem [{\citenamefont {Colangelo}\ \emph {et~al.}(2011)\citenamefont
  {Colangelo}, \citenamefont {Durr}, \citenamefont {Juttner} \emph
  {et~al.}}]{Colangelo:2010et}%
  \BibitemOpen
  \bibfield  {author} {\bibinfo {author} {\bibfnamefont {G.}~\bibnamefont
  {Colangelo}}, \bibinfo {author} {\bibfnamefont {S.}~\bibnamefont {Durr}},
  \bibinfo {author} {\bibfnamefont {A.}~\bibnamefont {Juttner}},  \emph
  {et~al.},\ }\bibfield  {title} {\enquote {\bibinfo {title} {{Review of
  lattice results concerning low energy particle physics}},}\ }\href {\doibase
  10.1140/epjc/s10052-011-1695-1} {\bibfield  {journal} {\bibinfo  {journal}
  {Eur.\ Phys.\ J.}\ }\textbf {\bibinfo {volume} {C71}},\ \bibinfo {pages}
  {1695} (\bibinfo {year} {2011})},\ \Eprint {http://arxiv.org/abs/1011.4408}
  {arXiv:1011.4408 [hep-lat]} \BibitemShut {NoStop}%
%%CITATION = 1011.4408;%%
\bibitem [{\citenamefont {Sutherland}(1966)}]{Sutherland:1966zz}%
  \BibitemOpen
  \bibfield  {author} {\bibinfo {author} {\bibfnamefont {D.~G.}\ \bibnamefont
  {Sutherland}},\ }\bibfield  {title} {\enquote {\bibinfo {title} {{Current
  algebra and the decay $\eta\to3\pi$}},}\ }\href@noop {} {\bibfield  {journal}
  {\bibinfo  {journal} {Phys.\ Lett.}\ }\textbf {\bibinfo {volume} {23}},\
  \bibinfo {pages} {384} (\bibinfo {year} {1966})}\BibitemShut {NoStop}%
\bibitem [{\citenamefont {Baur}\ \emph {et~al.}(1996)\citenamefont {Baur},
  \citenamefont {Kambor},\ and\ \citenamefont {Wyler}}]{Baur:1995gc}%
  \BibitemOpen
  \bibfield  {author} {\bibinfo {author} {\bibfnamefont {R.}~\bibnamefont
  {Baur}}, \bibinfo {author} {\bibfnamefont {J.}~\bibnamefont {Kambor}}, \ and\
  \bibinfo {author} {\bibfnamefont {D.}~\bibnamefont {Wyler}},\ }\bibfield
  {title} {\enquote {\bibinfo {title} {{Electromagnetic corrections to the
  decays $\eta\to3\pi$}},}\ }\href {\doibase 10.1016/0550-3213(95)00643-5}
  {\bibfield  {journal} {\bibinfo  {journal} {Nucl.\ Phys.}\ }\textbf {\bibinfo
  {volume} {B460}},\ \bibinfo {pages} {127--142} (\bibinfo {year} {1996})},\
  \Eprint {http://arxiv.org/abs/hep-ph/9510396} {arXiv:hep-ph/9510396}
  \BibitemShut {NoStop}%
\bibitem [{\citenamefont {Ditsche}\ \emph {et~al.}(2009)\citenamefont
  {Ditsche}, \citenamefont {Kubis},\ and\ \citenamefont
  {Meissner}}]{Ditsche:2008cq}%
  \BibitemOpen
  \bibfield  {author} {\bibinfo {author} {\bibfnamefont {C.}~\bibnamefont
  {Ditsche}}, \bibinfo {author} {\bibfnamefont {B.}~\bibnamefont {Kubis}}, \
  and\ \bibinfo {author} {\bibfnamefont {U.-G.}\ \bibnamefont {Meissner}},\
  }\bibfield  {title} {\enquote {\bibinfo {title} {{Electromagnetic corrections
  in $\eta\to3\pi$ decays}},}\ }\href {\doibase 10.1140/epjc/s10052-009-0863-z}
  {\bibfield  {journal} {\bibinfo  {journal} {Eur.\ Phys.\ J.}\ }\textbf
  {\bibinfo {volume} {C60}},\ \bibinfo {pages} {83--105} (\bibinfo {year}
  {2009})},\ \Eprint {http://arxiv.org/abs/0812.0344} {arXiv:0812.0344
  [hep-ph]} \BibitemShut {NoStop}%
\bibitem [{\citenamefont {Bell}\ and\ \citenamefont
  {Sutherland}(1968)}]{Bell:1996mi}%
  \BibitemOpen
  \bibfield  {author} {\bibinfo {author} {\bibfnamefont {J.~S.}\ \bibnamefont
  {Bell}}\ and\ \bibinfo {author} {\bibfnamefont {D.~G.}\ \bibnamefont
  {Sutherland}},\ }\bibfield  {title} {\enquote {\bibinfo {title} {{Current
  algebra and $\eta\to3\pi$}},}\ }\href {\doibase 10.1016/0550-3213(68)90316-7}
  {\bibfield  {journal} {\bibinfo  {journal} {Nucl.\ Phys.}\ }\textbf {\bibinfo
  {volume} {B4}},\ \bibinfo {pages} {315--325} (\bibinfo {year}
  {1968})}\BibitemShut {NoStop}%
\bibitem [{\citenamefont {Cronin}(1967)}]{Cronin:1967jq}%
  \BibitemOpen
  \bibfield  {author} {\bibinfo {author} {\bibfnamefont {J.~A.}\ \bibnamefont
  {Cronin}},\ }\bibfield  {title} {\enquote {\bibinfo {title}
  {{Phenomenological model of strong and weak interactions in chiral
  U(3)$\times$U(3)}},}\ }\href {\doibase 10.1103/PhysRev.161.1483} {\bibfield
  {journal} {\bibinfo  {journal} {Phys.\ Rev.}\ }\textbf {\bibinfo {volume}
  {161}},\ \bibinfo {pages} {1483--1494} (\bibinfo {year} {1967})}\BibitemShut
  {NoStop}%
\bibitem [{\citenamefont {Osborn}\ and\ \citenamefont
  {Wallace}(1970)}]{Osborn:1970nn}%
  \BibitemOpen
  \bibfield  {author} {\bibinfo {author} {\bibfnamefont {H.}~\bibnamefont
  {Osborn}}\ and\ \bibinfo {author} {\bibfnamefont {D.~J.}\ \bibnamefont
  {Wallace}},\ }\bibfield  {title} {\enquote {\bibinfo {title} {{Eta - x
  mixing, $\eta\to3\pi$ and chiral lagrangians}},}\ }\href {\doibase
  10.1016/0550-3213(70)90194-X} {\bibfield  {journal} {\bibinfo  {journal}
  {Nucl.\ Phys.}\ }\textbf {\bibinfo {volume} {B20}},\ \bibinfo {pages}
  {23--44} (\bibinfo {year} {1970})}\BibitemShut {NoStop}%
\bibitem [{\citenamefont {Gasser}\ and\ \citenamefont
  {Leutwyler}(1985{\natexlab{b}})}]{Gasser:1984pr}%
  \BibitemOpen
  \bibfield  {author} {\bibinfo {author} {\bibfnamefont {J.}~\bibnamefont
  {Gasser}}\ and\ \bibinfo {author} {\bibfnamefont {H.}~\bibnamefont
  {Leutwyler}},\ }\bibfield  {title} {\enquote {\bibinfo {title} {{$\eta\to
  3\pi$ to One Loop}},}\ }\href {\doibase 10.1016/0550-3213(85)90494-8}
  {\bibfield  {journal} {\bibinfo  {journal} {Nucl.\ Phys.}\ }\textbf {\bibinfo
  {volume} {B250}},\ \bibinfo {pages} {539} (\bibinfo {year}
  {1985}{\natexlab{b}})}\BibitemShut {NoStop}%
%%CITATION = NUPHA,B250,539;%%
\bibitem [{\citenamefont {Bijnens}\ and\ \citenamefont {Ghorbani}(2007)}]{BG}%
  \BibitemOpen
  \bibfield  {author} {\bibinfo {author} {\bibfnamefont {J.}~\bibnamefont
  {Bijnens}}\ and\ \bibinfo {author} {\bibfnamefont {K.}~\bibnamefont
  {Ghorbani}},\ }\bibfield  {title} {\enquote {\bibinfo {title} {{$\eta \to 3
  \pi$ at Two Loops In Chiral Perturbation Theory}},}\ }\href {\doibase
  10.1088/1126-6708/2007/11/030} {\bibfield  {journal} {\bibinfo  {journal}
  {JHEP}\ }\textbf {\bibinfo {volume} {11}},\ \bibinfo {pages} {030} (\bibinfo
  {year} {2007})},\ \Eprint {http://arxiv.org/abs/0709.0230} {arXiv:0709.0230
  [hep-ph]} \BibitemShut {NoStop}%
%%CITATION = 0709.0230;%%
\bibitem [{\citenamefont {Prakhov}\ \emph {et~al.}(2009)\citenamefont {Prakhov}
  \emph {et~al.}}]{MAMI-C}%
  \BibitemOpen
  \bibfield  {author} {\bibinfo {author} {\bibfnamefont {S.}~\bibnamefont
  {Prakhov}} \emph {et~al.} (\bibinfo {collaboration} {Crystal Ball at MAMI}),\
  }\bibfield  {title} {\enquote {\bibinfo {title} {{Measurement of the Slope
  Parameter $\alpha$ for the $\eta\to 3\pi^0$ decay with the Crystal Ball at
  MAMI-C}},}\ }\href {\doibase 10.1103/PhysRevC.79.035204} {\bibfield
  {journal} {\bibinfo  {journal} {Phys.\ Rev.}\ }\textbf {\bibinfo {volume}
  {C79}},\ \bibinfo {pages} {035204} (\bibinfo {year} {2009})},\ \Eprint
  {http://arxiv.org/abs/0812.1999} {arXiv:0812.1999 [hep-ex]} \BibitemShut
  {NoStop}%
%%CITATION = 0812.1999;%%
\bibitem [{\citenamefont {Ambrosino}\ \emph {et~al.}(2008)\citenamefont
  {Ambrosino} \emph {et~al.}}]{KLOE-etac}%
  \BibitemOpen
  \bibfield  {author} {\bibinfo {author} {\bibfnamefont {F.}~\bibnamefont
  {Ambrosino}} \emph {et~al.} (\bibinfo {collaboration} {KLOE}),\ }\bibfield
  {title} {\enquote {\bibinfo {title} {{Determination of
  $\eta\to\pi^+\pi^-\pi^0$ Dalitz plot slopes and asymmetries with the KLOE
  detector}},}\ }\href {\doibase 10.1088/1126-6708/2008/05/006} {\bibfield
  {journal} {\bibinfo  {journal} {JHEP}\ }\textbf {\bibinfo {volume} {0805}},\
  \bibinfo {pages} {006} (\bibinfo {year} {2008})},\ \Eprint
  {http://arxiv.org/abs/0801.2642} {arXiv:0801.2642 [hep-ex]} \BibitemShut
  {NoStop}%
%%CITATION = 0801.2642;%%
\bibitem [{\citenamefont {Ecker}\ \emph
  {et~al.}(1989{\natexlab{a}})\citenamefont {Ecker}, \citenamefont {Gasser},
  \citenamefont {Pich},\ and\ \citenamefont {de~Rafael}}]{Ecker:1988te}%
  \BibitemOpen
  \bibfield  {author} {\bibinfo {author} {\bibfnamefont {G.}~\bibnamefont
  {Ecker}}, \bibinfo {author} {\bibfnamefont {J.}~\bibnamefont {Gasser}},
  \bibinfo {author} {\bibfnamefont {A.}~\bibnamefont {Pich}}, \ and\ \bibinfo
  {author} {\bibfnamefont {E.}~\bibnamefont {de~Rafael}},\ }\bibfield  {title}
  {\enquote {\bibinfo {title} {{The Role of Resonances in Chiral Perturbation
  Theory}},}\ }\href {\doibase 10.1016/0550-3213(89)90346-5} {\bibfield
  {journal} {\bibinfo  {journal} {Nucl.\ Phys.}\ }\textbf {\bibinfo {volume}
  {B321}},\ \bibinfo {pages} {311} (\bibinfo {year}
  {1989}{\natexlab{a}})}\BibitemShut {NoStop}%
\bibitem [{\citenamefont {Ecker}\ \emph
  {et~al.}(1989{\natexlab{b}})\citenamefont {Ecker}, \citenamefont {Gasser},
  \citenamefont {Leutwyler} \emph {et~al.}}]{Ecker:1989yg}%
  \BibitemOpen
  \bibfield  {author} {\bibinfo {author} {\bibfnamefont {G.}~\bibnamefont
  {Ecker}}, \bibinfo {author} {\bibfnamefont {J.}~\bibnamefont {Gasser}},
  \bibinfo {author} {\bibfnamefont {H.}~\bibnamefont {Leutwyler}},  \emph
  {et~al.},\ }\bibfield  {title} {\enquote {\bibinfo {title} {{Chiral
  Lagrangians for Massive Spin 1 Fields}},}\ }\href {\doibase
  10.1016/0370-2693(89)91627-4} {\bibfield  {journal} {\bibinfo  {journal}
  {Phys.\ Lett.}\ }\textbf {\bibinfo {volume} {B223}},\ \bibinfo {pages} {425}
  (\bibinfo {year} {1989}{\natexlab{b}})}\BibitemShut {NoStop}%
\bibitem [{\citenamefont {Moussallam}\ and\ \citenamefont
  {Stern}(1994)}]{Moussallam:1994at}%
  \BibitemOpen
  \bibfield  {author} {\bibinfo {author} {\bibfnamefont {B.}~\bibnamefont
  {Moussallam}}\ and\ \bibinfo {author} {\bibfnamefont {J.}~\bibnamefont
  {Stern}},\ }\bibfield  {title} {\enquote {\bibinfo {title} {{Chiral symmetry
  aspects of the scalars}},}\ }\href@noop {} {\  (\bibinfo {year} {1994})},\
  \Eprint {http://arxiv.org/abs/hep-ph/9404353} {arXiv:hep-ph/9404353}
  \BibitemShut {NoStop}%
\bibitem [{\citenamefont {Knecht}\ and\ \citenamefont
  {Nyffeler}(2001)}]{Knecht:2001xc}%
  \BibitemOpen
  \bibfield  {author} {\bibinfo {author} {\bibfnamefont {M.}~\bibnamefont
  {Knecht}}\ and\ \bibinfo {author} {\bibfnamefont {A.}~\bibnamefont
  {Nyffeler}},\ }\bibfield  {title} {\enquote {\bibinfo {title} {{Resonance
  estimates of $O(p^6)$ low-energy constants and QCD short distance
  constraints}},}\ }\href {\doibase 10.1007/s100520100755} {\bibfield
  {journal} {\bibinfo  {journal} {Eur.\ Phys.\ J.}\ }\textbf {\bibinfo {volume}
  {C21}},\ \bibinfo {pages} {659--678} (\bibinfo {year} {2001})},\ \Eprint
  {http://arxiv.org/abs/hep-ph/0106034} {arXiv:hep-ph/0106034} \BibitemShut
  {NoStop}%
\bibitem [{\citenamefont {Kambor}\ \emph {et~al.}(1996)\citenamefont {Kambor},
  \citenamefont {Wiesendanger},\ and\ \citenamefont {Wyler}}]{KWW}%
  \BibitemOpen
  \bibfield  {author} {\bibinfo {author} {\bibfnamefont {J.}~\bibnamefont
  {Kambor}}, \bibinfo {author} {\bibfnamefont {C.}~\bibnamefont
  {Wiesendanger}}, \ and\ \bibinfo {author} {\bibfnamefont {D.}~\bibnamefont
  {Wyler}},\ }\bibfield  {title} {\enquote {\bibinfo {title} {{Final State
  Interactions and Khuri-Treiman Equations in $\eta\to 3\pi$ decays}},}\ }\href
  {\doibase 10.1016/0550-3213(95)00676-1} {\bibfield  {journal} {\bibinfo
  {journal} {Nucl.\ Phys.}\ }\textbf {\bibinfo {volume} {B465}},\ \bibinfo
  {pages} {215--266} (\bibinfo {year} {1996})},\ \Eprint
  {http://arxiv.org/abs/hep-ph/9509374} {arXiv:hep-ph/9509374} \BibitemShut
  {NoStop}%
%%CITATION = HEP-PH/9509374;%%
\bibitem [{\citenamefont {Anisovich}\ and\ \citenamefont
  {Leutwyler}(1996)}]{Anisovich:1996tx}%
  \BibitemOpen
  \bibfield  {author} {\bibinfo {author} {\bibfnamefont {A.~V.}\ \bibnamefont
  {Anisovich}}\ and\ \bibinfo {author} {\bibfnamefont {H.}~\bibnamefont
  {Leutwyler}},\ }\bibfield  {title} {\enquote {\bibinfo {title} {{Dispersive
  analysis of the decay $\eta\to3\pi$}},}\ }\href {\doibase
  10.1016/0370-2693(96)00192-X} {\bibfield  {journal} {\bibinfo  {journal}
  {Phys.\ Lett.}\ }\textbf {\bibinfo {volume} {B375}},\ \bibinfo {pages}
  {335--342} (\bibinfo {year} {1996})},\ \Eprint
  {http://arxiv.org/abs/hep-ph/9601237} {arXiv:hep-ph/9601237} \BibitemShut
  {NoStop}%
\bibitem [{\citenamefont {Colangelo}\ \emph {et~al.}(2009)\citenamefont
  {Colangelo}, \citenamefont {Lanz},\ and\ \citenamefont
  {Passemar}}]{Colangelo:2009db}%
  \BibitemOpen
  \bibfield  {author} {\bibinfo {author} {\bibfnamefont {G.}~\bibnamefont
  {Colangelo}}, \bibinfo {author} {\bibfnamefont {S.}~\bibnamefont {Lanz}}, \
  and\ \bibinfo {author} {\bibfnamefont {E.}~\bibnamefont {Passemar}},\
  }\bibfield  {title} {\enquote {\bibinfo {title} {{A New Dispersive Analysis
  of $\eta\to3\pi$}},}\ }\href@noop {} {\bibfield  {journal} {\bibinfo
  {journal} {PoS}\ }\textbf {\bibinfo {volume} {CD09}},\ \bibinfo {pages} {047}
  (\bibinfo {year} {2009})},\ \Eprint {http://arxiv.org/abs/0910.0765}
  {arXiv:0910.0765 [hep-ph]} \BibitemShut {NoStop}%
\bibitem [{\citenamefont {Colangelo}\ \emph {et~al.}(2006)\citenamefont
  {Colangelo}, \citenamefont {Gasser}, \citenamefont {Kubis},\ and\
  \citenamefont {Rusetsky}}]{Colangelo:2006va}%
  \BibitemOpen
  \bibfield  {author} {\bibinfo {author} {\bibfnamefont {G.}~\bibnamefont
  {Colangelo}}, \bibinfo {author} {\bibfnamefont {J.}~\bibnamefont {Gasser}},
  \bibinfo {author} {\bibfnamefont {B.}~\bibnamefont {Kubis}}, \ and\ \bibinfo
  {author} {\bibfnamefont {A.}~\bibnamefont {Rusetsky}},\ }\bibfield  {title}
  {\enquote {\bibinfo {title} {{Cusps in $K\to3\pi$ decays}},}\ }\href
  {\doibase 10.1016/j.physletb.2006.05.017} {\bibfield  {journal} {\bibinfo
  {journal} {Phys.\ Lett.}\ }\textbf {\bibinfo {volume} {B638}},\ \bibinfo
  {pages} {187--194} (\bibinfo {year} {2006})},\ \Eprint
  {http://arxiv.org/abs/hep-ph/0604084} {arXiv:hep-ph/0604084} \BibitemShut
  {NoStop}%
\bibitem [{\citenamefont {Gasser}\ \emph {et~al.}(2011)\citenamefont {Gasser},
  \citenamefont {Kubis},\ and\ \citenamefont {Rusetsky}}]{Gasser:2011ju}%
  \BibitemOpen
  \bibfield  {author} {\bibinfo {author} {\bibfnamefont {J.}~\bibnamefont
  {Gasser}}, \bibinfo {author} {\bibfnamefont {B.}~\bibnamefont {Kubis}}, \
  and\ \bibinfo {author} {\bibfnamefont {A.}~\bibnamefont {Rusetsky}},\
  }\bibfield  {title} {\enquote {\bibinfo {title} {{Cusps in $K\to3\pi$ decays:
  a theoretical framework}},}\ }\href {\doibase
  10.1016/j.nuclphysb.2011.04.013} {\bibfield  {journal} {\bibinfo  {journal}
  {Nucl.\ Phys.}\ }\textbf {\bibinfo {volume} {B850}},\ \bibinfo {pages}
  {96--147} (\bibinfo {year} {2011})},\ \Eprint
  {http://arxiv.org/abs/1103.4273} {arXiv:1103.4273 [hep-ph]} \BibitemShut
  {NoStop}%
\bibitem [{\citenamefont {Bissegger}\ \emph {et~al.}(2008)\citenamefont
  {Bissegger}, \citenamefont {Fuhrer}, \citenamefont {Gasser} \emph
  {et~al.}}]{Bissegger:2007yq}%
  \BibitemOpen
  \bibfield  {author} {\bibinfo {author} {\bibfnamefont {M.}~\bibnamefont
  {Bissegger}}, \bibinfo {author} {\bibfnamefont {A.}~\bibnamefont {Fuhrer}},
  \bibinfo {author} {\bibfnamefont {J.}~\bibnamefont {Gasser}},  \emph
  {et~al.},\ }\bibfield  {title} {\enquote {\bibinfo {title} {{Cusps in
  $K(L)\to 3\pi$ decays}},}\ }\href {\doibase 10.1016/j.physletb.2007.11.008}
  {\bibfield  {journal} {\bibinfo  {journal} {Phys.\ Lett.}\ }\textbf {\bibinfo
  {volume} {B659}},\ \bibinfo {pages} {576--584} (\bibinfo {year} {2008})},\
  \Eprint {http://arxiv.org/abs/0710.4456} {arXiv:0710.4456 [hep-ph]}
  \BibitemShut {NoStop}%
\bibitem [{\citenamefont {Gullstrom}\ \emph {et~al.}(2009)\citenamefont
  {Gullstrom}, \citenamefont {Kupsc},\ and\ \citenamefont
  {Rusetsky}}]{Gullstrom:2008sy}%
  \BibitemOpen
  \bibfield  {author} {\bibinfo {author} {\bibfnamefont {C.-O.}\ \bibnamefont
  {Gullstrom}}, \bibinfo {author} {\bibfnamefont {A.}~\bibnamefont {Kupsc}}, \
  and\ \bibinfo {author} {\bibfnamefont {A.}~\bibnamefont {Rusetsky}},\
  }\bibfield  {title} {\enquote {\bibinfo {title} {{Predictions for the cusp in
  $\eta\to 3\pi^0$ decay}},}\ }\href {\doibase 10.1103/PhysRevC.79.028201}
  {\bibfield  {journal} {\bibinfo  {journal} {Phys.\ Rev.}\ }\textbf {\bibinfo
  {volume} {C79}},\ \bibinfo {pages} {028201} (\bibinfo {year} {2009})},\
  \Eprint {http://arxiv.org/abs/0812.2371} {arXiv:0812.2371 [hep-ph]}
  \BibitemShut {NoStop}%
\bibitem [{\citenamefont {Schneider}\ \emph {et~al.}(2011)\citenamefont
  {Schneider}, \citenamefont {Kubis},\ and\ \citenamefont
  {Ditsche}}]{Schneider:2010hs}%
  \BibitemOpen
  \bibfield  {author} {\bibinfo {author} {\bibfnamefont {S.~P.}\ \bibnamefont
  {Schneider}}, \bibinfo {author} {\bibfnamefont {B.}~\bibnamefont {Kubis}}, \
  and\ \bibinfo {author} {\bibfnamefont {C.}~\bibnamefont {Ditsche}},\
  }\bibfield  {title} {\enquote {\bibinfo {title} {{Rescattering effects in
  $\eta\to3\pi$ decays}},}\ }\href {\doibase 10.1007/JHEP02(2011)028}
  {\bibfield  {journal} {\bibinfo  {journal} {JHEP}\ }\textbf {\bibinfo
  {volume} {1102}},\ \bibinfo {pages} {028} (\bibinfo {year} {2011})},\ \Eprint
  {http://arxiv.org/abs/1010.3946} {arXiv:1010.3946 [hep-ph]} \BibitemShut
  {NoStop}%
\bibitem [{\citenamefont {Kubis}(2010)}]{Kubis-talk}%
  \BibitemOpen
  \bibfield  {author} {\bibinfo {author} {\bibfnamefont {B.}~\bibnamefont
  {Kubis}},\ }\href {http://intern.universe-cluster.de/indico/getFile.py/
  access?contribId=22\&sessionId=4\&resId=0\&materialId=slides\&confId=1486}
  {\enquote {\bibinfo {title} {\emph{Rescattering effects in $\eta\to3\pi$
  decays}},}\ }\bibinfo {howpublished} {Talk given at EuroFlavour2010, Munich}
  (\bibinfo {year} {2010})\BibitemShut {NoStop}%
\bibitem [{\citenamefont {Kampf}\ \emph {et~al.}({\natexlab{a}})\citenamefont
  {Kampf}, \citenamefont {Knecht}, \citenamefont {Novotn\'y},\ and\
  \citenamefont {Zdr\'ahal}}]{kknzprep}%
  \BibitemOpen
  \bibfield  {author} {\bibinfo {author} {\bibfnamefont {K.}~\bibnamefont
  {Kampf}}, \bibinfo {author} {\bibfnamefont {M.}~\bibnamefont {Knecht}},
  \bibinfo {author} {\bibfnamefont {J.}~\bibnamefont {Novotn\'y}}, \ and\
  \bibinfo {author} {\bibfnamefont {M.}~\bibnamefont {Zdr\'ahal}},\ }\href@noop
  {} {\enquote {\bibinfo {title} {\emph{Dispersive construction of two-loop
  $P\to3\pi$ amplitudes beyond isospin limit}},}\ }\bibinfo {howpublished} {in
  preparation} \BibitemShut {NoStop}%
\bibitem [{\citenamefont {Zdr\'ahal}\ \emph
  {et~al.}(2009{\natexlab{a}})\citenamefont {Zdr\'ahal}, \citenamefont {Kampf},
  \citenamefont {Knecht},\ and\ \citenamefont {Novotn\'y}}]{nas-proceeding2}%
  \BibitemOpen
  \bibfield  {author} {\bibinfo {author} {\bibfnamefont {M.}~\bibnamefont
  {Zdr\'ahal}}, \bibinfo {author} {\bibfnamefont {K.}~\bibnamefont {Kampf}},
  \bibinfo {author} {\bibfnamefont {M.}~\bibnamefont {Knecht}}, \ and\ \bibinfo
  {author} {\bibfnamefont {J.}~\bibnamefont {Novotn\'y}},\ }\bibfield  {title}
  {\enquote {\bibinfo {title} {{Dispersive construction of two-loop $P\to3\pi$
  $(P=K,\eta)$ amplitudes}},}\ }\href@noop {} {\bibfield  {journal} {\bibinfo
  {journal} {PoS}\ }\textbf {\bibinfo {volume} {EFT09}},\ \bibinfo {pages}
  {063} (\bibinfo {year} {2009}{\natexlab{a}})},\ \Eprint
  {http://arxiv.org/abs/0905.4868} {arXiv:0905.4868 [hep-ph]} \BibitemShut
  {NoStop}%
\bibitem [{\citenamefont {Unverzagt}(2010)}]{Unverzagt:2009vm}%
  \BibitemOpen
  \bibfield  {author} {\bibinfo {author} {\bibfnamefont {M.}~\bibnamefont
  {Unverzagt}} (\bibinfo {collaboration} {Crystal Ball at MAMI}),\ }\bibfield
  {title} {\enquote {\bibinfo {title} {{$\eta$ and $\eta'$ Physics at MAMI}},}\
  }\href {\doibase 10.1016/j.nuclphysbps.2009.12.034} {\bibfield  {journal}
  {\bibinfo  {journal} {Nucl.\ Phys.\ Proc.\ Suppl.}\ }\textbf {\bibinfo
  {volume} {198}},\ \bibinfo {pages} {174--181} (\bibinfo {year} {2010})},\
  \Eprint {http://arxiv.org/abs/0910.1331} {arXiv:0910.1331 [hep-ex]}
  \BibitemShut {NoStop}%
\bibitem [{\citenamefont {Kampf}\ \emph {et~al.}({\natexlab{b}})\citenamefont
  {Kampf}, \citenamefont {Knecht}, \citenamefont {Novotn\'y},\ and\
  \citenamefont {Zdr\'ahal}}]{kknzprep2}%
  \BibitemOpen
  \bibfield  {author} {\bibinfo {author} {\bibfnamefont {K.}~\bibnamefont
  {Kampf}}, \bibinfo {author} {\bibfnamefont {M.}~\bibnamefont {Knecht}},
  \bibinfo {author} {\bibfnamefont {J.}~\bibnamefont {Novotn\'y}}, \ and\
  \bibinfo {author} {\bibfnamefont {M.}~\bibnamefont {Zdr\'ahal}},\ }\href@noop
  {} {\enquote {\bibinfo {title} {\emph{Analytical dispersive construction of
  $\eta\to3\pi$ amplitude: isospin breaking corrections}},}\ }\bibinfo
  {howpublished} {in preparation} \BibitemShut {NoStop}%
\bibitem [{\citenamefont {Kampf}\ and\ \citenamefont
  {Moussallam}(2009)}]{Kampf:2009tk}%
  \BibitemOpen
  \bibfield  {author} {\bibinfo {author} {\bibfnamefont {K.}~\bibnamefont
  {Kampf}}\ and\ \bibinfo {author} {\bibfnamefont {B.}~\bibnamefont
  {Moussallam}},\ }\bibfield  {title} {\enquote {\bibinfo {title} {{Chiral
  expansions of the $\pi^0$ lifetime}},}\ }\href {\doibase
  10.1103/PhysRevD.79.076005} {\bibfield  {journal} {\bibinfo  {journal}
  {Phys.\ Rev.}\ }\textbf {\bibinfo {volume} {D79}},\ \bibinfo {pages} {076005}
  (\bibinfo {year} {2009})},\ \Eprint {http://arxiv.org/abs/0901.4688}
  {arXiv:0901.4688 [hep-ph]} \BibitemShut {NoStop}%
\bibitem [{\citenamefont {Nakamura}\ \emph {et~al.}(2010)\citenamefont
  {Nakamura} \emph {et~al.}}]{PDG}%
  \BibitemOpen
  \bibfield  {author} {\bibinfo {author} {\bibfnamefont {K.}~\bibnamefont
  {Nakamura}} \emph {et~al.} (\bibinfo {collaboration} {Particle Data Group}),\
  }\bibfield  {title} {\enquote {\bibinfo {title} {{Review of particle
  physics}},}\ }\href {\doibase 10.1088/0954-3899/37/7A/075021} {\bibfield
  {journal} {\bibinfo  {journal} {J.\ Phys.}\ }\textbf {\bibinfo {volume}
  {G37}},\ \bibinfo {pages} {075021} (\bibinfo {year} {2010})}\BibitemShut
  {NoStop}%
\bibitem [{\citenamefont {Gormley}\ \emph {et~al.}(1970)\citenamefont {Gormley}
  \emph {et~al.}}]{Gormley:1970qz}%
  \BibitemOpen
  \bibfield  {author} {\bibinfo {author} {\bibfnamefont {M.}~\bibnamefont
  {Gormley}} \emph {et~al.},\ }\bibfield  {title} {\enquote {\bibinfo {title}
  {{Experimental determination of the dalitz-plot distribution of the decays
  $\eta\to\pi^+\pi^-\pi^0$ and $\eta\to\pi^+\pi^-\gamma$, and the branching
  ratio $\eta\to \pi^+\pi^-\gamma / \eta\to\pi^+\pi^-\pi^0$}},}\ }\href
  {\doibase 10.1103/PhysRevD.2.501} {\bibfield  {journal} {\bibinfo  {journal}
  {Phys.\ Rev.}\ }\textbf {\bibinfo {volume} {D2}},\ \bibinfo {pages}
  {501--505} (\bibinfo {year} {1970})}\BibitemShut {NoStop}%
%%CITATION = PHRVA,D2,501;%%
\bibitem [{\citenamefont {Layter}\ \emph {et~al.}(1973)\citenamefont {Layter}
  \emph {et~al.}}]{Layter:1973ti}%
  \BibitemOpen
  \bibfield  {author} {\bibinfo {author} {\bibfnamefont {J.~G.}\ \bibnamefont
  {Layter}} \emph {et~al.},\ }\bibfield  {title} {\enquote {\bibinfo {title}
  {{Study of dalitz-plot distributions of the decays $\eta\to\pi^+\pi^-\pi^0$
  and $\eta\to\pi^+\pi^-\gamma$}},}\ }\href {\doibase 10.1103/PhysRevD.7.2565}
  {\bibfield  {journal} {\bibinfo  {journal} {Phys.\ Rev.}\ }\textbf {\bibinfo
  {volume} {D7}},\ \bibinfo {pages} {2565--2568} (\bibinfo {year}
  {1973})}\BibitemShut {NoStop}%
%%CITATION = PHRVA,D7,2565;%%
\bibitem [{\citenamefont {Abele}\ \emph
  {et~al.}(1998{\natexlab{a}})\citenamefont {Abele} \emph
  {et~al.}}]{CrystalBarrel-etac}%
  \BibitemOpen
  \bibfield  {author} {\bibinfo {author} {\bibfnamefont {A.}~\bibnamefont
  {Abele}} \emph {et~al.} (\bibinfo {collaboration} {Crystal Barrel}),\
  }\bibfield  {title} {\enquote {\bibinfo {title} {{Momentum dependence of the
  decay $\eta\to \pi^+ \pi^- \pi^0$}},}\ }\href {\doibase
  10.1016/S0370-2693(97)01376-2} {\bibfield  {journal} {\bibinfo  {journal}
  {Phys.\ Lett.}\ }\textbf {\bibinfo {volume} {B417}},\ \bibinfo {pages}
  {197--201} (\bibinfo {year} {1998}{\natexlab{a}})}\BibitemShut {NoStop}%
%%CITATION = PHLTA,B417,197;%%
\bibitem [{\citenamefont {Abele}\ \emph
  {et~al.}(1998{\natexlab{b}})\citenamefont {Abele} \emph
  {et~al.}}]{CrystalBarrel-eta0}%
  \BibitemOpen
  \bibfield  {author} {\bibinfo {author} {\bibfnamefont {A.}~\bibnamefont
  {Abele}} \emph {et~al.} (\bibinfo {collaboration} {Crystal Barrel}),\
  }\bibfield  {title} {\enquote {\bibinfo {title} {{Decay dynamics of the
  process $\eta\to 3\pi^0$}},}\ }\href {\doibase 10.1016/S0370-2693(97)01377-4}
  {\bibfield  {journal} {\bibinfo  {journal} {Phys.\ Lett.}\ }\textbf {\bibinfo
  {volume} {B417}},\ \bibinfo {pages} {193--196} (\bibinfo {year}
  {1998}{\natexlab{b}})}\BibitemShut {NoStop}%
%%CITATION = PHLTA,B417,193;%%
\bibitem [{\citenamefont {Tippens}\ \emph {et~al.}(2001)\citenamefont {Tippens}
  \emph {et~al.}}]{CrystalBall}%
  \BibitemOpen
  \bibfield  {author} {\bibinfo {author} {\bibfnamefont {W.~B.}\ \bibnamefont
  {Tippens}} \emph {et~al.} (\bibinfo {collaboration} {Crystal Ball}),\
  }\bibfield  {title} {\enquote {\bibinfo {title} {{Determination of the
  quadratic slope parameter in $\eta\to 3\pi^0$ decay}},}\ }\href {\doibase
  10.1103/PhysRevLett.87.192001} {\bibfield  {journal} {\bibinfo  {journal}
  {Phys.\ Rev.\ Lett.}\ }\textbf {\bibinfo {volume} {87}},\ \bibinfo {pages}
  {192001} (\bibinfo {year} {2001})}\BibitemShut {NoStop}%
%%CITATION = PRLTA,87,192001;%%
\bibitem [{\citenamefont {Bashkanov}\ \emph {et~al.}(2007)\citenamefont
  {Bashkanov} \emph {et~al.}}]{Celsius}%
  \BibitemOpen
  \bibfield  {author} {\bibinfo {author} {\bibfnamefont {M.}~\bibnamefont
  {Bashkanov}} \emph {et~al.},\ }\bibfield  {title} {\enquote {\bibinfo {title}
  {{Measurement of the Slope Parameter for the $\eta\to 3\pi^0$ Decay in the
  $pp\to pp \eta$ Reaction}},}\ }\href {\doibase 10.1103/PhysRevC.76.048201}
  {\bibfield  {journal} {\bibinfo  {journal} {Phys.\ Rev.}\ }\textbf {\bibinfo
  {volume} {C76}},\ \bibinfo {pages} {048201} (\bibinfo {year} {2007})},\
  \Eprint {http://arxiv.org/abs/0708.2014} {arXiv:0708.2014 [nucl-ex]}
  \BibitemShut {NoStop}%
%%CITATION = 0708.2014;%%
\bibitem [{\citenamefont {Adolph}\ \emph {et~al.}(2009)\citenamefont {Adolph}
  \emph {et~al.}}]{WASAatCOSY}%
  \BibitemOpen
  \bibfield  {author} {\bibinfo {author} {\bibfnamefont {C.}~\bibnamefont
  {Adolph}} \emph {et~al.} (\bibinfo {collaboration} {WASA-at-COSY}),\
  }\bibfield  {title} {\enquote {\bibinfo {title} {{Measurement of the
  $\eta\to3\pi^0$ Dalitz Plot Distribution with the WASA Detector at COSY}},}\
  }\href {\doibase 10.1016/j.physletb.2009.03.063} {\bibfield  {journal}
  {\bibinfo  {journal} {Phys.\ Lett.}\ }\textbf {\bibinfo {volume} {B677}},\
  \bibinfo {pages} {24--29} (\bibinfo {year} {2009})},\ \Eprint
  {http://arxiv.org/abs/0811.2763} {arXiv:0811.2763 [nucl-ex]} \BibitemShut
  {NoStop}%
%%CITATION = 0811.2763;%%
\bibitem [{\citenamefont {Unverzagt}\ \emph {et~al.}(2009)\citenamefont
  {Unverzagt} \emph {et~al.}}]{MAMI-B}%
  \BibitemOpen
  \bibfield  {author} {\bibinfo {author} {\bibfnamefont {M.}~\bibnamefont
  {Unverzagt}} \emph {et~al.} (\bibinfo {collaboration} {Crystal Ball at
  MAMI}),\ }\bibfield  {title} {\enquote {\bibinfo {title} {{Determination of
  the Dalitz plot parameter alpha for the decay $\eta\to 3\pi^0$ with the
  Crystal Ball at MAMI-B}},}\ }\href {\doibase 10.1140/epja/i2008-10710-7}
  {\bibfield  {journal} {\bibinfo  {journal} {Eur.\ Phys.\ J.}\ }\textbf
  {\bibinfo {volume} {A39}},\ \bibinfo {pages} {169--177} (\bibinfo {year}
  {2009})},\ \Eprint {http://arxiv.org/abs/0812.3324} {arXiv:0812.3324
  [hep-ex]} \BibitemShut {NoStop}%
%%CITATION = 0812.3324;%%
\bibitem [{\citenamefont {Ambrosino}\ \emph {et~al.}(2010)\citenamefont
  {Ambrosino} \emph {et~al.}}]{KLOE-eta0}%
  \BibitemOpen
  \bibfield  {author} {\bibinfo {author} {\bibfnamefont {F.}~\bibnamefont
  {Ambrosino}} \emph {et~al.} (\bibinfo {collaboration} {KLOE}),\ }\bibfield
  {title} {\enquote {\bibinfo {title} {{Measurement of the $\eta\to 3\pi^{0}$
  slope parameter $\alpha$ with the KLOE detector}},}\ }\href {\doibase
  10.1016/j.physletb.2010.09.022} {\bibfield  {journal} {\bibinfo  {journal}
  {Phys.\ Lett.}\ }\textbf {\bibinfo {volume} {B694}},\ \bibinfo {pages}
  {16--21} (\bibinfo {year} {2010})},\ \Eprint {http://arxiv.org/abs/1004.1319}
  {arXiv:1004.1319 [hep-ex]} \BibitemShut {NoStop}%
%%CITATION = 1004.1319;%%
\bibitem [{\citenamefont {Adler}(1965{\natexlab{a}})}]{Adler:1965ab}%
  \BibitemOpen
  \bibfield  {author} {\bibinfo {author} {\bibfnamefont {S.~L.}\ \bibnamefont
  {Adler}},\ }\bibfield  {title} {\enquote {\bibinfo {title} {{Consistency
  Conditions on the Strong Interactions Implied by a Partially Conserved
  Axial-Vector Current}},}\ }\href {\doibase 10.1103/PhysRev.137.B1022}
  {\bibfield  {journal} {\bibinfo  {journal} {Phys.\ Rev.}\ }\textbf {\bibinfo
  {volume} {137}},\ \bibinfo {pages} {B1022--B1033} (\bibinfo {year}
  {1965}{\natexlab{a}})}\BibitemShut {NoStop}%
\bibitem [{\citenamefont {Adler}(1965{\natexlab{b}})}]{Adler:1965ga}%
  \BibitemOpen
  \bibfield  {author} {\bibinfo {author} {\bibfnamefont {S.~L.}\ \bibnamefont
  {Adler}},\ }\bibfield  {title} {\enquote {\bibinfo {title} {{Consistency
  Conditions on the Strong Interactions Implied by a Partially Conserved
  Axial-vector Current. II}},}\ }\href {\doibase 10.1103/PhysRev.139.B1638}
  {\bibfield  {journal} {\bibinfo  {journal} {Phys.\ Rev.}\ }\textbf {\bibinfo
  {volume} {139}},\ \bibinfo {pages} {B1638--B1643} (\bibinfo {year}
  {1965}{\natexlab{b}})}\BibitemShut {NoStop}%
%%CITATION = PHRVA,139,B1638;%%
\bibitem [{\citenamefont {Cabibbo}(2004)}]{Cabibbo:2004gq}%
  \BibitemOpen
  \bibfield  {author} {\bibinfo {author} {\bibfnamefont {N.}~\bibnamefont
  {Cabibbo}},\ }\bibfield  {title} {\enquote {\bibinfo {title} {{Determination
  of the $a_0 - a_2$ pion scattering length from $K^+\to \pi^+ \pi^0 \pi^0$
  decay}},}\ }\href {\doibase 10.1103/PhysRevLett.93.121801} {\bibfield
  {journal} {\bibinfo  {journal} {Phys.\ Rev.\ Lett.}\ }\textbf {\bibinfo
  {volume} {93}},\ \bibinfo {pages} {121801} (\bibinfo {year} {2004})},\
  \Eprint {http://arxiv.org/abs/hep-ph/0405001} {arXiv:hep-ph/0405001}
  \BibitemShut {NoStop}%
\bibitem [{\citenamefont {Batley}\ \emph {et~al.}(2006)\citenamefont {Batley}
  \emph {et~al.}}]{Batley:2005ax}%
  \BibitemOpen
  \bibfield  {author} {\bibinfo {author} {\bibfnamefont {J.~R.}\ \bibnamefont
  {Batley}} \emph {et~al.} (\bibinfo {collaboration} {NA48/2}),\ }\bibfield
  {title} {\enquote {\bibinfo {title} {{Observation of a cusp-like structure in
  the $\pi^0\pi^0$ invariant mass distribution from $K^\pm\to \pi^\pm \pi^0
  \pi^0$ decay and determination of the $\pi\pi$ scattering lengths}},}\ }\href
  {\doibase 10.1016/j.physletb.2005.11.087} {\bibfield  {journal} {\bibinfo
  {journal} {Phys.\ Lett.}\ }\textbf {\bibinfo {volume} {B633}},\ \bibinfo
  {pages} {173--182} (\bibinfo {year} {2006})},\ \Eprint
  {http://arxiv.org/abs/hep-ex/0511056} {arXiv:hep-ex/0511056} \BibitemShut
  {NoStop}%
\bibitem [{\citenamefont {Abouzaid}\ \emph {et~al.}(2008)\citenamefont
  {Abouzaid} \emph {et~al.}}]{Abouzaid:2008js}%
  \BibitemOpen
  \bibfield  {author} {\bibinfo {author} {\bibfnamefont {E.}~\bibnamefont
  {Abouzaid}} \emph {et~al.} (\bibinfo {collaboration} {KTeV}),\ }\bibfield
  {title} {\enquote {\bibinfo {title} {{Detailed Study of the $K(L)\to \pi^0
  \pi^0 \pi^0$ Dalitz Plot}},}\ }\href {\doibase 10.1103/PhysRevD.78.032009}
  {\bibfield  {journal} {\bibinfo  {journal} {Phys.\ Rev.}\ }\textbf {\bibinfo
  {volume} {D78}},\ \bibinfo {pages} {032009} (\bibinfo {year} {2008})},\
  \Eprint {http://arxiv.org/abs/0806.3535} {arXiv:0806.3535 [hep-ex]}
  \BibitemShut {NoStop}%
\bibitem [{\citenamefont {Cabibbo}\ and\ \citenamefont {Isidori}(2005)}]{CI}%
  \BibitemOpen
  \bibfield  {author} {\bibinfo {author} {\bibfnamefont {N.}~\bibnamefont
  {Cabibbo}}\ and\ \bibinfo {author} {\bibfnamefont {G.}~\bibnamefont
  {Isidori}},\ }\bibfield  {title} {\enquote {\bibinfo {title} {{Pion pion
  scattering and the $K\to3\pi$ decay amplitudes}},}\ }\href@noop {} {\bibfield
   {journal} {\bibinfo  {journal} {JHEP}\ }\textbf {\bibinfo {volume} {0503}},\
  \bibinfo {pages} {021} (\bibinfo {year} {2005})},\ \Eprint
  {http://arxiv.org/abs/hep-ph/0502130} {arXiv:hep-ph/0502130} \BibitemShut
  {NoStop}%
\bibitem [{\citenamefont {Di~Lella}(2007)}]{DiLella-talk}%
  \BibitemOpen
  \bibfield  {author} {\bibinfo {author} {\bibfnamefont {L.}~\bibnamefont
  {Di~Lella}},\ }\href
  {http://www.cern.ch/NA48/Welcome/images/talks/KAON07/luigi.dilella.pdf}
  {\enquote {\bibinfo {title} {\emph{Review of $\pi\pi$ scattering measurements
  in $K$ decays}},}\ }\bibinfo {howpublished} {Talk given at Kaon2007,
  Frascati} (\bibinfo {year} {2007})\BibitemShut {NoStop}%
\bibitem [{\citenamefont {Kampf}\ \emph {et~al.}(2009)\citenamefont {Kampf},
  \citenamefont {Knecht}, \citenamefont {Novotn\'y},\ and\ \citenamefont
  {Zdr\'ahal}}]{nas-proceeding1}%
  \BibitemOpen
  \bibfield  {author} {\bibinfo {author} {\bibfnamefont {K.}~\bibnamefont
  {Kampf}}, \bibinfo {author} {\bibfnamefont {M.}~\bibnamefont {Knecht}},
  \bibinfo {author} {\bibfnamefont {J.}~\bibnamefont {Novotn\'y}}, \ and\
  \bibinfo {author} {\bibfnamefont {M.}~\bibnamefont {Zdr\'ahal}},\ }\bibfield
  {title} {\enquote {\bibinfo {title} {{Dispersive representation of $K\to3\pi$
  amplitudes and cusps}},}\ }\href {\doibase 10.1016/j.nuclphysbps.2008.12.075}
  {\bibfield  {journal} {\bibinfo  {journal} {Nucl.\ Phys.\ Proc.\ Suppl.}\
  }\textbf {\bibinfo {volume} {186}},\ \bibinfo {pages} {334--337} (\bibinfo
  {year} {2009})},\ \Eprint {http://arxiv.org/abs/0810.1906} {arXiv:0810.1906
  [hep-ph]} \BibitemShut {NoStop}%
\bibitem [{\citenamefont {Bijnens}(2007)}]{Bijnens:2006zp}%
  \BibitemOpen
  \bibfield  {author} {\bibinfo {author} {\bibfnamefont {J.}~\bibnamefont
  {Bijnens}},\ }\bibfield  {title} {\enquote {\bibinfo {title} {{Chiral
  perturbation theory beyond one loop}},}\ }\href {\doibase
  10.1016/j.ppnp.2006.08.002} {\bibfield  {journal} {\bibinfo  {journal}
  {Prog.\ Part.\ Nucl.\ Phys.}\ }\textbf {\bibinfo {volume} {58}},\ \bibinfo
  {pages} {521--586} (\bibinfo {year} {2007})},\ \Eprint
  {http://arxiv.org/abs/hep-ph/0604043} {arXiv:hep-ph/0604043} \BibitemShut
  {NoStop}%
%%CITATION = HEP-PH/0604043;%%
\bibitem [{\citenamefont {Bijnens}\ and\ \citenamefont
  {Jemos}(2009{\natexlab{a}})}]{Bijnens:2009zd}%
  \BibitemOpen
  \bibfield  {author} {\bibinfo {author} {\bibfnamefont {J.}~\bibnamefont
  {Bijnens}}\ and\ \bibinfo {author} {\bibfnamefont {I.}~\bibnamefont
  {Jemos}},\ }\bibfield  {title} {\enquote {\bibinfo {title} {{Relations at
  Order $p^6$ in Chiral Perturbation Theory}},}\ }\href {\doibase
  10.1140/epjc/s10052-009-1145-5} {\bibfield  {journal} {\bibinfo  {journal}
  {Eur.\ Phys.\ J.}\ }\textbf {\bibinfo {volume} {C64}},\ \bibinfo {pages}
  {273--282} (\bibinfo {year} {2009}{\natexlab{a}})},\ \Eprint
  {http://arxiv.org/abs/0906.3118} {arXiv:0906.3118 [hep-ph]} \BibitemShut
  {NoStop}%
\bibitem [{\citenamefont {Cirigliano}\ \emph {et~al.}(2003)\citenamefont
  {Cirigliano}, \citenamefont {Ecker}, \citenamefont {Neufeld},\ and\
  \citenamefont {Pich}}]{Cirigliano:2003yq}%
  \BibitemOpen
  \bibfield  {author} {\bibinfo {author} {\bibfnamefont {V.}~\bibnamefont
  {Cirigliano}}, \bibinfo {author} {\bibfnamefont {G.}~\bibnamefont {Ecker}},
  \bibinfo {author} {\bibfnamefont {H.}~\bibnamefont {Neufeld}}, \ and\
  \bibinfo {author} {\bibfnamefont {A.}~\bibnamefont {Pich}},\ }\bibfield
  {title} {\enquote {\bibinfo {title} {{Meson resonances, large N(c) and chiral
  symmetry}},}\ }\href@noop {} {\bibfield  {journal} {\bibinfo  {journal}
  {JHEP}\ }\textbf {\bibinfo {volume} {0306}},\ \bibinfo {pages} {012}
  (\bibinfo {year} {2003})},\ \Eprint {http://arxiv.org/abs/hep-ph/0305311}
  {arXiv:hep-ph/0305311} \BibitemShut {NoStop}%
\bibitem [{\citenamefont {Kampf}\ and\ \citenamefont
  {Moussallam}(2006)}]{Kampf:2006bn}%
  \BibitemOpen
  \bibfield  {author} {\bibinfo {author} {\bibfnamefont {K.}~\bibnamefont
  {Kampf}}\ and\ \bibinfo {author} {\bibfnamefont {B.}~\bibnamefont
  {Moussallam}},\ }\bibfield  {title} {\enquote {\bibinfo {title} {{Tests of
  the naturalness of the coupling constants in ChPT at order $p^6$}},}\ }\href
  {\doibase 10.1140/epjc/s2006-02606-7} {\bibfield  {journal} {\bibinfo
  {journal} {Eur.\ Phys.\ J.}\ }\textbf {\bibinfo {volume} {C47}},\ \bibinfo
  {pages} {723--736} (\bibinfo {year} {2006})},\ \Eprint
  {http://arxiv.org/abs/hep-ph/0604125} {arXiv:hep-ph/0604125} \BibitemShut
  {NoStop}%
\bibitem [{\citenamefont {Amoros}\ \emph {et~al.}(2001)\citenamefont {Amoros},
  \citenamefont {Bijnens},\ and\ \citenamefont {Talavera}}]{Amoros:2001cp}%
  \BibitemOpen
  \bibfield  {author} {\bibinfo {author} {\bibfnamefont {G.}~\bibnamefont
  {Amoros}}, \bibinfo {author} {\bibfnamefont {J.}~\bibnamefont {Bijnens}}, \
  and\ \bibinfo {author} {\bibfnamefont {P.}~\bibnamefont {Talavera}},\
  }\bibfield  {title} {\enquote {\bibinfo {title} {{QCD isospin breaking in
  meson masses, decay constants and quark mass ratios}},}\ }\href {\doibase
  10.1016/S0550-3213(01)00121-3} {\bibfield  {journal} {\bibinfo  {journal}
  {Nucl.\ Phys.}\ }\textbf {\bibinfo {volume} {B602}},\ \bibinfo {pages}
  {87--108} (\bibinfo {year} {2001})},\ \Eprint
  {http://arxiv.org/abs/hep-ph/0101127} {arXiv:hep-ph/0101127} \BibitemShut
  {NoStop}%
\bibitem [{\citenamefont {Bijnens}\ and\ \citenamefont
  {Jemos}(2009{\natexlab{b}})}]{Bijnens:2009hy}%
  \BibitemOpen
  \bibfield  {author} {\bibinfo {author} {\bibfnamefont {J.}~\bibnamefont
  {Bijnens}}\ and\ \bibinfo {author} {\bibfnamefont {I.}~\bibnamefont
  {Jemos}},\ }\bibfield  {title} {\enquote {\bibinfo {title} {{Determination of
  Low Energy Constants and testing Chiral Perturbation Theory at order $p^6$
  (NNLO)}},}\ }\href@noop {} {\bibfield  {journal} {\bibinfo  {journal} {PoS}\
  }\textbf {\bibinfo {volume} {CD09}},\ \bibinfo {pages} {087} (\bibinfo {year}
  {2009}{\natexlab{b}})},\ \Eprint {http://arxiv.org/abs/0909.4477}
  {arXiv:0909.4477 [hep-ph]} \BibitemShut {NoStop}%
\bibitem [{\citenamefont {Bijnens}\ and\ \citenamefont
  {Jemos}(2012)}]{Bijnens:2011tb}%
  \BibitemOpen
  \bibfield  {author} {\bibinfo {author} {\bibfnamefont {J.}~\bibnamefont
  {Bijnens}}\ and\ \bibinfo {author} {\bibfnamefont {I.}~\bibnamefont
  {Jemos}},\ }\bibfield  {title} {\enquote {\bibinfo {title} {{A new global fit
  of the $L^r_i$ at next-to-next-to-leading order in Chiral Perturbation
  Theory}},}\ }\href {\doibase 10.1016/j.nuclphysb.2011.09.013} {\bibfield
  {journal} {\bibinfo  {journal} {Nucl.\ Phys.}\ }\textbf {\bibinfo {volume}
  {B854}},\ \bibinfo {pages} {631--665} (\bibinfo {year} {2012})},\ \Eprint
  {http://arxiv.org/abs/1103.5945} {arXiv:1103.5945 [hep-ph]} \BibitemShut
  {NoStop}%
\bibitem [{\citenamefont {Stern}\ \emph {et~al.}(1993)\citenamefont {Stern},
  \citenamefont {Sazdjian},\ and\ \citenamefont {Fuchs}}]{Stern:1993rg}%
  \BibitemOpen
  \bibfield  {author} {\bibinfo {author} {\bibfnamefont {J.}~\bibnamefont
  {Stern}}, \bibinfo {author} {\bibfnamefont {H.}~\bibnamefont {Sazdjian}}, \
  and\ \bibinfo {author} {\bibfnamefont {N.~H.}\ \bibnamefont {Fuchs}},\
  }\bibfield  {title} {\enquote {\bibinfo {title} {{What $\pi$ - $\pi$
  scattering tells us about chiral perturbation theory}},}\ }\href {\doibase
  10.1103/PhysRevD.47.3814} {\bibfield  {journal} {\bibinfo  {journal} {Phys.\
  Rev.}\ }\textbf {\bibinfo {volume} {D47}},\ \bibinfo {pages} {3814--3838}
  (\bibinfo {year} {1993})},\ \Eprint {http://arxiv.org/abs/hep-ph/9301244}
  {arXiv:hep-ph/9301244} \BibitemShut {NoStop}%
%%CITATION = HEP-PH/9301244;%%
\bibitem [{\citenamefont {Knecht}\ \emph {et~al.}(1995)\citenamefont {Knecht},
  \citenamefont {Moussallam}, \citenamefont {Stern},\ and\ \citenamefont
  {Fuchs}}]{Knecht:1995tr}%
  \BibitemOpen
  \bibfield  {author} {\bibinfo {author} {\bibfnamefont {M.}~\bibnamefont
  {Knecht}}, \bibinfo {author} {\bibfnamefont {B.}~\bibnamefont {Moussallam}},
  \bibinfo {author} {\bibfnamefont {J.}~\bibnamefont {Stern}}, \ and\ \bibinfo
  {author} {\bibfnamefont {N.~H.}\ \bibnamefont {Fuchs}},\ }\bibfield  {title}
  {\enquote {\bibinfo {title} {{The Low-energy $\pi\pi$ amplitude to one and
  two loops}},}\ }\href {\doibase 10.1016/0550-3213(95)00515-3} {\bibfield
  {journal} {\bibinfo  {journal} {Nucl.\ Phys.}\ }\textbf {\bibinfo {volume}
  {B457}},\ \bibinfo {pages} {513--576} (\bibinfo {year} {1995})},\ \Eprint
  {http://arxiv.org/abs/hep-ph/9507319} {arXiv:hep-ph/9507319} \BibitemShut
  {NoStop}%
%%CITATION = HEP-PH/9507319;%%
\bibitem [{\citenamefont {Zdr\'ahal}\ and\ \citenamefont
  {Novotn\'y}(2008)}]{Zdrahal:2008bd}%
  \BibitemOpen
  \bibfield  {author} {\bibinfo {author} {\bibfnamefont {M.}~\bibnamefont
  {Zdr\'ahal}}\ and\ \bibinfo {author} {\bibfnamefont {J.}~\bibnamefont
  {Novotn\'y}},\ }\bibfield  {title} {\enquote {\bibinfo {title} {{Dispersive
  Approach to Chiral Perturbation Theory}},}\ }\href {\doibase
  10.1103/PhysRevD.78.116016} {\bibfield  {journal} {\bibinfo  {journal}
  {Phys.\ Rev.}\ }\textbf {\bibinfo {volume} {D78}},\ \bibinfo {pages} {116016}
  (\bibinfo {year} {2008})},\ \Eprint {http://arxiv.org/abs/0806.4529}
  {arXiv:0806.4529 [hep-ph]} \BibitemShut {NoStop}%
\bibitem [{\citenamefont {Bronzan}\ and\ \citenamefont
  {Kacser}(1963)}]{BronzanKacser}%
  \BibitemOpen
  \bibfield  {author} {\bibinfo {author} {\bibfnamefont {J.~B.}\ \bibnamefont
  {Bronzan}}\ and\ \bibinfo {author} {\bibfnamefont {C.}~\bibnamefont
  {Kacser}},\ }\bibfield  {title} {\enquote {\bibinfo {title} {{The
  Khuri-Treiman Representation and Perturbation Theory}},}\ }\href@noop {}
  {\bibfield  {journal} {\bibinfo  {journal} {Phys.\ Rev.}\ }\textbf {\bibinfo
  {volume} {132}},\ \bibinfo {pages} {2703} (\bibinfo {year}
  {1963})}\BibitemShut {NoStop}%
\bibitem [{\citenamefont {Kacser}(1963)}]{Kacser}%
  \BibitemOpen
  \bibfield  {author} {\bibinfo {author} {\bibfnamefont {C.}~\bibnamefont
  {Kacser}},\ }\bibfield  {title} {\enquote {\bibinfo {title} {{Analytic
  Structure of Partial-wave Amplitudes for Production and Decay Processes}},}\
  }\href@noop {} {\bibfield  {journal} {\bibinfo  {journal} {Phys.\ Rev.}\
  }\textbf {\bibinfo {volume} {132}},\ \bibinfo {pages} {2712--2721} (\bibinfo
  {year} {1963})}\BibitemShut {NoStop}%
\bibitem [{\citenamefont {Chew}\ and\ \citenamefont
  {Mandelstam}(1960)}]{Chew:1960iv}%
  \BibitemOpen
  \bibfield  {author} {\bibinfo {author} {\bibfnamefont {G.~F.}\ \bibnamefont
  {Chew}}\ and\ \bibinfo {author} {\bibfnamefont {S.}~\bibnamefont
  {Mandelstam}},\ }\bibfield  {title} {\enquote {\bibinfo {title} {{Theory of
  low-energy pion pion interactions}},}\ }\href {\doibase
  10.1103/PhysRev.119.467} {\bibfield  {journal} {\bibinfo  {journal} {Phys.\
  Rev.}\ }\textbf {\bibinfo {volume} {119}},\ \bibinfo {pages} {467--477}
  (\bibinfo {year} {1960})}\BibitemShut {NoStop}%
\bibitem [{\citenamefont {Zdr\'ahal}\ \emph
  {et~al.}(2009{\natexlab{b}})\citenamefont {Zdr\'ahal}, \citenamefont {Kampf},
  \citenamefont {Knecht},\ and\ \citenamefont {Novotn\'y}}]{nas-proceeding3}%
  \BibitemOpen
  \bibfield  {author} {\bibinfo {author} {\bibfnamefont {M.}~\bibnamefont
  {Zdr\'ahal}}, \bibinfo {author} {\bibfnamefont {K.}~\bibnamefont {Kampf}},
  \bibinfo {author} {\bibfnamefont {M.}~\bibnamefont {Knecht}}, \ and\ \bibinfo
  {author} {\bibfnamefont {J.}~\bibnamefont {Novotn\'y}},\ }\bibfield  {title}
  {\enquote {\bibinfo {title} {{Construction of the $\eta \to3\pi$ (and $K
  \to3\pi$) amplitudes using dispersive approach}},}\ }\href@noop {} {\bibfield
   {journal} {\bibinfo  {journal} {PoS}\ }\textbf {\bibinfo {volume} {CD09}},\
  \bibinfo {pages} {122} (\bibinfo {year} {2009}{\natexlab{b}})},\ \Eprint
  {http://arxiv.org/abs/0910.1721} {arXiv:0910.1721 [hep-ph]} \BibitemShut
  {NoStop}%
\bibitem [{\citenamefont {Descotes-Genon}\ \emph {et~al.}(2002)\citenamefont
  {Descotes-Genon}, \citenamefont {Fuchs}, \citenamefont {Girlanda},\ and\
  \citenamefont {Stern}}]{DescotesGenon:2001tn}%
  \BibitemOpen
  \bibfield  {author} {\bibinfo {author} {\bibfnamefont {S.}~\bibnamefont
  {Descotes-Genon}}, \bibinfo {author} {\bibfnamefont {N.H.}\ \bibnamefont
  {Fuchs}}, \bibinfo {author} {\bibfnamefont {L.}~\bibnamefont {Girlanda}}, \
  and\ \bibinfo {author} {\bibfnamefont {J.}~\bibnamefont {Stern}},\ }\bibfield
   {title} {\enquote {\bibinfo {title} {{Analysis and interpretation of new
  low-energy $\pi\pi$ scattering data}},}\ }\href {\doibase
  10.1007/s10052-002-0965-3} {\bibfield  {journal} {\bibinfo  {journal} {Eur.\
  Phys.\ J.}\ }\textbf {\bibinfo {volume} {C24}},\ \bibinfo {pages} {469--483}
  (\bibinfo {year} {2002})},\ \Eprint {http://arxiv.org/abs/hep-ph/0112088}
  {arXiv:hep-ph/0112088} \BibitemShut {NoStop}%
\bibitem [{\citenamefont {Kupsc}\ \emph {et~al.}(2009)\citenamefont {Kupsc},
  \citenamefont {Rusetsky},\ and\ \citenamefont {Gullstrom}}]{Kupsc:2009zza}%
  \BibitemOpen
  \bibfield  {author} {\bibinfo {author} {\bibfnamefont {A.}~\bibnamefont
  {Kupsc}}, \bibinfo {author} {\bibfnamefont {A.}~\bibnamefont {Rusetsky}}, \
  and\ \bibinfo {author} {\bibfnamefont {C.-O.}\ \bibnamefont {Gullstrom}},\
  }\bibfield  {title} {\enquote {\bibinfo {title} {{A step towards systematic
  studies of the cusp in $\eta\to3\pi^0$ decay}},}\ }\href@noop {} {\bibfield
  {journal} {\bibinfo  {journal} {Acta Phys.\ Polon.\ Supp.}\ }\textbf
  {\bibinfo {volume} {2}},\ \bibinfo {pages} {169--176} (\bibinfo {year}
  {2009})}\BibitemShut {NoStop}%
\bibitem [{\citenamefont {Achasov}\ \emph {et~al.}(2006)\citenamefont
  {Achasov}, \citenamefont {Aulchenko}, \citenamefont {Beloborodov} \emph
  {et~al.}}]{Achasov:2006dv}%
  \BibitemOpen
  \bibfield  {author} {\bibinfo {author} {\bibfnamefont {M.~N.}\ \bibnamefont
  {Achasov}}, \bibinfo {author} {\bibfnamefont {V.~M.}\ \bibnamefont
  {Aulchenko}}, \bibinfo {author} {\bibfnamefont {K.~I.}\ \bibnamefont
  {Beloborodov}},  \emph {et~al.},\ }\bibfield  {title} {\enquote {\bibinfo
  {title} {{Study of the $e^+e^- \to \eta \gamma$ process with SND detector at
  the VEPP-2M $e^+e^-$ collider}},}\ }\href {\doibase
  10.1103/PhysRevD.74.014016} {\bibfield  {journal} {\bibinfo  {journal}
  {Phys.\ Rev.}\ }\textbf {\bibinfo {volume} {D74}},\ \bibinfo {pages} {014016}
  (\bibinfo {year} {2006})},\ \Eprint {http://arxiv.org/abs/hep-ex/0605109}
  {arXiv:hep-ex/0605109} \BibitemShut {NoStop}%
\bibitem [{\citenamefont {Kaplan}\ and\ \citenamefont
  {Manohar}(1986)}]{Kaplan:1986ru}%
  \BibitemOpen
  \bibfield  {author} {\bibinfo {author} {\bibfnamefont {D.~B.}\ \bibnamefont
  {Kaplan}}\ and\ \bibinfo {author} {\bibfnamefont {A.~V.}\ \bibnamefont
  {Manohar}},\ }\bibfield  {title} {\enquote {\bibinfo {title} {{Current Mass
  Ratios of the Light Quarks}},}\ }\href {\doibase 10.1103/PhysRevLett.56.2004}
  {\bibfield  {journal} {\bibinfo  {journal} {Phys.\ Rev.\ Lett.}\ }\textbf
  {\bibinfo {volume} {56}},\ \bibinfo {pages} {2004} (\bibinfo {year}
  {1986})}\BibitemShut {NoStop}%
\bibitem [{\citenamefont {Oller}\ and\ \citenamefont
  {Roca}(2007)}]{Oller:2006xb}%
  \BibitemOpen
  \bibfield  {author} {\bibinfo {author} {\bibfnamefont {J.~A.}\ \bibnamefont
  {Oller}}\ and\ \bibinfo {author} {\bibfnamefont {L.}~\bibnamefont {Roca}},\
  }\bibfield  {title} {\enquote {\bibinfo {title} {{Non-Perturbative Study of
  the Light Pseudoscalar Masses in Chiral Dynamics}},}\ }\href {\doibase
  10.1140/epja/i2006-10516-7} {\bibfield  {journal} {\bibinfo  {journal} {Eur.\
  Phys.\ J.}\ }\textbf {\bibinfo {volume} {A34}},\ \bibinfo {pages} {371--386}
  (\bibinfo {year} {2007})},\ \Eprint {http://arxiv.org/abs/hep-ph/0608290}
  {arXiv:hep-ph/0608290} \BibitemShut {NoStop}%
\bibitem [{\citenamefont {Leutwyler}(2009)}]{Leutwyler:2009jg}%
  \BibitemOpen
  \bibfield  {author} {\bibinfo {author} {\bibfnamefont {H.}~\bibnamefont
  {Leutwyler}},\ }\bibfield  {title} {\enquote {\bibinfo {title} {{Light quark
  masses}},}\ }\href@noop {} {\bibfield  {journal} {\bibinfo  {journal} {PoS}\
  }\textbf {\bibinfo {volume} {CD09}},\ \bibinfo {pages} {005} (\bibinfo {year}
  {2009})},\ \Eprint {http://arxiv.org/abs/0911.1416} {arXiv:0911.1416
  [hep-ph]} \BibitemShut {NoStop}%
\bibitem [{\citenamefont {Zdr\'ahal}(2011)}]{Zdrahal:2011ta}%
  \BibitemOpen
  \bibfield  {author} {\bibinfo {author} {\bibfnamefont {M.}~\bibnamefont
  {Zdr\'ahal}},\ }\bibfield  {title} {\enquote {\bibinfo {title}
  {{Determination of the $m_u$ and $m_d$ quark masses from $\eta \to 3\pi$
  decay}},}\ }\href@noop {} {\  (\bibinfo {year} {2011})},\ \Eprint
  {http://arxiv.org/abs/1109.1835} {arXiv:1109.1835 [hep-ph]} \BibitemShut
  {NoStop}%
\bibitem [{\citenamefont {Ananthanarayan}\ and\ \citenamefont
  {Moussallam}(2004)}]{Ananthanarayan:2004qk}%
  \BibitemOpen
  \bibfield  {author} {\bibinfo {author} {\bibfnamefont {B.}~\bibnamefont
  {Ananthanarayan}}\ and\ \bibinfo {author} {\bibfnamefont {B.}~\bibnamefont
  {Moussallam}},\ }\bibfield  {title} {\enquote {\bibinfo {title} {{Four-point
  correlator constraints on electromagnetic chiral parameters and resonance
  effective Lagrangians}},}\ }\href {\doibase 10.1088/1126-6708/2004/06/047}
  {\bibfield  {journal} {\bibinfo  {journal} {JHEP}\ }\textbf {\bibinfo
  {volume} {0406}},\ \bibinfo {pages} {047} (\bibinfo {year} {2004})},\ \Eprint
  {http://arxiv.org/abs/hep-ph/0405206} {arXiv:hep-ph/0405206} \BibitemShut
  {NoStop}%
\bibitem [{\citenamefont {Kastner}\ and\ \citenamefont
  {Neufeld}(2008)}]{Kastner:2008ch}%
  \BibitemOpen
  \bibfield  {author} {\bibinfo {author} {\bibfnamefont {A.}~\bibnamefont
  {Kastner}}\ and\ \bibinfo {author} {\bibfnamefont {H.}~\bibnamefont
  {Neufeld}},\ }\bibfield  {title} {\enquote {\bibinfo {title} {{The
  $K_{\ell3}$ scalar form factors in the standard model}},}\ }\href {\doibase
  10.1140/epjc/s10052-008-0703-6} {\bibfield  {journal} {\bibinfo  {journal}
  {Eur.\ Phys.\ J.}\ }\textbf {\bibinfo {volume} {C57}},\ \bibinfo {pages}
  {541--556} (\bibinfo {year} {2008})},\ \Eprint
  {http://arxiv.org/abs/0805.2222} {arXiv:0805.2222 [hep-ph]} \BibitemShut
  {NoStop}%
\end{thebibliography}%

\end{document}